# Roadmap on STIRAP applications

Klaas Bergmann[1,28], Hanns-Christoph Nägerl[2], Cristian Panda[3], Gerald Gabrielse[3], Eduard Miloglyadov[4], Martin Quack[4], Georg Seyfang[4], Gunther Wichmann[4], Silke Ospelkaus[5], Axel Kuhn[6], Stefano Longhi[7], Alexander Szameit[8], Philipp Pirro[1], Burkard Hillebrands[1], Xue-Feng Zhu[9], Jie Zhu[10], Michael Drewsen[11], Winfried K Hensinger[12], Sebastian Weidt[12], Thomas Halfmann[13], Hailin Wang[14], G. S. Paraoanu[15], Nikolay V. Vitanov[16], J. Mompart[17], Th. Busch[18], Timothy J. Barnum[19], David D. Grimes[20,21,22], Robert W. Field[19], Mark G. Raizen[23], Edvardas Narevicius[24], Marcis Auzinsh[25], Dmitry Budker[26], Adriana Pálffy[27] and Christoph H. Keitel[27]

[1] Fachbereich Physik and Landesforschungszentrum OPTIMAS, Technische Universität Kaiserslautern, 67663 Kaiserslautern, Germany
[2] University of Innsbruck, Austria
[3] Harvard University, Northwestern University, USA
[4] Laboratorium für Physikalische Chemie, ETH Zürich, Switzerland
[5] Institut für Quantenoptik, Leibniz Universität Hannover, Germany
[6] Clarendon Laboratory, University of Oxford, UK
[7] Politecnico di Milano, IFN-CNR, Italy
[8] University of Rostock, Germany
[9] Huazhong University of Science and Technology, China
[10] The Hong Kong Polytechnic University, Hong Kong, China
[11] Aarhus University, Denmark
[12] Sussex Centre for Quantum Technologies, University of Sussex, Brighton, BN1 9QH, UK
[13] Institute of Applied Physics, Technical University of Darmstadt, Germany
[14] University of Oregon, USA
[15] Aalto University, Finland
[16] Faculty of Physics, St Kliment Ohridski University of Sofia, Bulgaria
[17] Universitat Autònoma de Barcelona, Bellaterra, Spain
[18] 2OIST Graduate University, Okinawa, Japan
[19] Department of Chemistry, Massachusetts Institute of Technology, Cambridge, USA
[20] Department of Chemistry and Chemical Biology, Harvard University, Cambridge, USA
[21] Department of Physics, Harvard University, Cambridge, USA
[22] Harvard-MIT Center for Ultracold Atoms, Cambridge, USA
[23] University of Texas at Austin, USA
[24] Weizmann Institute of Science, Israel
[25] University of Latvia
[26] Helmholtz Institute, Johannes Gutenberg University, Mainz, Germany and University of California, Berkeley, USA
[27] Max Planck Institute for Nuclear Physics, Saupfercheckweg 1, D-69117 Heidelberg, Germany

E-mail: bergmann@rhrk.uni-kl.de

[28] Guest editor of the Roadmap

# Abstract

STIRAP (Stimulated Raman Adiabatic Passage) is a powerful laser-based method, usually involving two photons, for efficient and selective transfer of population between quantum states. A particularly interesting feature is the fact that the coupling between the initial and the final quantum states is via an intermediate state even though the lifetime of the latter can be much shorter than the interaction time with the laser radiation. Nevertheless, spontaneous emission from the intermediate state is prevented by quantum interference. Maintaining the coherence between the initial and final state throughout the transfer process is crucial.

STIRAP was initially developed with applications in chemical dynamics in mind. That is why the original paper of 1990 was published in *The Journal of Chemical Physics*. However, as of about the year 2000, the unique capabilities of STIRAP and its robustness with respect to small variations of some experimental parameters stimulated many researchers to apply the scheme in a variety of other fields of physics. The successes of these efforts are documented in this collection of articles.

In Part A the experimental success of STIRAP in manipulating or controlling molecules, photons, ions or even quantum systems in a solid-state environment is documented. After a brief introduction to the basic physics of STIRAP, the central role of the method in the formation of ultra-cold molecules is discussed, followed by a presentation of how precision experiments (measurement of the upper limit of the electric dipole moment of the electron or detecting the consequences of parity violation in chiral molecules) or chemical dynamics studies at ultra-low temperatures benefit from STIRAP. Next comes the STIRAP-based control of photons in cavities followed by a group of three contributions which highlight the potential of the STIRAP concept in classical physics by presenting data on the transfer of waves (photonic, magnonic and phononic) between respective wave guides. The works on ions or ion-strings discuss options for applications e.g. in quantum information. Finally, the success of STIRAP in the controlled manipulation of quantum states in solid-state systems, which are usually hostile towards coherent processes, is presented, dealing with data storage in rare-earth ion doped crystals and in NV-centers or even in superconducting quantum circuits. The works on ions and those involving solid-state systems emphasize the relevance of the results for quantum information protocols.

Part B deals with theoretical work including further concepts relevant for quantum information or invoking STIRAP for the manipulation of matter waves. The subsequent articles discuss experiments underway to demonstrate the potential of STIRAP for populating otherwise inaccessible high-lying Rydberg states of molecules, or controlling and cooling the translational motion of particles in a molecular beam or the polarization of angular momentum states. The series of articles concludes with a more speculative application of STIRAP in nuclear physics which, if suitable radiation fields become available, could lead to spectacular results.

# Contents

## A EXPERIMENTS

### A1 Introduction

### A2  Molecules

### A3  Photons, Magnons and Phonons

# A EXPERIMENTS

## A1 Introduction

### A1.1 Basics of the STIRAP process

Klaas Bergmann

Fachbereich Physik der Technischen Universität Kaiserslautern and OPTIMAS Research Center

**Status**

The basic task of STIRAP (stimulated Raman adiabatic passage) is to transfer population within a quantum system efficiently and selectivity from a state 1 to an initially unpopulated state 3. This is done by means of a two-photon process involving the radiation fields P and S (see Figure 1). The dipole coupling is provided via an intermediate state 2. However, if STIRAP is properly implemented, state 2 is never populated and thus no losses because of spontaneous emission or other decay processes occur. The frequencies of the P- or S-field are tuned to resonance (Δ = 0), or near resonance, with the respective one-photon transitions. It is essential to maintain two-photon resonance (δ = 0) throughout the process. A characteristic, at first glance most surprising, features is the fact that the coupling of the unpopulated states 2 and 3 by the S-field begins earlier than the coupling of states 1 and 2 by the P-field. Furthermore, the S-field interaction also ends prior to the P-field interaction (see Fig. 2). Because STIRAP is an adiabatic process constraints regarding the time-evolution of the intensity of the S- and P-fields apply.

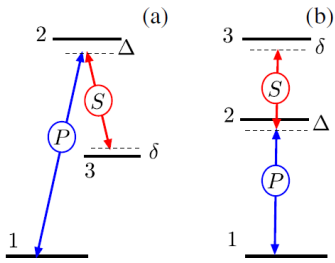

Fig. 1: Typical three-level STIRAP linkage pattern in form of (a) a lambda-system and (b) a ladder system. For STIRAP to be successful the two-photon resonance, i.e. δ = 0, must be maintained throughout the process. In most cases, STIRAP works best for Δ = 0.

STIRAP was initially developed for studying chemical dynamics of small molecules in highly excited vibrational levels of the electronic ground state. That is why the original paper [1] was published in The Journal of Chemical Physics. The subsequent early theoretical work of Peter Zoller et.al. [2] alerted in particular the quantum optics community to the potential of the method for manipulating population distributions over quantum states. It was only after about the year 2000 that a broader audience took notice and STIRAP was introduced in many different areas, as documented in this roadmap article. The development of the field until 2016 is reviewed in [3 – 6].

Some very basic features of the STIRAP process are summarized in what follows. The coupling strength is given by the Rabi frequencies

$$\Omega_P(t) = -\, d_{12}\, E_P(t)/\hbar \quad \text{and} \quad \Omega_S(t) = -\, d_{13}\, E_S(t)/\hbar \qquad (1)$$

where $d_{ik}$ is the transition dipole moment for the transition between states i and k and $E_{P,S}$ is the respective electric field. The Hamiltonian in the rotating wave approximation, which is valid in most cases of interest, reads

$$H(t) = \hbar \begin{bmatrix} \mathbf{0} & \frac{1}{2}\,\Omega_P(t) & \mathbf{0} \\ \frac{1}{2}\,\Omega_P(t) & \Delta & \frac{1}{2}\,\Omega_S(t) \\ \mathbf{0} & \frac{1}{2}\,\Omega_S(t) & \delta \end{bmatrix} \qquad (2)$$

The three eigenvalues of this Hamiltonian for δ = 0 are

$$\varepsilon_\pm(t) = \tfrac{1}{2}\,[\Delta \pm \sqrt{\Delta^2 + \Omega_{rms}(t)^2}\,]\quad \text{and} \quad \varepsilon_0 = 0 \qquad (3)$$

with $\Omega_{rms}(t) = \sqrt{\Omega_P(t)^2 + \Omega_S(t)^2}$, while the eigenstates $\Phi_j$ (the adiabatic states or "dressed states") are linear superpositions of the eigenstates $\psi_k$, k = 1, 2 and 3, of the bare quantum system:

$$\Phi_+(t) = \psi_1 \sin\vartheta(t)\sin\varphi(t) + \psi_2 \cos\varphi(t) + \psi_3 \cos\vartheta(t)\sin\varphi(t)$$

$$\Phi_0(t) = \psi_1 \cos\vartheta(t)\; -\; \psi_3 \sin\vartheta(t) \qquad \text{(the dark state)} \qquad (4)$$

$$\Phi_-(t) = \psi_1 \sin\vartheta(t)\cos\varphi(t) - \psi_2 \sin\varphi(t) + \psi_3 \cos\vartheta(t)\cos\varphi(t)$$

The mixing angles $\vartheta(t)$ or $\varphi(t)$ are given by the *ratio* of the coupling strengths or the *ratio* of coupling strength $\Omega_{rms}$ and detuning Δ, respectively, according to

$$\tan\vartheta(t) = \frac{\Omega_P(t)}{\Omega_S(t)} \quad \text{and} \quad \tan 2\varphi(t) = \frac{\Omega_{rms}(t)}{\Delta} \qquad (5)$$

For on-resonance tuning, Δ = 0, we have φ = 45°. According to eq. (4) the evolution of the adiabatic states can be visualized as the rotation of vectors $\Phi_j(t)$ in the Hilbert space spanned by the $\psi_k$. When $\Omega_P = 0$ while $|\Omega_S| > 0$ we have $\vartheta = 0$ and thus $\Phi_0 \equiv \psi_1$. When $\Omega_S = 0$ while $|\Omega_P| > 0$ we have $\vartheta = 90°$ and thus $\Phi_0 \equiv \psi_3$. If this "rotation" is adiabatic, i.e. slow enough, $\Phi_0$ remains in the plane spanned by $\psi_1$ and $\psi_3$ and thus never acquires a component of the "leaky" state $\psi_2$.

The evolution is exactly adiabatic only for a smooth, infinitely long interaction time. Thus, any transfer process completed in a finite time, will be accompanied by non-adiabatic coupling from the states $\Phi_0$ to the states $\Phi_\pm$ which do include a component of the leaky state $\psi_2$. However, such diabatic coupling is negligibly small when the radiative coupling is strong enough as given by:

$$\Omega_{rms}(t) \gg \left| \frac{d\vartheta}{dt} \right| = \frac{\Omega_S(t)\,d\Omega_P(t)/dt - \Omega_P(t)\,d\Omega_S(t)/dt}{\Omega_P(t)^2 + \Omega_S(t)^2} \qquad (6)$$

Equation (6) is called the "local" adiabatic conditions as it allows evaluating the r.h.s at any time t during the transfer process. For pulses with a smooth envelop it often suffices to look at the "global" adiabatic condition. Integration over the entire duration T of the transfer process, where T is the duration for which we have both $\Omega_P > 0$ *and* $\Omega_S > 0$, leads to

$$\Omega_{rms} T \gg 1 \qquad (7)$$

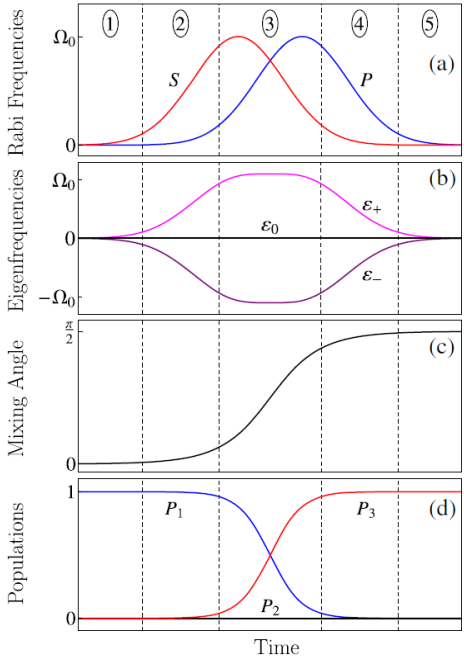

Fig. 2: (a) Typical variation of the S- and P-laser intensity at the location of the quantum system with the related variation of (b) the eigenvalues for on-resonance tuning ($\Delta = 0$), (c) the mixing angle and (d) the population of the three states. In interval (1), we have $\Omega_S > 0$ but $\Omega_P = 0$, the separation of $\varepsilon_\pm$ is given by the Autler-Towns splitting driven by $\Omega_S$. In interval (2) we have $\Omega_S \gg \Omega_P$ with the absorption of the P-field prevented through the S-field by the phenomena called electromagnetically induced transparency (EIT) [7]. The population transfer occurs in interval (3), where we have $d\Omega_S/dt < 0$ and $d\Omega_P/dt > 0$. The intervals (4) and (5) repeat the processes of (2) and (1), respectively, however with the role of $\Omega_S$ and $\Omega_P$ interchanged.

The above discusses STIRAP in the time domain with complete population transfer in mind. There is, however, also significant interest in an approach, called "fractional STIRAP", which leads to an only partial population transfer and results in a coherent superposition of the wave-function amplitudes of the initial and final states. Implementing fractional STIRAP requires controlling the evolution of the mixing angle $\vartheta(t)$. Rather than designing the pulses $\Omega_{S,P}(t)$ such that $\vartheta(t \to \infty)$ approaches $\pi$, see eq. 5, the pulses are shaped to have the ratio $R_{S,P}(t) = \Omega_S(t) / \Omega_P(t)$ reach a constant value towards the end of the interaction, i.e. $\vartheta(t \to \infty) = \vartheta_{fS} < \pi$. The numerical value of $\vartheta_{fS}$, and thus the fraction of population which reaches the final state, is given by $R_{S,P}(t \to \infty)$.

It shall also be noted in passing, that the previous discussion in the time domain can be mapped easily onto the spatial domain when invoking propagation. One example is particles in a directed beam crossing suitably spatially displaced laser beams with the intensity (and profile) of the latter being constant in time. Another example is the propagation of e.g. light, spin- or even acoustic or matter waves along a set of three suitable wave guides with the strength of the coupling between them arranged STIRAP-like. Related experiments (or proposals) are discussed in Sections A 3.2 – A 3.4 and B 2.1, respectively.

## Current and Future Challenges

Although by now the STIRAP method is well established and widely applied, certain challenges remain: For many applications a transfer efficiency of the order of 90% suffices. However, when an efficiency approaching 100% is wanted, the adiabatic condition requires that the radiation intensity is sufficiently high and/or the interaction period is sufficiently long (see eq. 7). In any case, the spectral linewidth must be strictly transform-limited, which requires a smooth variation of the laser intensity over time. Depending on the requirements for a specific process, a spectral linewidth of 1 kHz or even less may be needed. In other words: although the absolute phase difference between the S- and P-field does not matter, it is essential that phase fluctuations during the interaction time with the quantum system are negligibly small.

Meeting the requirement for adiabatic evolution may call for pulse-amplification of cw radiation which, however, often result in a chirped laser pulse. A frequency chirp during the evolution of the pulse is detrimental because it prevents maintaining the two-photon resonance ($\delta = 0$) unless the chirp of one of the S- or P-pulses is exactly compensated by the chirp of the other pulse.

Application of STIRAP to molecules or other quantum systems with a high energy-level density remains a challenges because nearby states will cause Stark-shifts that vary with the laser intensity and thus prevent maintaining two-photon resonance. In these cases low laser intensity is wanted. As a consequence, the adiabatic condition demands long pulses, imposing stringent limits on the laser linewidth. The latter must be much smaller than the level spacing.

An early experimental study of the consequences of high density of energy levels can be found in [8]. The largest molecule to which STIRAP has so far been successfully applied is $SO_2$ [9]. A detailed theoretical study of the variation of the transfer efficiency with level density is presented in [10].

The lack of (near) transform limited pulses of sufficient intensity in the UV or VUV region of the spectrum has so far prevented application of STIRAP to molecules such as $H_2$, $O_2$, or $N_2$ (see also the appendix of [5]).

## Advances in Science and Technology to Meet Challenges

Applications of STIRAP will benefit from the further advances in laser technology and schemes for laser stabilization and linewidth reductions. Whether the coherence properties of new accelerator-based light sources, in particular free electron lasers, providing radiation over a very large range of the spectrum (including light in the VUV and even shorter wavelength region) are suitable for STIRAP remains to be seen.

## Concluding Remarks

This brief introduction to STIRAP mentions only the very basic aspects and equations. The many extension to e.g. multilevel system and the preparation of coherent superposition states are discussed in detail in the referenced review articles. Most importantly the following articles highlight the many, very diverse and promising applications of the STIRAP process.

## Acknowledgments

The development of the STIRAP method in the Bergmann-group at Kaiserslautern benefitted a lot from the contribution of graduate students, postdocs and visiting scientists, among those (to mention only some) are Uli Gaubatz, Axel Kuhn, Jürgen Martin, Heiko Theuer, Frank Vewinger, George Coulston, Piotr Rudecki, Bruce W. Shore, Razmik Unanyan, Leonid Yatsenko, Mathew Fewell and Nikolay Vitanov.

# A2 Molecules

## A2.1 STIRAP in Ultracold Molecule Formation

*Hanns-Christoph Nägerl*
University of Innsbruck

### Status

Trapped samples of ultracold molecules in the regime of quantum degeneracy present a novel platform for exploring various phenomena in quantum many-body and condensed-matter physics. In particular, polar molecules with their strong, long-range, and anisotropic electric dipole-dipole interaction promise the realization of various strongly correlated quantum phases together with the corresponding quantum phase transitions. A multitude of interacting spin models has been proposed, some of which might shed light on the fundamental question of the mechanism that gives rise to high-temperature superconductivity. Others can be used for quantum simulation and quantum computation purposes.

The majority of the proposals to realize novel types of quantum matter on the basis of polar molecules requires full quantum control over the external (motional) *and* internal (ro-vibronic) degrees of freedom of the molecules. For atomic species, full control is possible, essentially by combining laser cooling and optical pumping, and this has led to spectacular achievements such as the formation of gaseous atomic Bose-Einstein condensates (BEC) and atomic degenerate Fermi gases. For molecules, similar matter-wave control is much harder to achieve, if at all. Molecules are not readily laser cooled, largely due to optical pumping into unwanted ro-vibrational states, and whether forced evaporative cooling will work on ultracold molecules is an open question.

For dimer molecules, however, a powerful preparation strategy has been developed: Laser cooled atoms are associated to molecules in a controlled way by means of a Feshbach scattering resonance [1] near the collision threshold, and the newly-formed weakly-bound molecules are transferred to lower-lying states and even to the ro-vibronic ground state by STIRAP with very high efficiency [2-8]. The central idea is to achieve maximum phase-space density for the initial atomic samples (usually by first laser cooling and subsequently evaporatively cooling the atomic sample), and then to maintain this density by employing a coherent and efficient transfer process such as STIRAP. In a lambda-type configuration, STIRAP has the favorable property that it connects the weakly-bound initial molecular state close to the atomic threshold (called Feshbach molecule) to the deeply-bound molecular target state without populating the electronically excited state that is needed for state coupling but which is likely to decay due to spontaneous emission. It is ideally suited for maintaining full quantum control over the external and internal degrees of freedom of the molecules.

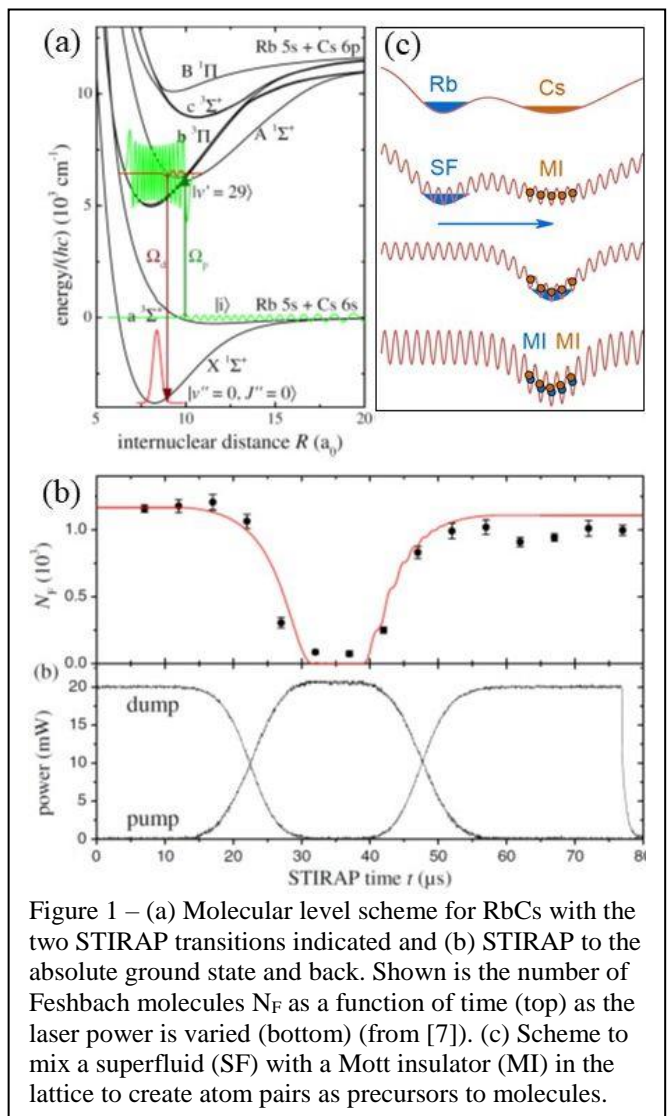

Figure 1 – (a) Molecular level scheme for RbCs with the two STIRAP transitions indicated and (b) STIRAP to the absolute ground state and back. Shown is the number of Feshbach molecules $N_F$ as a function of time (top) as the laser power is varied (bottom) (from [7]). (c) Scheme to mix a superfluid (SF) with a Mott insulator (MI) in the lattice to create atom pairs as precursors to molecules.

STIRAP on ultracold molecules was first tested with homonuclear molecules such as $Rb_2$ [2] and $Cs_2$ [3] about 12 years ago. Given a quantum degenerate bulk gaseous sample of Rb or Cs, up to 30% of the atoms could typically be converted into Feshbach molecules by sweeping the magnetic field over the magnetically induced Feshbach resonance [1]. These molecules were then STIRAPed into the ro-vibronic ground state. Almost simultaneously, STIRAP was demonstrated on ultracold heteronuclear, polar molecules [4], specifically on KRb. By choosing the fermionic isotope $^{40}K$ (for Rb, one only has the option of a bosonic isotope such as $^{87}Rb$), the molecule is a fermion. Other combinations have followed, such as bosonic RbCs [6,7] (Fig.1), fermionic NaK, and bosonic NaRb. STIRAP transfer efficiencies to the molecular absolute ground state have exceeded 90%. By virtue of the initial atomic state preparation, the molecules are prepared in a specific hyperfine sublevel. The samples are very cold, i.e. significantly colder than 1 µK, and their spatial densities are comparatively high, i.e. on the order of $10^{11}$ molecules/cm$^3$ for trapped bulk samples. However, even though the phase-space densities were high, it became soon clear that entering the regime of quantum degeneracy was not an easy exercise.

## Current and Future Challenges

Creating a quantum degenerate sample of interacting gaseous particles in thermal equilibrium requires that the elastic collisional properties dominate over the inelastic ones. For ultracold atoms, two-body hyperfine relaxation is weak or can even be avoided by preparing the sample in the absolute internal (hyperfine) ground state [1]. Higher-order processes, such as three-body relaxation, are sufficiently slow, and their detrimental effects can be held at reasonable levels by lowering the spatial density. Atomic BECs are thus stable on the timescale of 10 s and beyond. However, molecules have additional channels for relaxation. For example, vibrational and rotational relaxation occurs at universal rates when the molecules are not prepared in their ro-vibrational ground state. Immediately after formation by means of the Feshbach resonance, the molecules are prone to relaxation due to collisions with atoms or with other newly-formed molecules. And even in their very (hyperfine) ground state the molecules can potentially be subject to two-body loss due to the transient formation of comparatively long-lived two-molecule complexes.

Stabilizing the molecular samples can be done in various ways. Fermionic molecules naturally avoid collisions due to the Pauli exclusion principle. Alternatively, the molecules can be shielded from close encounter by localizing them at the wells of a three-dimensional (3D) lattice potential. The molecules can then still interact via their long-range dipole-dipole interaction. Confining dipolar molecules to lower dimensions so that the molecules are only allowed to collide with a specific orientation of their dipoles is another option for achieving stability. Finally, mixing-in of higher rotational states that experience repulsive potentials upon encounter via microwave excitation is a further possibility that could work for some of the molecular species that are or soon will be available in the laboratory.

Experiments to stabilize the ground-state molecules and, in particular, to shield them from collisions during the initial formation phase and during the STIRAP process started early on with the work on homonuclear Cs$_2$ [4]. In that work, a BEC of Cs atoms was loaded into a 3D lattice potential and the superfluid-to-Mott-insulator quantum phase transition was driven to create a maximally filled two-atom Mott shell. Given the harmonic confinement of the samples, nearly 50% of the atoms in the BEC could be converted into Feshbach molecules, and about 80% of these (unpublished) could be transferred into the ro-vibrational ground state by a 4-photon STIRAP process, creating a very cold and very dense sample of 20.000 individually trapped homonuclear ground-state molecules.

The next crucial step was (and still is) to generalize this method to the case of binary atom mixtures. For the case of KRb, a fermionic band insulator and a one-atom-per-site Mott insulator were mixed in the presence of a 3D lattice near a zero crossing for the interspecies interaction strength in the vicinity of a Feshbach resonance to create K-Rb double occupancies at the wells of the lattice [9]. Subsequent STIRAP created a sample of dipolar KRb molecules with a 30% filling fraction in the lattice. By means of a similar procedure, namely by mixing a Rb superfluid and a Cs Mott insulator (Fig.1(c)), our group in Innsbruck has been able to create more 5000 RbCs Feshbach molecules with a filling fraction close to 40% [10]. This dense sample yet awaits STIRAP transfer to the absolute ground state. As a recent highlight, the group at JILA has been able to put nearly 50000 KRb molecules into the absolute ground state, creating a quantum degenerate Fermi gas of ground-state molecules [11].

The formation of quantum degenerate samples of molecules by means of the atom-association technique demands that every step is taken at maximum efficiency. It is desirable to push the STIRAP efficiencies towards 100%. Due to the weak Franck-Condon overlap between the molecular states that are involved and given the finite excited-state lifetime one is forced to use comparatively long STIRAP pulse times. This in turn requires lasers with high phase stability. In order to maintain the Raman resonance condition one has to assure that random level shifts for the initial and final molecular states (e.g. due to fluctuating magnetic fields) remain low.

## Advances in Science and Technology to Meet Challenges

STIRAP in ultracold molecules will benefit from various improvements on the laser sources and the experimental setups. Phase noise in e.g. diode lasers must be reduced to levels that are much below to what is usually routine in high-resolution spectroscopy experiments. One possibility is to use the transmitted (filtered) light of a high-finesse cavity. Novel high-bandwidth electro-optic modulators in addition allow for high-bandwidth servo loops. In order to achieve high molecular densities, e.g. in the presence of a 3D lattice, the lattice filling with heteronuclear atom pairs as precursors to dimer molecules has to be pushed towards unity. One option to avoid atomic three-body losses would be to merge parallel planar samples of atoms along the out-of-plane direction.

## Concluding Remarks

STIRAP is the enabling technology to create molecular samples in the regime of quantum degeneracy. It in particular allows the creation of quantum gases of dipolar molecules. Exciting times are ahead since molecular densities are now high enough so that strong dipolar interaction effects can be observed.

**Acknowledgments –** The author would like to acknowledge important contributions to the Innsbruck STIRAP experiments by, in particular, Johann Danzl and Tetsu Takekoshi.

### A2.2 STIRAP and the precise measurement of the electron's electric dipole moment

*Cristian Panda and Gerald Gabrielse*
Harvard University, Northwestern University

### Status

Precision measurements of the electric dipole moment of the electron (eEDM), $d_e$, probe physics beyond the standard model [1,2]. To date, no electric dipole moments of fundamental particles have been observed. The standard model predicts $d_e < 10^{-38}$ e cm [3], many orders of magnitude below current experimental bounds. However, theories beyond the standard model commonly posit the existence of massive particles, which can interact with the electron and can give rise to an eEDM slightly smaller than current experimental bounds [4,5]. These theories attempt to answer questions that arise from cosmological observations, such as the nature of dark matter and why matter dominates over antimatter throughout the Universe. In addition, a nonzero value of the eEDM is a source of violation of parity and time-reversal symmetry [6], which could explain the cosmological matter-antimatter asymmetry. eEDM searches with increased sensitivity continue to probe deeper into the energy range available for new particles, typically at or above the TeV energy scale [4,5].

Searches for the eEDM are designed to measure a small energy shift, $U = -d_e \mathcal{E}$, which occurs due to the interaction between $d_e$ and an electric field, $\mathcal{E}$. Recent experimental efforts have greatly improved sensitivity by using the extremely high internal electric field ($\mathcal{E}_{\text{eff}}$) of heavy polar molecules [6]. Additional degrees of freedom inherently present in molecules allow for states with opposite parity that are closely spaced in energy. Molecules with this type of structure can be fully spin aligned in small laboratory electric fields of a few V/cm, which typically enhances the achievable $\mathcal{E}_{\text{eff}}$ by $2 - 3$ orders of magnitude above atomic systems. The eEDM is typically extracted by measuring the difference in energy between states where the molecule polarization is aligned and anti-aligned with the applied electric field.

The ACME Collaboration improved the eEDM limit by two orders of magnitude over previous experiments by measuring spin precession in the $H^3\Delta_1$ state of thorium monoxide (ThO) [1,8]. ThO provides $\mathcal{E}_{\text{eff}} = 78$ GV/cm [9,10], three orders of magnitude larger than typically achieved with atoms. The $H$ state has a low magnetic moment $\mu_H = 0.004 \, \mu_B$ (where $\mu_B$ is the Bohr magneton), which reduces sensitivity to residual ambient magnetic fields. In addition, the lifetime of the $H$ state is longer than the typical transit time of the molecular beam through the experimental apparatus.

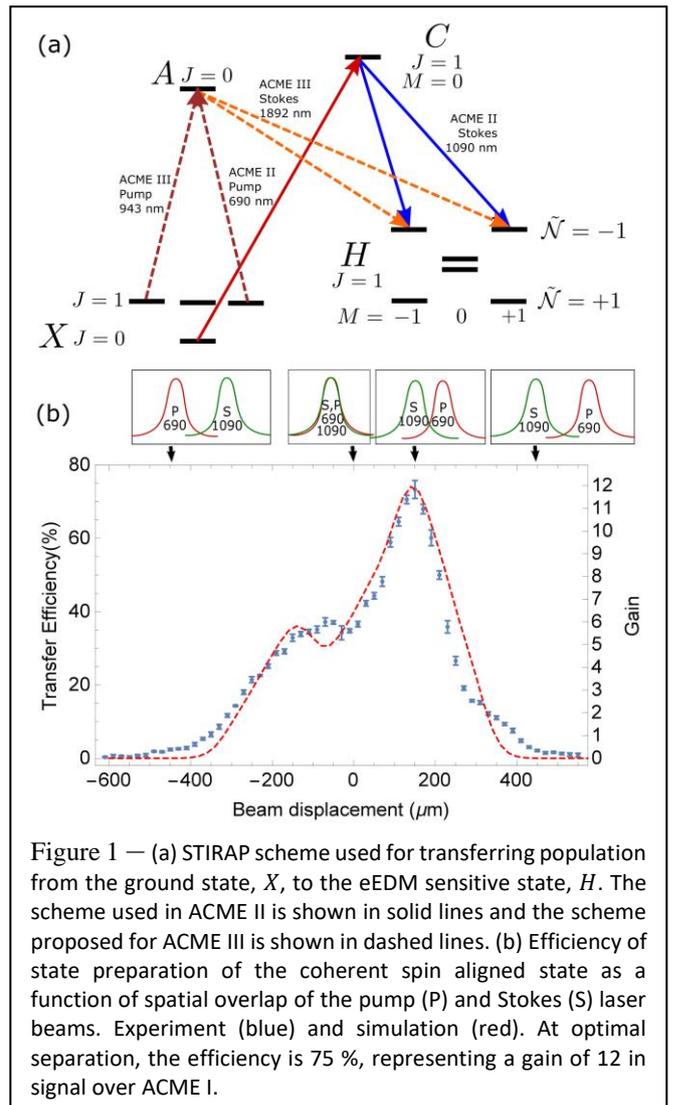

Figure 1 — (a) STIRAP scheme used for transferring population from the ground state, $X$, to the eEDM sensitive state, $H$. The scheme used in ACME II is shown in solid lines and the scheme proposed for ACME III is shown in dashed lines. (b) Efficiency of state preparation of the coherent spin aligned state as a function of spatial overlap of the pump (P) and Stokes (S) laser beams. Experiment (blue) and simulation (red). At optimal separation, the efficiency is 75 %, representing a gain of 12 in signal over ACME I.

The reasons that make the ThO molecule a good laboratory for performing the eEDM measurement also make it challenging to use experimentally. The multitude of molecular quantum states and complex coupling increases the challenge of precise quantum control. In the ACME experiment, ThO molecules are produced by a cryogenic buffer gas beam source. The molecular population follows a Boltzmann distribution with a temperature of 4 K and therefore resides in the ground electronic state, $X$. For the population to be useful for the eEDM measurement, it first requires transfer to the $H$ state (Fig. 1a). Standard transfer schemes, such as optical pumping, suffer from the fact that the intermediary state can decay to multiple lower energy states, which leads to low transfer efficiency.

The first-generation ACME I experiment used an optical pumping scheme to prepare the dark state required for the spin precession measurement with only 6 % efficiency [8]. In ACME II, we implemented STIRAP, which offers direct preparation of the coherent superposition [2]. Using STIRAP, we have achieved 75 % efficiency, which increased our signal by a factor of 12 (Fig. 1b). This along with other improvements allowed ACME II to measure the eEDM with an order of magnitude lower uncertainty [1] than the previous best result, obtained in ACME I [8].

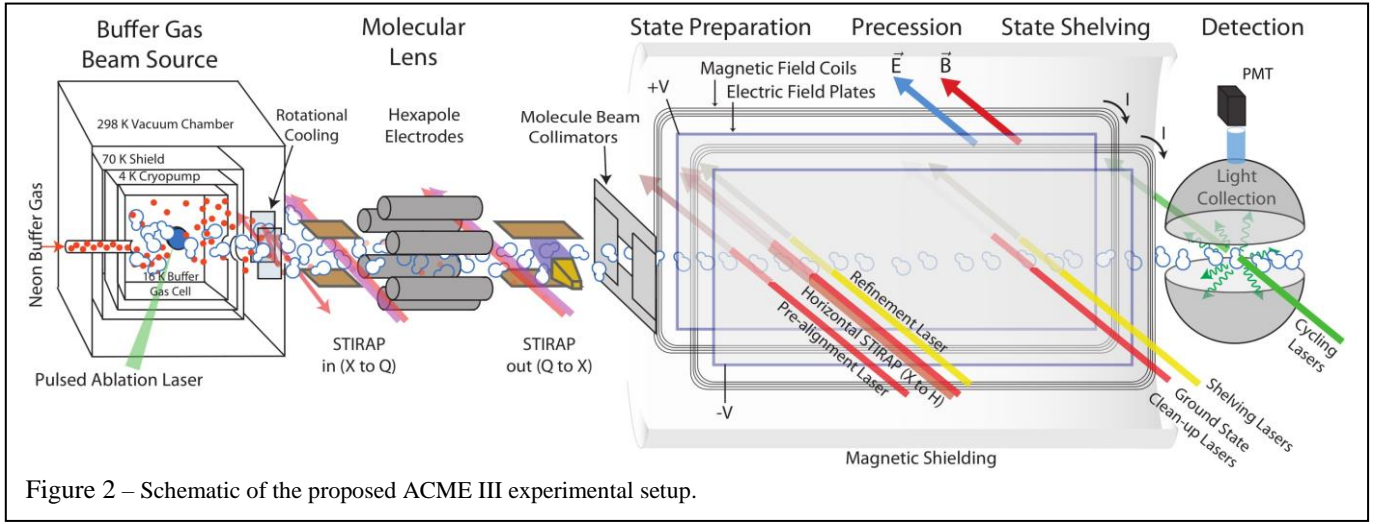

Figure 2 – Schematic of the proposed ACME III experimental setup.

**Current and Future Challenges**

In the ACME experiment, we transfer molecular ensembles with large phase-space distributions via weak molecular transitions. Satisfying the two-photon resonance condition and adiabatic criterion requires high intensity and smooth laser beam intensity profiles. The phase coherence of the lasers must be carefully controlled to achieve highly efficient STIRAP. Furthermore, STIRAP must be integrated well into the rest of the ACME apparatus, where optical access is limited by the presence of magnetic shielding, magnetic coils and electric field plates. The STIRAP optical setup and laser systems must be sufficiently robust to allow reliable day-to-day operation for the typical 1-2 year experiment runs. We have met such challenges in ACME II by using robust external cavity diode lasers and mechanically robust laser beam shaping optics.

Performing STIRAP in ACME II by sending laser light in between the parallel field plates proved susceptible to noise that we reduced using a refinement optical pumping laser beam. In addition, that geometry fixed the orientation of the prepared spin-aligned state along an axis that is perpendicular to both the laser propagation and electric field directions. We can recover the ability to freely rotate the orientation of the spin-aligned state by sending the lasers through the field plates. Doing so allows us to reduce one of the largest systematic effects in ACME I and ACME II, which was related to residual birefringence in the glass the state preparation lasers pass through.

Sending the STIRAP lasers through the field plates will require the use of lower intensity than used in ACME II to avoid damage of the indium tin oxide (ITO) coating. We are currently investigating new intermediary states that have stronger coupling to $X$ and $H$. A lead option is $X - A - H$ (see *Fig. 1*), where our measurements suggest that less than 200 mW of power on each leg (943 nm for $X - A$ and 1892 nm for $A - H$) will be sufficient to saturate the STIRAP transition, low enough to avoid significantly heating the ITO coating.

**Advances in Science and Technology to Meet Challenges**

We are also investigating using STIRAP in future ACME sensitivity upgrades (Fig. 2). In ACME II, only about 0.04 % of the produced molecules reached the detection area and were read out. We expect that guiding the molecules with an electrostatic lens will increase this fraction by more than an order of magnitude. We have explored this option by using the ground electronic state, $X$, where the focusing force is relatively weak due to the quadratic Stark shift. The metastable electronic state $Q^3\Delta_2$ has a linear Stark shift that makes it better suitable for a more efficient lens. Population will be transferred via STIRAP from $X$ to $Q$ before the lens and then back from $Q$ to $X$ after the lens via the $C$ state. We have demonstrated efficiencies above 90 % in a test setup. We are preparing tests of STIRAP in the context of the electrostatic lens, where significant fringing field gradients could affect the STIRAP transfer, but simulations promise minimal reduction in transfer efficiency.

**Concluding Remarks**

STIRAP preparation of the spin-aligned eEDM state was one of the leading upgrades in ACME II, which achieved an order of magnitude improved limit of the eEDM. This could be improved in the future by using a new intermediary state that would require lower intensity lasers. The future ACME III measurement (see Fig. 2) could use STIRAP to transfer population in and out of the molecular lensing $Q$ state. All these experimental steps will be implemented using robust, low phase noise, diode laser systems. Other ACME III improvements include better photon collection efficiency by using optical cycling and longer interaction time to take advantage of the long lifetime of the $H$ state. These improvements promise to allow ACME III to perform a new measurement of the eEDM with another order of magnitude improved precision.

**Acknowledgments –** The work described here was performed as part of the ACME Collaboration, to whom we are grateful for its contributions, and was supported by the NSF.

# Precision experiments for parity violation in chiral molecules: The role of STIRAP

Eduard Miloglyadov, Martin Quack, Georg Seyfang, Gunther Wichmann, Physical Chemistry, ETH Zürich, CH-8093 Zürich, Switzerland, Martin@Quack.ch

## Status

Precision experiments measuring the extremely small energy difference $\Delta_{pv}E$ between the enantiomers of chiral molecules, predicted to be on the order of 100 aeV to 1 feV (depending on the molecule) are among the greatest challenges in physical-chemical stereochemistry relating also to the Standard Model of particle physics (SMPP) [1-7]. So far, no successful experiments have been reported and it turns out that STIRAP [8] may contribute importantly in enabling such precision experiments.

Following the discovery of parity violation in nuclear and particle physics in 1956/57 it has been surmised qualitatively, that the ground state energies of the enantiomers of chiral molecules are slightly different, as are also their absorption frequencies in the infrared or other spectral ranges. Thus parity violation is of fundamental importance for our understanding of the structure and dynamics of chiral molecules with potential implications for the evolution of biomolecular homochirality, which has been an enigma of stereochemistry for more than a century (for in depth reviews see [1-4]). A scheme of how to measure the parity violating energy difference $\Delta_{pv}E$ between the ground state of the enantiomers of chiral molecules (and also $\Delta_{pv}E^*$ between corresponding excited rovibronic states) has been proposed in 1986 [5]. At that time, however, the spectroscopic ground work of high-resolution analyses of rovibronic spectra of chiral molecules was not available and appeared very difficult. Also theories available at that time were qualitatively incorrect, predicting values too low by a factor 100 for typical prototype molecules (see review [3]), and the proposed experiment on $\Delta_{pv}E$ appeared correspondingly almost impossible.

The situation has changed importantly over the last decades. Theoretical approaches developed by many groups over the last decades following our discovery in 1995 of the new orders of magnitude agree to converge today to the larger values. At the same time, there has been substantial progress in the high-resolution spectroscopy of chiral molecules, as well as in laser technology enabling efficient and selective population transfer making therefore the observation of $\Delta_{pv}E$ a realistic goal for our current experiments (see [3,6]).

The basic experimental scheme is shown in Fig. 1. In brief, in a first step laser excitation of a chiral molecule (either enantiopure or simply from a racemate) leads to a preparation of an excited rovibronic state of well defined parity, which is therefore achiral. Such a state can either arise from an excited electronic state with an achiral (for instance planar) equilibrium structure or it can arise from rovibrational tunneling sublevels in the electronic ground state, which are near or above the potential barrier for interconversion between the enantiomers. Such tunneling sublevels can therefore satisfy the condition that the tunneling splitting $\Delta E_{\pm}^{*}$ between sublevels of well defined parity in that excited state is much larger than $\Delta_{\mathrm{pv}}E^{*}$ ,

$$\Delta E_{\pm}^{*} >> \Delta_{\mathrm{pv}}E^{*} \qquad (1)$$

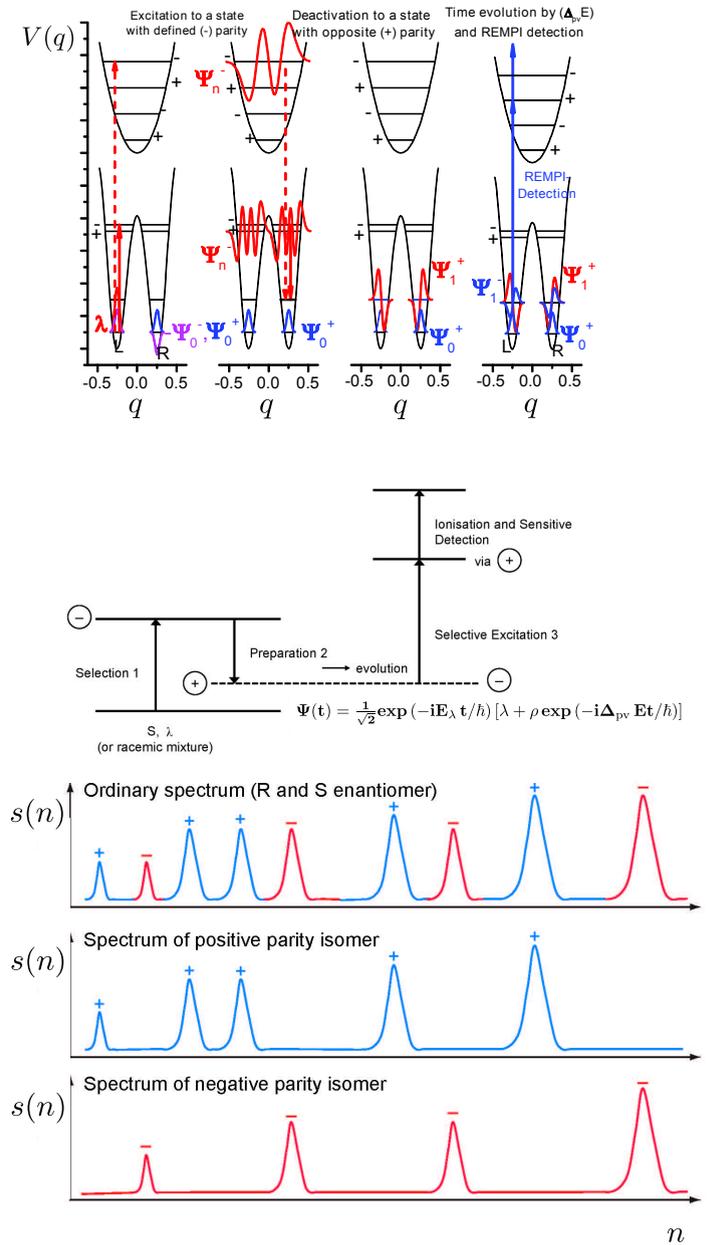

Figure 1: Scheme of the preparation and detection steps for the time resolved experiment to measure $\Delta_{pv}E$. *Top*: The transitions to the intermediate states are indicated together with the corresponding wave functions for an excited state with well defined parity close to the barrier of a double minimum potential (full line) or an achiral electronically excited state (dashed line) as an intermediate. The right hand part shows the sensitive detection step with REMPI. *Middle*: Summary scheme for the three steps. *Bottom*: The spectra of the normal enantiomers (top) and of the selected positive (blue) and negative (red) parity isomers (modified after [4-6]). n is a reduced frequency difference $(\nu - \nu_0)/\nu_0$, where the frequency spacings between lines are of the order of MHz in order to separate lines connecting states of different parity (+ or -) in the rovibronic, resolved spectrum. The high resolution (Hz to subHz) needed for $\Delta_{pv}E$ is obtained from measuring the time evolution of the spectrum in the middle towards the spectrum at the bottom at very high sensitivity on the ms time scale.

This allows then for a spectroscopic selection of states of well defined parity. In a second step in the scheme of Fig. 1 one prepares a state of well defined parity in the ground state (or some other low energy state), which satisfies the condition

$$\Delta_{\mathrm{pv}}E >> \Delta E_{\pm} \qquad (2)$$

The parity selection arises from the electric dipole selection rule connecting levels of different parity. Thus, if in the first step one has selected a state of some given parity, in the second step one prepares a state of the opposite parity. Such a state is a superposition of the energy eigenstates of the two enantiomers separated by $\Delta_{\mathrm{pv}}E$ and therefore shows a periodic time evolution with a period

$$\tau_{\mathrm{pv}} = \frac{h}{\Delta_{\mathrm{pv}}E} \qquad (3)$$

In such a state parity evolves in time due to parity violation and parity is not a constant of the motion. The probability of finding a given parity ($p^{+}$ for positive parity and $p^{-}$ for negative parity) is given by equation (4)

$$p^{-}(t) = 1 - p^{+}(t) = \sin^{2}\left(\frac{\pi \Delta_{\mathrm{pv}}E t}{h}\right) \qquad (4)$$

In the third step the initially 'forbidden' population of negative parity $p^{-}(t)$ is probed very sensitively, for example by resonantly enhanced multiphoton ionization (REMPI). This is possible, because the line spectra of positive and negative parity isomers are different, with lines that are well separated at high resolution (Fig. 1 and Ref. [7]). In this fashion it is not necessary to wait for a whole period, but it is sufficient to probe the initial time evolution at very early times. The sensitivity in the probe step determines in essence, how small a value of $\Delta_{\mathrm{pv}}E$ can be measured. In a recent test experiment with a current experimental set up in our laboratory on the achiral molecule ammonia, $NH_3$, it was estimated that an energy difference as small as 100 aeV should be measurable with the existing current experiment.

The original proposal of 1986 [5] preceded the invention of STIRAP [8], and therefore assumed population transfer using pulse shaping or chirp by rapid adiabatic passage. It is clear, however, that STIRAP is an ideal technique for generating population transfer in a well controlled fashion.

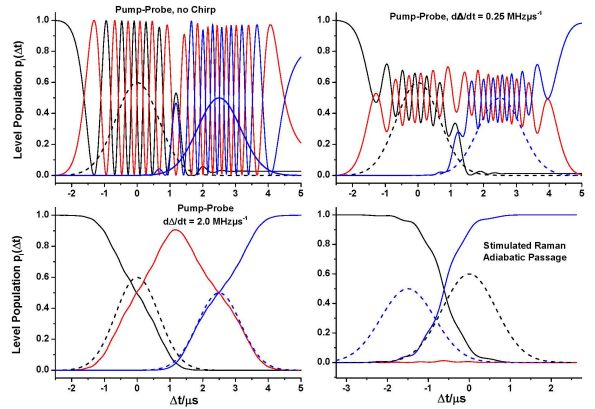

Figure 2: Time evolution of a three level system exposed to two laser pulses nearly resonant with $|1\rangle \rightarrow |2\rangle$ and with $|2\rangle \rightarrow |3\rangle$ transition for different pulse conditions: Pump - Dump (Stokes), no frequency chirp (upper left), Pump - Dump(Stokes), small frequency chirp (0.25 MHzμs$^{-1}$, upper right), Pump - Dump (Stokes), larger frequency chirp (2.0 MHzμs$^{-1}$, lower left), STIRAP Dump (Stokes) - Pump (lower right). Time dependent level populations: $|1\rangle$: black, $|2\rangle$: red, $|3\rangle$: blue. The laser pulses are indicated by the dashed line (Pump: black, Dump (Stokes): blue). Experimental conditions: vibrational transition moments: $\mu_{12} = \mu_{23} = 0.0262$ D, laser power: Pump: 0.6 W, Stokes: 0.5 W, pulse duration: $\tau = 1.31$ μs (after Ref. [6])

## Current and Future Challenges

Figure 2 shows simulations of population transfer using various methods including STIRAP (see Ref. [6]). In the experiment using RAP we could demonstrate a population transfer efficiency for the combined two step procedure of about 60%. Because of the better and more flexible control of experimental parameters in the STIRAP process [8], it should be possible to achieve a transfer efficiency near to 100%. The modification needed to implement the STIRAP process in the current experimental set up are relatively straightforward, although not trivial. The major current and future challenges are related to the much greater complexity of the rovibrational-tunneling spectra of chiral molecules as compared to the test molecule $NH_3$ with the well known spectra. However first spectroscopic investigations on two candidate molecules proved promising (1,2 Dithiine $C_4H_4S_2$ [9] and Trisulfane HSSSH [10]).

## Advances in Science and Technology to Meet Challenges

The current cw-OPO laser systems (coupled to a frequency comb) cover only spectral ranges above about 2500 cm-1 in the infrared. This limits the choice of molecules. Further development in laser technology, for instance of quantum cascade lasers with power and resolution meeting our needs in the future might make other molecules accessible, for instance the simpler molecule ClOOCl, for which complete theoretical simulations of the experiment have been achieved already [7].

**Concluding Remarks**

While the experiment to measure $\Delta_{pv}E$ might have appeared impossible, when it was first proposed in 1986 [5] the current outlook on a successful experiment is excellent. Indeed, provided that adequate funding for the continuation of the current project is guaranteed and required spectroscopic analyses can be completed, most significant results can be expected for any of two possible outcomes:

1. Either one finds experimentally the theoretically predicted values for $\Delta_{pv}E$ , then one can analyze the results of the precision experiments in terms of the Standard Model (SMPP) in a range not yet tested by previous experiments.

2. Or else one finds values for $\Delta_{pv}E$ different from the theoretical predictions. This then will lead to a fundamental revision of current theories for $\Delta_{pv}E$ with the potential also for new physics.

In addition, the experimental results will have implications for our understanding the evolution of biomolecular homochirality.

**Acknowledgments** − We acknowledge recent support from generous ETH grants and from an ERC Advanced Grant. The contributions of our coworkers in previous stages of the project can be seen from the references and reviews cited, and we are grateful for current help and support from Frédéric Merkt and Daniel Zindel.

## A2.4 Ultracold chemistry in STIRAP prepared ultracold molecular samples

*Silke Ospelkaus*
Leibniz Universität Hannover, Germany

### Status

Ultracold molecular gases, that is molecules cooled to below 1 millionth of a degree above absolute zero, are considered prime candidates for the study of true quantum chemistry. Unprecedented control over internal and motional quantum degrees of freedom in ultracold molecular quantum gases allows one to study chemical reactivity with reactants entering in well-defined single internal (and even motional) quantum states. In this regime, chemical reactivity can be studied without averaging over quantum states or motional degrees of freedom thermal.

The preparation of ultracold molecular gases is, however, challenging. The many degrees of freedom prevent the application of cooling techniques used to cool atoms to ultracold temperatures. Although various cooling schemes for molecules have been developed up to now only a single experimental approach was successful in the preparation of molecular ensembles at ultracold temperatures: This technique is based on STIRAP induced quantum state transfer resulting in molecular ensembles at high phase space densities close to quantum degeneracy. The preparation typically proceeds along a three step process: Starting from the preparation of quantum degenerate or near-quantum degenerate ensembles of atoms or of a two-species atomic mixtures, scattering resonances lead to the association of pairs of ultracold atoms into ultracold Feshbach molecules (extremely weakly bound molecules typically prepared in the least bound vibrational state of the electronic ground molecular ensemble). It is at this point that STIRAP provides the final step towards all ground state ultracold molecular ensembles. The ensemble of Feshbach molecules is mapped onto an ensemble of single hyperfine rovibrational ground state molecules using a single two-photon transfer step and bridging an energy gap of eV from the molecular dissociation threshold to the bottom of the electronic ground state potential. The result is an ensembles of single-quantum state ultracold molecules with well-defined external motional states (such as for example defined vibrational states in an optical lattice potential).

The preparation of high-phase space density samples of all-ground state molecules along the described path has first been demonstrated in pioneering experiments in 2008 with KRb [1], $Rb_2$ [2], and $Cs_2$ [3] molecules, resulting in ensembles of up to several 10 000 molecules. During the last ten years, the scheme has been further refined and applied to more molecular species including LiNa, LiK, NaK, and RbCs molecules. Most recently, even a quantum degenerate gas of KRb molecules has been reported [4].

The study of molecular quantum gases allows new insight into molecular collisions and chemical reactions with unprecedented energy resolution. In traditional chemistry, averaging over thermal energy and quantum state blur many details of the reaction dynamics. In ultracold chemistry reactants are in a single quantum state, thermal energy is frozen out and the basic laws of quantum mechanics determine the chemical reaction dynamics. The chemical reaction rate is controlled by the tunneling of molecules given by the heights of chemical reaction barriers. Furthermore, long-range reaction barriers can be tailored by precise control over the internal quantum state of the scattering partners, by means of control over dipolar interactions between the reactants or restricted geometries resulting from external confinement. This has been impressively demonstrated in a first series of experiments with an ultracold STIRAP prepared ensembles of KRb molecules [5] studying the chemical reaction KRb+KRb-→$Rb_2$+$K_2$. This reaction is exothermic by about 10K. The reaction path involves crossing a very deep potential energy well of several thousand K. With ultracold KRb molecules prepared in the absolute lowest state quantum state, any detected two body loss process indicates the occurrence of the chemical reaction. It's reaction rate can be determined by measuring the two-body loss rate of molecules from the trap.[1]

Experiments with KRb molecules revealed that at ultracold temperature the chemical reaction rate is entirely determined by simple quantum mechanical laws and quantum statistics. Molecules in the experiments are a fermionic Potassium and a bosonic Rubidium atom, yielding fermionic molecules. Dealing with two identical spin-polarized fermions, the Pauli exclusion principle requires the two-body wavefunction in a collision to be anti-symmetrized with respect to particle exchange, restricting the collisional channels to odd partial waves with the p-wave channel determining the low-temperature threshold behavior. The p-wave channel involves angular momentum and therefore a centrifugal energy barrier. For sufficiently close approach for a chemical reaction to occur, molecules have to tunnel through the p-wave barrier with a height of several 10μK (see Fig. 1). The quantum mechanical probability for tunneling determines entirely the chemical reaction rate. With decreasing temperature of the molecules, tunneling of the molecules through the barrier becomes less likely and the chemical reaction rate is suppressed reflecting the Bethe-Wigner threshold law for p-wave collisions.

---

[1] Note that since all other chemical reaction paths are forbidden due to energetic arguments: Chemical reactions of the form KRb+KRb-→ $K_2$Rb+Rb or KRb+Krb→ $KRb_2$+K have been calculated to be endothermic by several 10 K and therefore forbidden at ultracold temperatures.

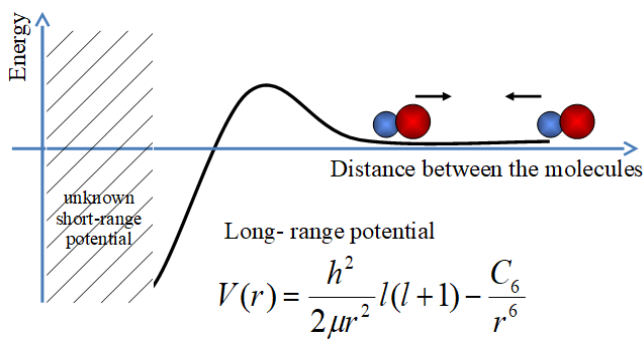

Energy

Distance between the molecules

unknown
short-range
potential

Long- range potential

$$V(r) = \frac{h^2}{2\mu r^2} l(l+1) - \frac{C_6}{r^6}$$

*Figure 1: Potential energy barrier resulting from the competition between the attractive Van-der-Waals interaction and the centrifugal energy in the collision.*

The dramatic consequences of quantum statistics are observed when the molecules are prepared as a 50:50 mixture of two different hyperfine states. In this case molecular collisions can occur in partial wave scattering channels of even symmetry, the lowest being the s-wave collision lacking a centrifugal energy barrier. Molecules are therefore free to approach each other closely and undergo a chemical reaction. By flipping the nuclear spin of half of the molecules enhancement of the reactions rate by 1 to 2 orders of magnitude has been observed (see Fig. 2a and [5]).

Apart from chemical reaction barriers resulting from quantum statistical properties, dipolar long-range interactions between polar molecules have been used to control the chemical reaction rate. Dipolar particles, including polar molecules, can collide in two relative orientations: Side-by-side or head-to-tail. Whereas the interaction for dipolar side-by-side collisions is repulsive, it is attractive for head-to-tail collisions. Working with a single-quantum state ensemble of fermionic molecules, it is then possible to control the height of the p-wave collisional barrier by controlling the direction and size of the molecular dipole moment in a collision. Whereas the effective energy barrier in a collision is being raised by dipolar interactions in a side-by-side collision, it is lowered in a head-to-tail configuration. When preparing molecules in an optical dipole trap and applying an electric field to induce the molecular dipole moment in the laboratory frame, molecules can collide in both channels. An increase of the reaction rate with the sixth power of the dipole moment has been observed (see Fig. 2 b and [6]). The Pauli-exclusion principle can again be invoked to explain the suppression of the reaction rate in the single quantum state ensemble of dipolar fermionic molecules. The trick is to make use of a pancake-like 2-dimensional restricted geometry and precise control of the molecules' motional state along the tightly confining axis to exclude attractive head-to-tail collisions. Due to the reduced number of collisional channels suppression of chemical reactions has been observed even when the molecular dipole moment is zero. Once a dipole moment is induced chemical reaction rates have been observed to be suppressed even beyond the zero dipole moment limit due to the rising p-wave barrier of the molecular collisions in dipolar side-by-side collisions (see Fig. 2B and [7]).

These first simple experiment already give a glimpse of the rich physics that can be observed with ultracold molecules with STIRAP being the key to prepare the molecules in a single quantum state. Chemical reactions can be studied with unprecedented energy resolution and the reaction rate can be chosen by deliberately shaping the reaction barrier paving the way to quantum controlled chemistry. Exciting physics is also expected to be revealed in the study of molecules where the chemical reaction of the above form is endothermic and therefore forbidden at ultracold temperatures. First experiments with NaK at MIT and in Munich and RbCs in Innsbruck hint at the formation of long-lived molecular complexes in so-called sticky collisions as first proposed by John Bohn and coworkers at JILA.

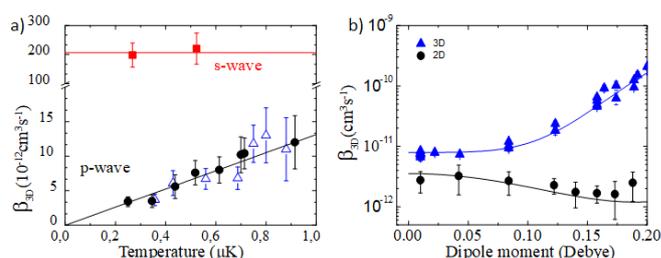

*Figure 2: a) Dramatic change of the chemical reaction rates for p-wave vs. s-wave collisions [5] b) Chemical reaction rate as a function of dipole moment for averaged head-to-tail and side-by-side collisions in 3d vs. side-by-side collisions only in 2d [6,7] .*

## Current and Future Challenges

Up to now chemical reactions of ground state molecules at temperatures close to absolute zero have only been indirectly detected through loss measurements of the incoming reactants. The outgoing products have neither been observed nor have they been characterized leaving half of the process completely in the dark. However, in particular physical chemists would be particularly interested in a full characterization of the process, since it might provide answers to many yet to be addressed questions. Does the atom exchange reaction of the form $XY+XY \rightarrow X_2+Y_2$ really happen? Through which transition state does it proceed? How long lived is this state? What happens in terms of energy and momentum transfer in the fully quantized system? Which role is played by fine or hyperfine interaction in the chemical reaction? Is there a novel class of chemical reactions induced by the very weak fine and hyperfine interactions?

These questions can only be addressed by studying the full process from single quantum and motional state prepared reactants to a precise analysis of the products and their quantum states. This is challenging. First of all, it is clear that due to the exothermic nature of many chemical reactions the products will have a much larger kinetic energy than the incoming particles. They are therefore not trapped in typical traps of quantum gas experiments. Second, it is inherently difficult to detect molecules, since scattering of many photons on closed cycling transitions, as it is required for fluorescence imaging, is prohibited by the complex molecular structure.

## Advances in Science and Technology to Meet Challenges

There is currently intense work to develop a novel generation of quantum gas apparatus combining molecule-formation with quantum state sensitive molecular detection techniques. J. Hecker Denschlag's group in Ulm has combined an atomic quantum gas apparatus with state selective ionization of molecular decay products and subsequent storage of the resulting ions in ion traps to demonstrate the detection of the full quantum state distribution of molecules resulting from a three-body decay process of the Rb quantum gas [8]. Kang-Kuen Ni's group in Harvard aims at applying state selective ionization techniques to detect the chemical reaction products of colliding KRb molecules. Furthermore, frequency comb based techniques have been developed recently for real-time detection of molecules allowing for the detection of an entire molecular reaction process starting from reactants via transients to the final products [9].

## Concluding Remarks

STIRAP-based formation of quantum gases of molecules has opened a vast, completely unknown playground in quantum chemistry with many exciting fundamental questions to be answered. New challenges lie ahead which might result in even more STIRAP applications including STIRAP controlled state-to-state chemical reactions.

# Quantum Interfacing of Stationary and Flying Qubits

*Axel Kuhn, University of Oxford,*

*Clarendon Laboratory, Parks Road, Oxford, OX1 3PU, UK*

**STATUS** – Originally, STIRAP was devised to prepare quantum states of atoms or molecules with the light field modelled semi-classically. Not too long after its invention, it was found that the technique equally applies when the field is quantized [1], thus enabling the controlled interfacing of single photons with a wide range of stationary quantum systems.

The quantised field is considered inside a cavity, with $|n\rangle$ denoting the $n$-photon Fock state, raising and lowering operators $\hat{a}^\dagger$ and $\hat{a}$, and the cavity coupling states $|g\rangle$ and $|x\rangle$ of a stationary system at strength $g$. The Jaynes-Cummings model [2] describes this coupled system. The dressed states $|x, n-1\rangle$ and $|g, n\rangle$ form the basis of the $n$-excitation manifold of the Jaynes-Cummings ladder, mutually coupled at the effective Rabi frequency $2g\sqrt{n}$.

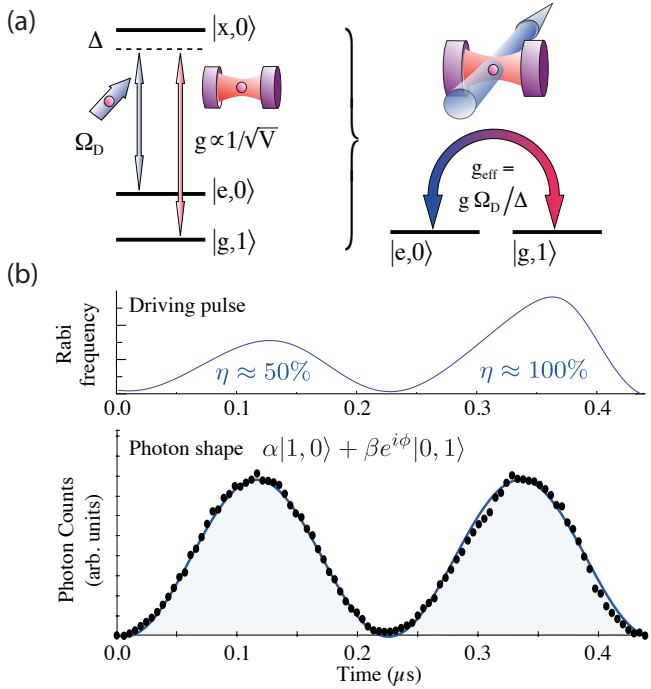

FIG. 1. (a) Raman-resonant atom-cavity coupling in the $n = 1$ manifold with $\Delta = \Delta_C = \Delta_D$. The coupling strength $g$ is proportional to $1/\sqrt{V}$, where $V$ is the cavity's mode volume. If $\Delta$ outweighs $\Omega_D$ and $g$, $|x, 0\rangle$ can be eliminated adiabatically, yielding an effective coupling $g_{\text{eff}}$ between $|e, 0\rangle$ and $|g, 1\rangle$. For small $\Delta$, STIRAP controls the process. (b) Photon shaping and time-bin encoding. The upper trace shows a driving pulse generating a single photon in an equal-amplitude superposition of two time bins, depicted in the lower trace. The photon is symmetric in each half, but not the driving pulse.

STIRAP-like control of the population flow in this system, which now includes the intra-cavity photon number, is achieved by a laser that couples $|x, n-1\rangle$ to a third state, $|e, n-1\rangle$, thus setting up a Λ-system as shown in fig. 1(a). The laser drives this transition at Rabi frequency $\Omega_D$, with its field being in a coherent state. For this laser, a semi-classical model is then adequate. In the rotating wave approximation, the time evolution of the coupled system is then governed by the Hamiltonian $\hat{H}$, with

$$\hat{H}/\hbar = (\Omega_D\hat{\sigma}_{ex} + \Omega_D^*\hat{\sigma}_{xe})/2 + g(\hat{a}^\dagger\hat{\sigma}_{gx} + \hat{a}\hat{\sigma}_{xg})$$
$$+ \Delta_C\,\hat{a}^\dagger\hat{a} + \Delta_D\,\hat{\sigma}_{ee}$$
$$- i\kappa\hat{a}^\dagger\hat{a} - i\gamma\hat{\sigma}_{xx}$$

where $\hat{\sigma}_{ab} = |a\rangle\langle b|$, with $\Delta_C$ and $\Delta_D$ denoting the cavity and driving laser detuning from the atomic resonances, and $\kappa$ and $\gamma$ the cavity-field and $|x\rangle$-amplitude decay rates. The last two terms phenomenologically take the decay into account.

Each $n$-manifold presents a three level system, where driving laser and cavity field provide the Raman coupling between $|e, n-1\rangle$ and $|g, n\rangle$, with $|x, n-1\rangle$ the intermediate state. Of particular interest is the $n = 1$ manifold, which can be used to generate or absorb single photons. For generation, $|e, 0\rangle$, $|x, 0\rangle$ and $|g, 1\rangle$ correspond to $|1\rangle$, $|2\rangle$ and $|3\rangle$ in STIRAP, with Rabi frequencies $\Omega_P \equiv \Omega_D$ and $\Omega_S \equiv 2g$. For absorption, $|e, 0\rangle$ and $|g, 1\rangle$ swap roles, and so do the Rabi frequencies.

For single-photon generation [1] we start from $|e, 0\rangle$, with the atom interacting with the cavity and the driving laser off. This yields $2g \gg \Omega_D$, and thus the initial STIRAP condition $\Omega_S \gg \Omega_P$ is met. On resonance, the system is therefore in the dark state

$$|\phi_0\rangle = \left(2g|e, 0\rangle - \Omega_D|g, 1\rangle\right) \Big/ \sqrt{4g^2 + \Omega_D^2}\;.$$

If $\Omega_D$ increases slowly, the state vector follows $|\phi_0\rangle$ and the atom evolves adiabatically into $|g\rangle$, whilst a photon is placed into the cavity. For STIRAP, one would normally require $\Omega_S$ to tail off to bring the system into $|g, 1\rangle$. Here, this can't be done because $g$ is constant. Fortunately the photon emission from the cavity has the same effect. $|g, 1\rangle$ decays at rate $2\kappa$ and eventually the system evolves into $|g, 0\rangle$, i.e. the atom de-couples from any further interaction, whilst a single free-running photon leaves the cavity.

**CURRENT AND FUTURE CHALLENGES** – More recent developments show that a suitable variation of the amplitude and the phase of the driving pulse allows for the deterministic shaping of the temporal envelope and phase function of the photon produced [3]. Fig. 1(b) illustrates how this can be employed to encode quantum bits in the temporal mode of a single photon [4]. Dividing the driving laser pulse into sub-pulses that provide different efficiencies and which differ in their relative phases is an excellent tool to prepare single photons in a superposition of $d$ time bins, and thus embed quantum-$d$-bits into their spatio-temporal mode profile.

Strongly coupled atom-cavity systems find another prominent application in universal quantum networking. For instance, direct atom-photon-atom quantum state mapping and entanglement swapping as proposed in [5] relies on the time reversal of the emission process to store an incoming single photon in a single atom, fig. 2(a). The receiving atom is supposed to start from $|g, 0\rangle$ with a single photon impinging onto one cavity mirror. The photon only enters the cavity if the atom is driven such that full impedance matching is met. One way of doing so is to generate photons symmetric in time, so that a time-reversed driving pulse stores them in a second cavity. Another possibility is to design driving pulses required to absorb incoming photons of particular shape [6]. Experiments demonstrating photon storage have been successfully conducted [7], and very recently a more sophisticated approach has been proposed to obtain driving pulses that achieve even higher efficiencies [8].

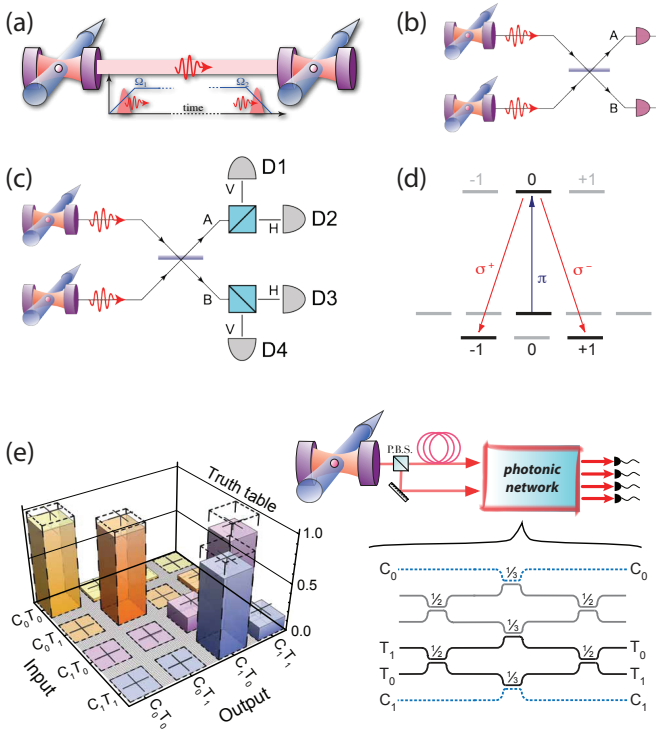

FIG. 2. Quantum networking schemes – (a) State mapping and entanglement swapping [5]. (b) Photon-detection induced atom-atom entanglement into $(|eg\rangle \pm |ge\rangle)/\sqrt{2}$. (c) Bell-state analysis to probabilistically project the emitters into a spin-entangled state upon the detection of two photons, using the (d) level scheme. (e) Linear optics quantum computing with cavity photons. From a pair of successively emitted photons, the first is delayed and both enter a photonic CNOT gate simultaneously. In dual-rail encoding, one photon is the target and the other the control qubit. The truth table of the gate exhibits a similarity with expectations better than 98%.

For heralding successful entanglement, other networking schemes, such as those shown in fig. 2(b) and (c) have been proposed. Here, probabilistic entanglement of remote emitters is achieved upon photon detection. This is the approach taken within the UK's quantum technology hub NQIT to implement a large-scale networked quantum-computing architecture. The detection is made in the output of a beam splitter so that the origin of the photon is not revealed. For this reason, the first photon detection projects the two emitters into $(|eg\rangle \pm |ge\rangle)/\sqrt{2}$. Whilst this seems straight-forward, non-negligible photon losses and detector efficiencies below unity reduce the success rate. Therefore the more sophisticated scheme depicted in fig. 2(c) and (d) seems more adequate. Detecting photons from both emitters eventually prepares the atoms in one of the maximally spin-entangled Bell states. No post-selection or other heralding scheme is required in this case.

Photons emitted from cavities have also been used successfully for all-optical quantum computing [9], with photons emitted into a delay network which eventually feeds them simultaneously into an integrated photonic chip. For a CNOT gate operating in the coincidence basis, shown in fig. 2(e), similarities with expectations exceeding 98% have been observed, underpinning that the coherence properties of the cavity photons are ideally suited for this purpose.

**ADVANCES IN SCIENCE AND TECHNOLOGY TO MEET CHALLENGES –** At first glance, it seems that the feasibility of the above applications has been successfully demonstrated and that these might be replicated easily for a large variety of purposes. However, increasing the number of simultaneously operating atom-cavity systems to a useful scale, e.g. to implement simple quantum-computing or communication protocols, remains a major challenge. Ideally the emitter should be stationary, e.g. a trapped atom or ion, which has proven to be a difficult task in combination with optical cavities. Issues that need to be addressed include the dynamic variation of the atomic level structure induced by the trapping potential, the large impact dielectric mirrors have on ion traps, cavity mode matching, cavity birefringence, etc. Furthermore, inevitable photon losses upon coupling into fibres, photonic networks or into less-than 100% efficient detectors result in severe constraints to the scalability of any technical implementation, and tend to render an inherently deterministic scheme eventually probabilistic.

Various technological advances will help mitigate most of these constraints. Amongst these are the fabrication of bespoke mirrors on arbitrary substrates, which should allow for the on-chip integration of cavity arrays to couple photons into photonic networks with negligible losses, highly efficient single-photon counters, e.g. using super-conductive nanowires, new interfacing schemes for photon generation exploiting cavity birefringence and magnetic substructures in atoms, and cavity-compliant trap designs.

A slightly different pathway would be the application of our coupling schemes to artificial atoms in solid-state systems, such as quantum dots or colour centres coupled to Bragg-stack cavities, or Josephson junctions in the super-conductive circuit-qed regime [10].

**CONCLUDING REMARKS –** STIRAP is a key enabling technique for the faithful interfacing of light and matter in various modern applications, such as distributed quantum computing, quantum networking and communication. To harness the light and to ensure photon coupling to a well-defined field mode, replacing the stimulating field by the vacuum-field mode of a high-finesse cavity is now an established approach that applies to a wide range of very different emitters. Whilst the feasibility has been proven, major challenges remain. At present, no large-scale quantum network comprising more than two cavities has been demonstrated. Nonetheless the limiting factors have been identified, and techniques are under development which are addressing most issues.

**ACKNOWLEDGMENTS –** Using STIRAP for interfacing light and matter has been pioneered in the group of Gerhard Rempe and was later refined by my own team. Major contributions were made by M. Hennrich, T. Legero, T. Wilk, S. Nußmann, B. Weber, M. Hijlkema, S. Webster, P. B. R. Nisbet-Jones, J. Dilley, M. Himsworth, D. Ljunggren, G. Vasilev, A. Holleczek, O. Barter, B. Yuen, D. Stuart, and T. D. Barrett. Integrated photonic networking was made possible by our partners in Bristol, J. C. F. Matthews, A. Rubenok, K. Poulios and J. L. O'Brien. The projects presented here were supported by the European Union, the DFG, and the EPSRC within the UK's quantum technology programme.

---

### A3.2 The STIRAP concept in Optical Wave Guides

*Stefano Longhi[1] and Alexander Szameit[2]*

[1] Politecnico di Milano, IFN-CNR
[2] University of Rostock

**Status**

Analogies between photonic transport in waveguide optics and coherent processes in atomic and molecular physics [1] have opened in the past decade a wealth of entirely new opportunities to control the flow of light and to design new classes of integrated photonic devices. Light propagation in evanescently-coupled waveguide structures is described by a matrix Hamiltonian which can be engineered to realize a wealth of transfer schemes such as Rabi oscillations, Landau-Zener and rapid adiabatic passage, STIRAP, coherent destruction of tunneling, etc. [1]. Many of such schemes, being generally robust to imperfections and perturbations, are very appealing in the design of integrated waveguide devices whose functionality is tolerant to manufacturing imperfections.

Pioneering experiments on photonic STIRAP date back to ten years ago [2,3]. Optical STIRAP was demonstrated using three weakly-curved coupled waveguides where the inner waveguide is not excited and the light pattern adiabatically evolves in the dark state of the three-state system (Fig.1a,b), thus realizing perfect transfer of photons between the outer waveguides with negligible excitation of the intermediate one. This makes the excitation transfer process efficient even though the inner (auxiliary) waveguide is lossy. Later on, photonic STIRAP was demonstrated in more general settings, such as via dressed states or via a continuum [4]. In the latter case, the dark state of STIRAP is a bound state embedded in the continuum, which exhibits topological protection and thus enables robust transfer of light (Fig1c). The experimental demonstration of a dark state in the continuum via STIRAP [4], based on two waveguides side-coupled to a photonic lattice (a tight-binding continuum), thus provided the first paradigmatic observation of a photonic bound state in the continuum - currently a very active area of research in photonics and beyond.

From a more applied viewpoint, STIRAP has inspired novel designs of integrated photonic devices, with the demonstration of several functionalities such as polychromatic beam splitting based on fractional STIRAP [5] (see the contribution by K Bergmann for a definition of fractional STIRAP), spectral filters [6], photon pair generation in integrated quantum photonic circuits [7], high-density waveguides with suppressed cross-talk [8], ultra-compact directional couplers and low-loss Si-based modulators, to mention a few. Finally, STIRAP has facilitated interesting applications in polarization control, optical isolators and nonlinear frequency conversion, where different schemes have been suggested and experimentally demonstrated. Also, a nonlinear version of STIRAP can be implemented in nonlinear optical systems [3]. The plethora of applications that STIRAP and adiabatic schemes have found in integrated waveguide photonics is expected to continuously inspire fresh and new ideas in the near future with great potential for controlling the flow of classical and quantum light at the micro and nanoscale.

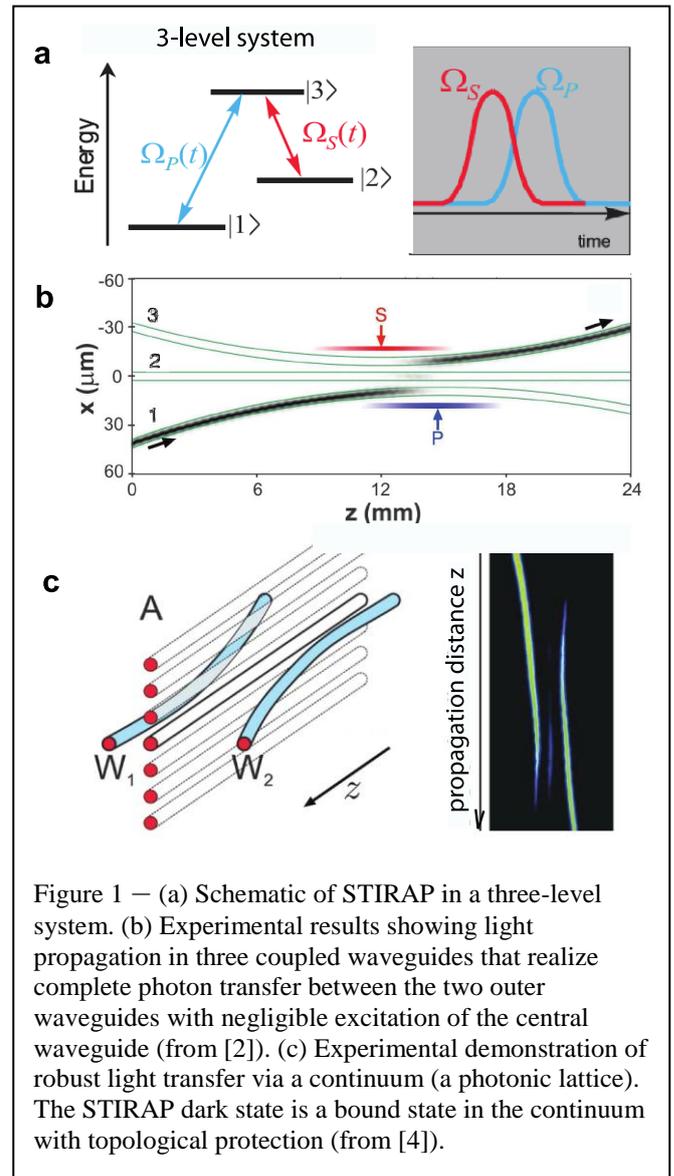

Figure 1 — (a) Schematic of STIRAP in a three-level system. (b) Experimental results showing light propagation in three coupled waveguides that realize complete photon transfer between the two outer waveguides with negligible excitation of the central waveguide (from [2]). (c) Experimental demonstration of robust light transfer via a continuum (a photonic lattice). The STIRAP dark state is a bound state in the continuum with topological protection (from [4]).

**Current and Future Challenges**

STIRAP and other adiabatic protocols provide robust methods for light splitting, coupling and filtering in waveguide optics, however in conventional low-contrast-index dielectric waveguides a current limitation is the need of sufficiently long structures (typically from a few mm to a few cm length) to ensure adiabaticity conditions. This suggests that, for applications requiring fully integrated and compact structures, one should either resort to high-contrast-index and packed waveguides or to variant methods such as the shortcut to adiabaticity protocols, which represent a current active and challenging area of research. A recent experiment [8] demonstrated that adiabatic elimination is a powerful method to strongly reduce crosstalk between tightly packed high-index waveguides (Fig.2). Hence, one should expect that, with such a technology, extremely compact STIRAP functionalities could be realized with sample length less than 1 mm. Another current challenging issue is the extension of STIRAP and adiabatic protocols to photonic devices with gain and loss, which are inherently non-Hermitian. Non-Hermitian terms could provide a means for shortcut to adiabaticity [9]. However, the adiabatic theorem in non-Hermitian systems can rapidly fail and can show an interesting asymmetric behavior. For example, recent experiments [10] demonstrated that dynamically encircling a non-Hermitian degeneracy (so-called exceptional point) leads to asymmetric behavior depending on the circulation direction. This provides a further potentiality of STIRAP to photonic systems with gain and loss. Application of STIRAP and adiabatic protocols to non-Hermitian systems remains largely unexplored, and could be a fertile area of research in the near future. Since photonic devices like lasers are inherently non-Hermitian systems, one could envisage novel adiabatic control methods in which non-Hermiticity plays a major role. Another interesting feature of photonic STIRAP is its close connection with topological photonics. For example, multilevel STIRAP in a waveguide lattice [11] is based on the existence of a zero-energy dark mode, which is topologically robust as the coupling constants between the waveguides are varied. Adiabatic light transfer in such a waveguide lattice, where the system remains in the slowly evolving dark state, realizes what is known as topological (Thouless) pumping in a topological insulators context. This feature, which has been overlooked so far, could provide a fertile ground to bring topological concepts into the STIRAP community in photonics and beyond. Finally, a rather unexplored area, where concepts of STIRAP and rapid adiabatic passage could be useful, is that of structured light. For example, a challenging task could be to design reconfigurable optical metasurfaces to structure light beams (e.g. to introduce topological charges or spin-angular momentum coupling) based on robust adiabatic protocols.

**Advances in Science and Technology to Meet Challenges**

Advances in waveguide fabrication, especially at the nanoscale, are a major objective to make optical functionalities based on adiabatic protocols scalable down to micrometer sizes. For example, directional couplers based on silicon nanowires have already been demonstrated, however, it remains challenging to manufacture bent wires with controlled distances, a major requirement to implement STIRAP and adiabatic protocols for light control. Also, application of adiabatic protocols to plasmonic-based waveguides is a rather unexplored possibility, which could become feasible in the near future once low-loss metal-dielectric guides will be available.

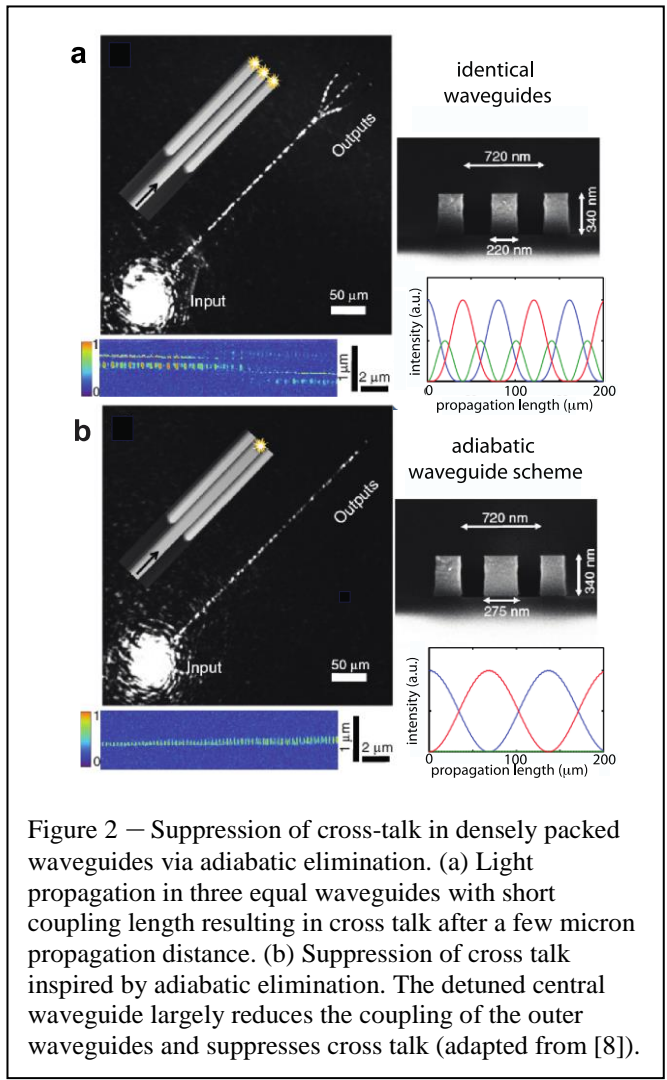

Figure 2 — Suppression of cross-talk in densely packed waveguides via adiabatic elimination. (a) Light propagation in three equal waveguides with short coupling length resulting in cross talk after a few micron propagation distance. (b) Suppression of cross talk inspired by adiabatic elimination. The detuned central waveguide largely reduces the coupling of the outer waveguides and suppresses cross talk (adapted from [8]).

**Concluding Remarks**

Photonic STIRAP and adiabatic protocols provide powerful means for robust manipulation of photons on integrated photonic platforms. Future areas of research that could promise major advances are the application of STIRAP to novel guiding structures (such as nanowires, plasmonic waveguides, etc.) and the link of adiabatic methods to the rapidly emerging fields of non-Hermitian and topological photonics.

### A3.3 The magnonic STIRAP process

*Philipp Pirro and Burkard Hillebrands*
Fachbereich Physik and Landesforschungszentrum OPTIMAS, Technische Universität Kaiserslautern, 67663 Kaiserslautern, Germany

### Status

Magnons are the bosonic quanta of spin waves, the elementary low energy excitations of an ordered magnetic system. The field of magnonics [1,2] aims to create novel applications based on magnons, similar like the field of photonics uses photons. The aim of this section is to demonstrate the way how STIRAP and related methods can be implemented into magnonics. The most striking difference to photons is the high flexibility of the magnon dispersion relation, which includes, among others, contributions of the exchange, dipolar, Zeeman and anisotropy energies. Since these energies give qualitatively and quantitatively different contributions to the dispersion relation, a choice of suitable magnetic materials, magnon waveguide dimensions as well as the strength and orientation of the external magnetic bias field allow designing the dispersion relations according to the particular needs [1,3,4]. This designability and reconfigurability, as well as easy accessibility of nonlinear processes, makes spin waves an ideal model system for wave physics in general. Fig. 1a shows the example of a dispersion relation in a nanostructured spin-wave waveguide. Due to the low group velocities compared to photons, spin waves can be easily detected and manipulated with electronic as well as optical means which allows for example for studying the time evolution of a spin-wave packet in a system with time dependent dispersion relations [5].

In terms of wave-based applications, spin waves are attractive because the magnon wavelength can be made easily up to five orders of magnitude lower than the wavelength of the photon at the same frequency. At the commonly used magnon frequencies in the GHz range, this means that spin waves with wavelengths in the submicron range can be readily created which facilitates miniaturization of devices significantly. The excitation of spin waves in the GHz range is usually realized by a local and coherent source, such as the magnetic field of a local microwave current in a suitable antenna structure. Since the frequency, wavelength and phase of the magnons are well defined, information carried by magnons can be readily encoded in the amplitude and phase. This way, interference phenomena can be utilized on the nanometer scale which makes spin waves a promising candidate for wave-based computing [1]. Another particular feature of spin waves is their intrinsic nonlinearity which stems from the particular form of the governing equation of motion, the Landau-Lifshitz equation. Thus, nonlinear frequency shifts and multi-magnon processes can be realized in any kind of magnonic system and are especially easy to achieve in materials which show good magnon propagation characteristics, such as the ferrimagnetic insulator material Yttrium Iron Garnet (YIG) or metallic Heusler compounds. Thus, spin waves are predestined to serve as information carriers in novel computing concepts based on nonlinear waves which could realize task like neuromorphic computing in a very energy efficient way [1].

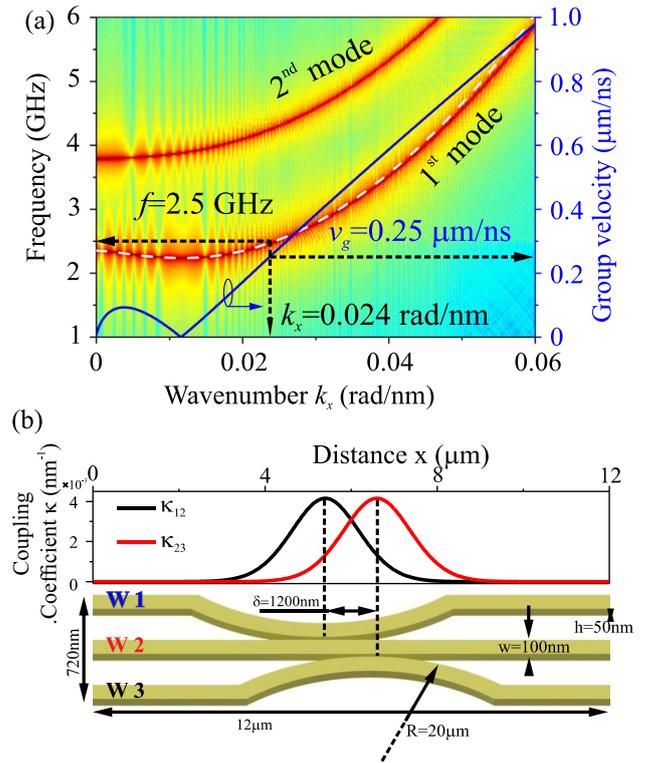

*Figure 1(a) Dispersion relation of an isolated magnonic waveguide made of YIG (100 nm width, 50 nm thickness, without external bias field) (b) Scheme of the magnonic STIRAP demonstrator based on three coupled magnonic waveguides. The upper part shows the coupling strength calculated according to [6].*

Since magnons carry energy and spin, it is possible to interlink the magnonic system with the electronic system via spintronic effects. These effects, which allow for example enhancing or completely suppressing the damping in the magnonic systems via DC currents, are studied in the still rapidly growing field of magnon spintronics [1].

In recent years, advances in the miniaturization of the ultralow magnon damping material YIG have led to a breakthrough in miniaturized magnonic devices since these YIG nanostructures allow circumventing one of the major drawbacks of magnonics which is the finite lifetime of magnons due to the intrinsic damping. This development has led to the concept of coupled magnonic waveguides which are used to create nanoscale magnonic directional couplers [6], similar to the concept used in photonics. The coupling between the waveguides is mediated via the dipole-dipole interaction, thus via the magnetic stray fields created by the spin waves. Consequently, the coupling strength depends strongly on the used wavelength, the thickness and width of the waveguides and the gap between them. In good approximation, the coupling strength decays exponentially with increasing gap and can be precisely calculated using the theory presented in [6].

Based on these results, we propose the concept of a magnonic STIRAP process which is inspired by the work on quantum-optical analogies presented in [7] and section A3.2 of this article. The goal is to use the STIRAP process and other related adiabatic methods and quantum classical-wave analogies to design magnonic networks with advanced functionalities and higher tolerance to defects. Analogous to the concept in photonics, we replace the time coordinate in the original STIRAP process by the spatial coordinate $x$ along the waveguide. Our demonstrator consists of three, partially curved magnonic waveguides W1, W2, W3, as depicted in Fig. 1b. The coupling strength between the waveguides varies strongly as a function of position and has a Gaussian distribution as depicted in the upper part of Fig 1b. From left to right, the point of maximum coupling to the middle waveguide W2 is first reached for W1 and then for W3. Thus, a propagation of spin waves from the left excited in W1 corresponds to the intuitive scheme whereas an initial propagation in W3 represents the counter-intuitive scheme. Using an analytical theory similar to the matrix Hamiltonian description in [6], we predict the spin-wave intensities in the waveguides and validate our model using a full micromagnetic simulation of the spin wave propagation (using *mumax3* [8]). Fig. 2a shows the counter-intuitive coupling case where the middle waveguide is almost not populated during the transfer from W3 to W1, thus representing the analogy of an adiabatic dark state. For the intuitive coupling case depicted in Fig. 2b however, W2 is strongly excited. The model as well as the simulation include the intrinsic damping of the spin waves which leads to a slight decay of the overall intensity along the propagation. A comparison between the model and the simulation shows a good general agreement if one neglects the standing wave pattern formed in the coupled region. This pattern stems from reflections of the spin waves due to the small impedance mismatch in the coupled region and is not present in the analytical model which only considers forward propagating waves.

**Current and Future Challenges**

The proposed scheme to realize magnonic STIRAP can be produced using state-of-the-art nanopattering techniques developed for nano magnonics [9]. Challenges for a further development are the implementation of a tunable control of the decay in the intermediate state (W2) which can be realized using spintronic effects or parametric spin-wave amplification [10]. Also, a study of the influence of nonlinear effects is straightforward due to the strong intrinsic nonlinearity of the spin-wave system. In addition, the magnonic system can be time-modulated to enhance the coupling between non-degenerated waveguides, similar to the transition between atomic levels in the original STIRAP concept.

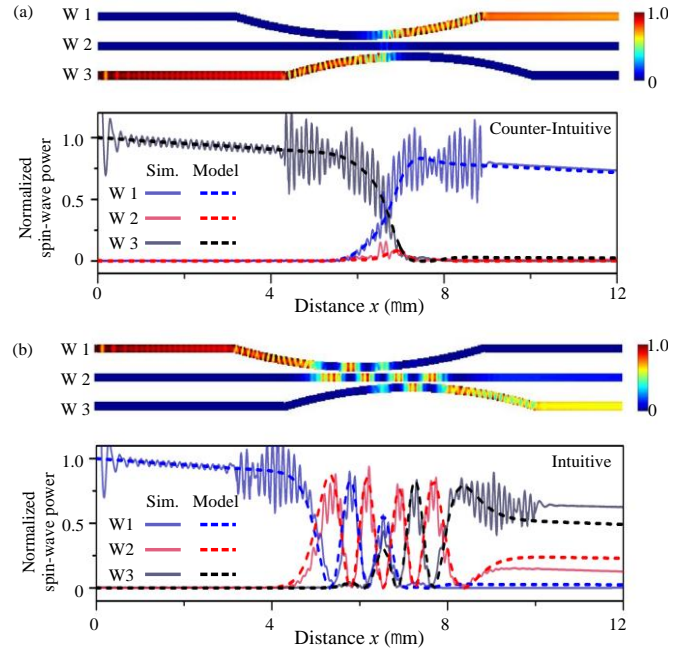

*Figure 2 Colour coded spin-wave intensity distribution from micromagnetic simulations (upper panels) and comparison to the analytical model based on the Hamiltonian model (lower panels). (a) Counter-intuitive coupling scheme where the intermediate waveguide W2 is only weakly excited. (b) Intuitive coupling scheme resulting in a strong excitation of W2.*

**Advances in Science and Technology to Meet Challenges**

A more compact and versatile realization of magnonic STIRAP at a length scale below 1 μm seems to be feasible but will require a better control of the coupling to still fulfill the adiabatic conditions. The coupling can be enhanced by fabricating waveguides with an additional magnetic material inside the gaps which could be either paramagnetic or ferromagnetic, and by taking advantage of exchange interaction. A further, non-volatile control can be envisioned by stray fields of magnetic patches in the vicinity of the device, whose magnetization direction is controlled externally.

**Concluding Remarks**

We have shown, that the population of magnons can be transferred between two waveguides via an intermediate waveguide, which is not excited, thus resembling the quantum-classical analogy of a dark state in the STIRAP process. We feel that our results bear high potential for future magnonic device functionalities and designs by bringing together the wealth of quantum-classical analogy phenomena with the wealth of means to control wave propagation in magnonic systems. Especially in the field of wave-based computing, STIRAP based magnon devices might become key in future all-magnon logic designs.

## Acknowledgments

We would like to thank Q. Wang, T. Brächer, A. Chumak and M. Fleischhauer for their valuable contributions and support of this work and the SFB/TRR 173 SPIN+X of the DFG for financial support.

## A3.4 The STIRAP concept in coupled acoustic cavities

Xue-Feng Zhu[1] and Jie Zhu[2]

[1]Huazhong University of Science and Technology
[2]The Hong Kong Polytechnic University

### Status

This work discusses the extension of the STIRAP concept [1] via quantum-classical analogs [2] to classical wave physics. Wave propagation in time-varying systems can be described by an asymmetric matrix Hamiltonian, indicating non-reciprocity. In past years, non-reciprocal transport of classical waves has been studied intensively. Various schemes were proposed to implement optic/acoustic diodes [3,4] and optic/acoustic topological insulators [5,6], which are devices with one-way responses and topological robustness against various defects. However, until very recently [7], the concept of STIRAP in acoustic systems was not considered.

Acoustic STIRAP can be realized with a one-dimensional (1D) array the coupling of which varies with time. As shown in Fig. 1(a), in analogy with STIRAP in a three-level quantum system, the three acoustic cavities **A-B-C** and the time-varying coupling respectively take the role of discrete states and laser pulses. Adiabatic passage from cavity **A** to cavity **C** occurs when the coupling $C_{BC}$ between cavities **B** and **C** is activated before the one between cavities **A** and **B**, with a certain overlap of $C_{BC}$ and $C_{AB}$. If the timing of the couplings $C_{BC}$ and $C_{AB}$ is properly implemented, the acoustic energy is transferred completely from cavity **A** to cavity **C**, with the cavity **B** being dark (or silent, corresponding to a zero-eigenvalue eigenstate), as shown in Fig. 1(b). The adiabatic passage from cavity **A** to cavity **C** is independent of the properties of cavity **B**, even if the latter is lossy. Thus, in such a time-varying acoustic system, the wave propagation can be designed to be nonreciprocal. We can tailor the overlap between the cavity couplings $C_{BC}$ and $C_{AB}$ to a specific value, where the adiabatic passage from cavity **A** to cavity **C** still holds, but acoustic waves input at cavity **C** won't reach cavity **A**. If cavity **B** is lossy, all the energy will be dissipated away.

At present, studies of acoustic STIRAP are still in its infancy. However, the unique phenomena presented here document promising potential for versatile applications involving single-pass acoustic communication, one-way sound absorption, and unidirectional matching layer design, to mention a few. For example, the physical model of acoustic STIRAP in

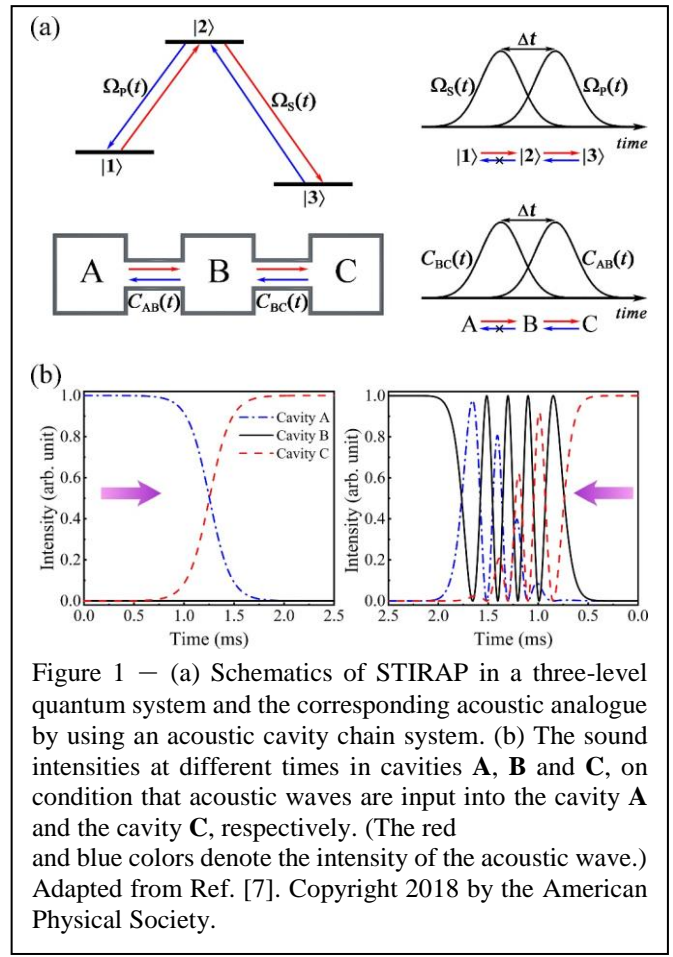

Figure 1 — (a) Schematics of STIRAP in a three-level quantum system and the corresponding acoustic analogue by using an acoustic cavity chain system. (b) The sound intensities at different times in cavities **A**, **B** and **C**, on condition that acoustic waves are input into the cavity **A** and the cavity **C**, respectively. (The red and blue colors denote the intensity of the acoustic wave.) Adapted from Ref. [7]. Copyright 2018 by the American Physical Society.

Fig. 1 provides a prototype of acoustic diode whose functionality would be robust against imperfections.

However, experimental realization of adiabatically time-varying couplings in the 1D acoustic cavity chain is challenging. The current solution is to map the time dimension into a space dimension [7]. The implementation, see Fig. 2, is through a static model realized by three parallel channels **A**, **B**, and **C** with couplings between them realized by a set of evenly spaced small tubes of varying diameters. The diameter of these tubes determines the coupling strength.

Experimental results obtained with the static three-pipe setup are shown in Fig. 2. Acoustic waves launched in channel **A** are completely transferred to channel **C**. Furthermore, acoustic waves sent back from channel **C** are localized in channel **B** and do not reach channel **A**, in agreement with the analogy of STIRAP in a three-level quantum system.

### Current and Future Challenges

Implementation of acoustic STIRAP with adiabatic time-varying couplings, as shown in Fig. 1, remains a significant challenge. This is also true for realizing the acoustic-STIRAP analog of multilevel-STIRAP in quantum systems either in the time- or in the spatial domain.

If the concept of acoustic STIRAP can be realized for high frequency ultrasound waves, interesting applications in medical ultrasonography and underwater acoustics may result. The challenge is, though, that solid structures can no longer be regarded as being rigid once they are in water. Acoustic STIRAP will fail in such an environment due to nontrivial sound leakage through fluid-solid coupling. Furthermore, when ultrasonic waves propagate inside narrow channels, viscosity and friction due to the presence of a wall will lead to quick conversion of acoustic to thermal energy. Similar challenges exist for the effort to implement STIRAP in elastic wave systems. In solid media, reflections and scattering of elastic waves at boundaries are accompanied by hard-to-control mode conversion. Research efforts to find ways how to suppress or control such mode conversions in coupled elastic resonator array are required and worthwhile.

## Advances in Science and Technology to Meet Challenges

Recent progress in the context of dynamic one-dimensional phononic crystal sheds light on the chances for the implementation of adiabatically time-varying couplings of acoustic waves. For example, a mass-spring chain of repelling magnets modulated by externally driven coils has time-varying local elastic properties [8]. In air-based acoustics, coupling strengths between resonators can be tailored by the flow velocity. It is realized through chiral-structured rotors driven by digitally precisely controlled motors [6]. The fluid-solid coupling problem in underwater acoustics can also be solved by using superhydrophobic surfaces, where a thin and stable air layer forms on superhydrophobic surface, providing a sufficiently large impedance mismatch. For elastic waves, a hybrid elastic solid was proposed to present a design of elastic metamaterial that supports compressional waves in a finite frequency range [9]. The latter results are obtained from encouraging theoretical studies. Realizing such novel metamaterials would allow tackling the mode conversion issue and make elastic-solid STIRAP feasible in the future.

## Concluding Remarks

Acoustic STIRAP provides a powerful way for robust and nonreciprocal manipulation of vibration on various platforms that include air-based acoustics, underwater acoustics and elastic waves in solids. Future research along this direction will aim at the implementation of STIRAP-based devices with nonreciprocal functionality, with many potential key applications in acoustics such as duct noise control, acoustic signal processing and communication, medical ultrasonography and constructing novel surface acoustics waves (SAW) devices.

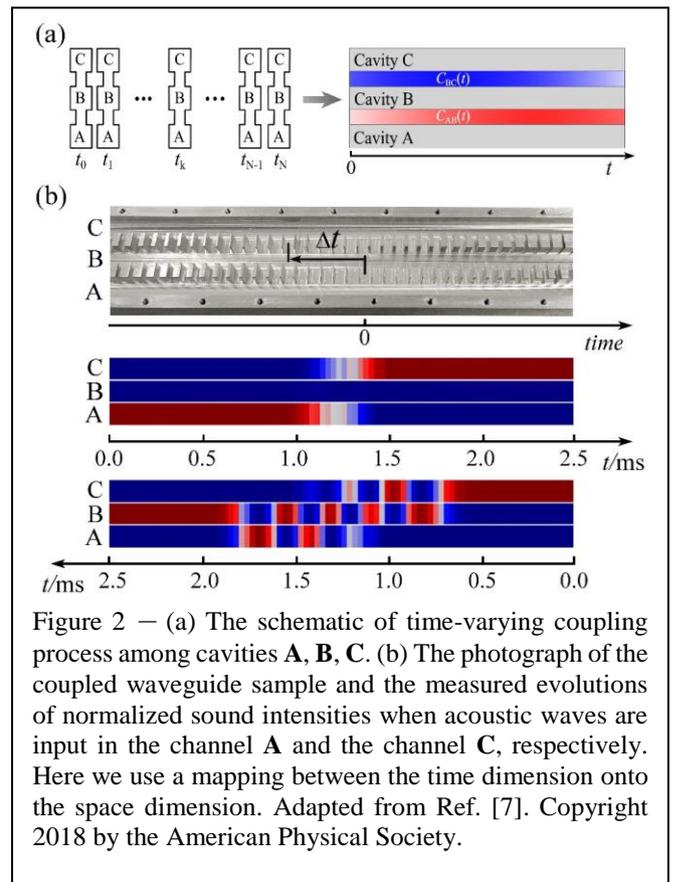

Figure 2 − (a) The schematic of time-varying coupling process among cavities **A**, **B**, **C**. (b) The photograph of the coupled waveguide sample and the measured evolutions of normalized sound intensities when acoustic waves are input in the channel **A** and the channel **C**, respectively. Here we use a mapping between the time dimension onto the space dimension. Adapted from Ref. [7]. Copyright 2018 by the American Physical Society.

In addition, future studies of acoustic STIRAP will promote the physics research in acoustic systems, such as acoustic diodes [4], Chern topological insulators [6], and Floquet topological insulators [10]. It will open up a vast virgin land in flexible and robust controls of nonreciprocal sound propagation.

**Acknowledgments** − We thank Ya-Xi Shen and Yu-Gui Peng for useful discussions. The project presented here was supported by the National Natural Science Foundation of China (Grant Nos. 11674119, 11774297, 11690030, 11690032). Jie Zhu acknowledges the support from the Early Career Scheme (ECS) of Hong Kong RGC (Grant No. PolyU 252081/15E).

## A4.1 STIRAP in Ions and Ion Strings –

Michael Drewsen, Aarhus University

### Status

Having the atomic or molecular system of interest trapped and cooled can lead to long term spatial localization, which significantly expands the laser parameter space for efficient STIRAP. In this context trapped and laser cooled atomic ions represent a nearly ideal case since such systems can be cooled to microkelvin temperatures and spatially localized within 100 nm for times of at least several seconds.

The first STIRAP-experiments with trapped atomic ions were carried out with nine $^{40}$Ca$^+$ ions Doppler laser-cooled into to a string-like equilibrium configuration [1]. As shown in Fig. 1(a), the STIRAP configuration was here of the $\Lambda$-type with $|1> = 3d\ ^2D_{3/2}$ ($|M_J|=3/2$), $|2> = 4p\ ^2P_{3/2}$ ($|M_J|=3/2$), and $|3> = 3d\ ^2D_{5/2}$ ($|M_J|=3/2$). The ions were all laser-cooled to a few millikelvin and optically pumped into state $|1>$ prior to applying few-$\mu$s STIRAP pulses driving the $|1> - |2>$ and $|3> - |2>$ transitions at 850 nm and 854 nm, respectively. This is a standard STIRAP-transfer into the state $|3>$. The STIRAP pulses were created by electrooptical switching of CW laser beams with a few mW of output power focused to a waist of ~50 $\mu$m. The quantum state of the nine ions was read out individually by a CCD-camera resolving the fluorescence at 397 nm from the individual ions when subjected to light fields being simultaneously resonant with the $|1>$ - $|E> = 4p\ ^2P_{1/2}$ and $|E> - |G> = 4s\ ^2S_{1/2}$ transitions (see Fig. 1(a)). Successful population transfer by STIRAP to the $|3>$ state hence results in a lack of fluorescence. Figure 1(b) presents CCD pictures for individual experimental runs with different delays of the STIRAP pulses. From these five pictures one clearly sees that the fluorescence measurement is a quantum state projection measurement (the individual ions are either dark or bright). The optimal STIRAP delay for the chosen laser pulse parameters is around 3 $\mu$s, where only one ion is fluorescing. By averaging over several runs, STIRAP transfer efficiency close to 95% was achieved using non-phase locked lasers. More recently, STIRAP has been applied to single $^{88}$Sr$^+$ ions in a ladder configuration ($|1> = 4d\ ^2D_{5/2} -> (|2> = 6p\ ^2P_{3/2} -> (|3> = 42s\ ^2S_{1/2})$ with the same level of efficiency to promote Rydberg state excitations [2].

A slightly more elaborated "inverted" STIRAP-scheme, involving a single state coupled to three higher lying states by individual laser fields, has been utilized to realize purely geometrically (or holonomically) single ion qubit gates [3]. The experiments were based on a single trapped and ground state laser sideband-cooled $^{40}$Ca$^+$ ion with $|u> = 4s\ ^2S_{1/2}$ ($M_J$=-1/2) being the unoccupied lower lying state to which the three states $|0> = 3d\ ^2D_{5/2}$ ($M_J$=-3/2), $|1> = 3d\ ^2D_{5/2}$ ($M_J$=-1/2) and $|2> = 3d\ ^2D_{5/2}$ ($M_J$=-5/2) are coupled by three phase-locked lasers resonant with the respective quadrupole transitions around 729 nm as sketched in Fig. 2 (a). Through a pulse sequence sketched in Fig. 2 (b), arbitrary single qubit rotation can be performed to a superposition of the $|0>$ and $|1>$ states by controlling the ratio of the Rabi frequencies $\Omega_0$ and $\Omega_1$, as well as a change in the phase $\phi_2$ of the field coupling the $|u>$ and $|2>$ states at the point ($\Omega_2$=0) where the first of two STIRAP sequences have progressed. In Ref. [3], different single qubit gate operations were realized with fidelities around 95%. Besides being an interesting result from a quantum technology perspective, this experiment proves that with laser-cooled and tightly confined trapped ions one can even perform STIRAP by addressing weak quadrupole transitions.

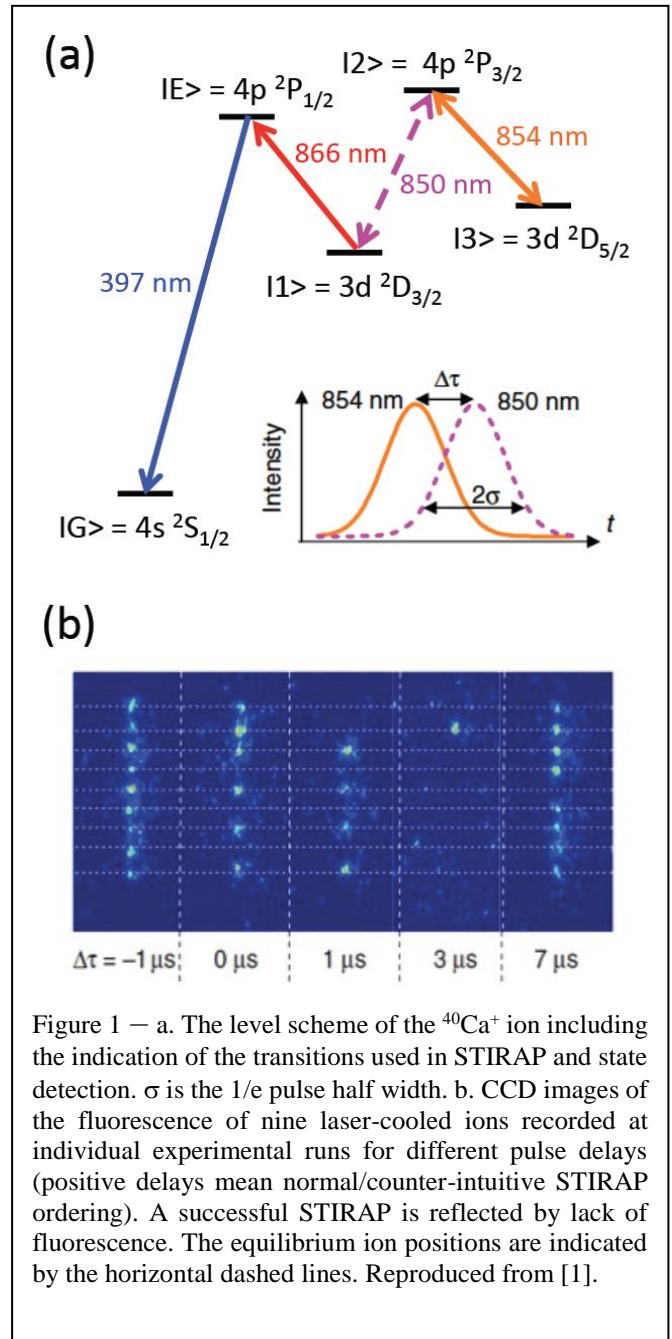

Figure 1 — a. The level scheme of the $^{40}$Ca$^+$ ion including the indication of the transitions used in STIRAP and state detection. $\sigma$ is the 1/e pulse half width. b. CCD images of the fluorescence of nine laser-cooled ions recorded at individual experimental runs for different pulse delays (positive delays mean normal/counter-intuitive STIRAP ordering). A successful STIRAP is reflected by lack of fluorescence. The equilibrium ion positions are indicated by the horizontal dashed lines. Reproduced from [1].

Finally, STIRAP has also been considered in various cavity QED settings with atomic ions. The first investigations have been towards deterministic single photon production from a single ion by having one of the two STIRAP transitions coupled strongly to one of the initially empty modes of an optical cavity, while the other transition is addressed by a freely propagating classical laser field [4,5]. In case of sufficiently strong coupling to the cavity mode, and proper pulse-shape of the classical laser field, one can achieve a STIRAP transfer of population triggering the generation of a single cavity photon. While several groups have produced single photons from single ions [4,5], fully deterministically single photon sources have yet not been demonstrated. However, with ensembles of cold ions in the form of larger Coulomb crystals sufficiently strong collective coupling to a cavity mode for potential single photon generation by STIRAP have been demonstrated [6]. The demonstration of EIT in the same system [7] looks promising for the realization of a quantum memory for single photons [8], as well as a photon number detector [9].

## Current and Future Challenges

STIRAP has already been used in a series of proof-of-principle experiments within the scope of quantum technology. Before reaching the device-level, a few issues like scalability for use in holonomical quantum computing [3] and extremely high fidelity in storage and retrieval of quantum states of individual photons have to be demonstrated. The challenges seem though more likely to be associated with issues other than fundamental limitations set by STIRAP. The application of STIRAP to cold molecular ion research is still to be explored. Based on the very early demonstration of STIRAP in molecular beam experiments, one will imagine STIRAP to become an important tool for, e.g., state-to-state cold and ultracold chemistry [10]. In this context, application of STIRAP to molecular ions beyond diatomics will remain a challenge.

## Advances in Science and Technology to Meet Challenges

Today, where optical frequency combs are available for phase locking of different lasers ranging from near infrared to deep ultraviolet, on the laser technology side, light sources suitable to STIRAP with ions are not any longer a serious issue. The advances needed to exploit the full potential of STIRAP for ion based research and quantum technology seem more to lie within ion trap technology and cooling. In particular, imperfect ion trapping fields and field noises are outstanding issues.

## Concluding Remarks

While STIRAP have already been applied within cold ion research, it is fair to state that STIRAP is still in its infancy in this context. The quality of the few experiments carried out so far indicate, however, a strong potential that will presents its full power only when technical issues related to ion trapping and cooling are resolved. Since currently such issues are at the heart of the current development of ion based quantum technologies, it seems very likely that STIRAP has a bright future within cold atomic as well as molecular ion research.

**Acknowledgments** – The author acknowledges support from the Danish Council of Independent Research through the Sapere Aude Advanced grant.

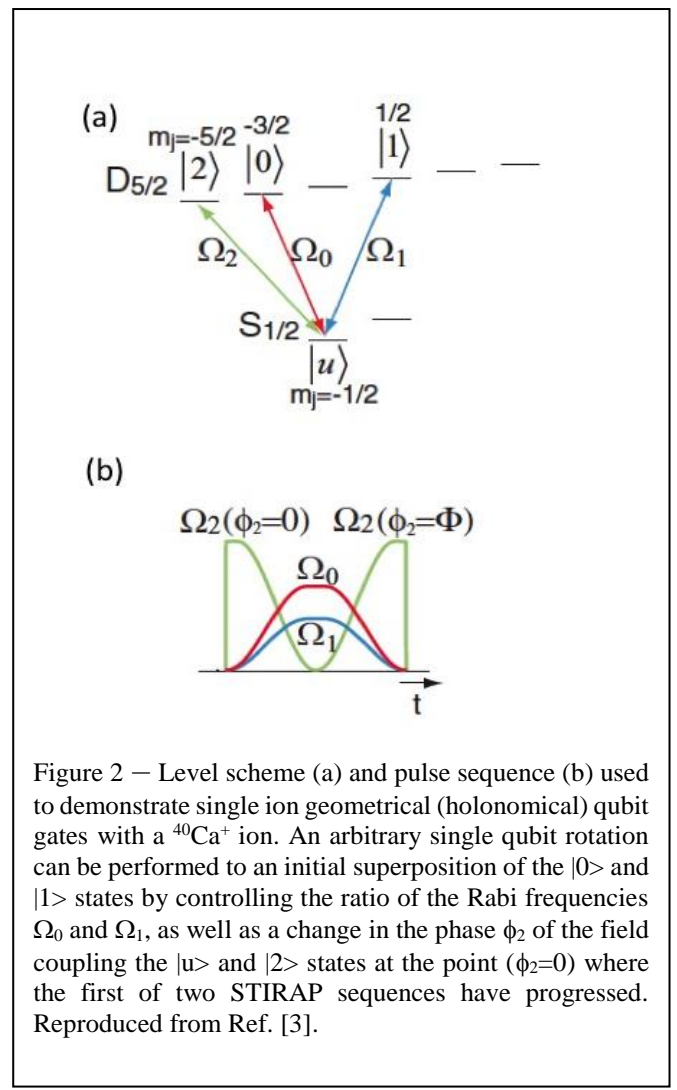

Figure 2 – Level scheme (a) and pulse sequence (b) used to demonstrate single ion geometrical (holonomical) qubit gates with a $^{40}Ca^+$ ion. An arbitrary single qubit rotation can be performed to an initial superposition of the |0> and |1> states by controlling the ratio of the Rabi frequencies $\Omega_0$ and $\Omega_1$, as well as a change in the phase $\phi_2$ of the field coupling the |u> and |2> states at the point ($\phi_2=0$) where the first of two STIRAP sequences have progressed. Reproduced from Ref. [3].

## A4.2 STIRAP for Quantum Information Processing with Global Radiation Fields

*Winfried K Hensinger and Sebastian Weidt*
University of Sussex

### Status

Building practical quantum computers is considered one of the holy grails of modern science. Trapped ions constitute a promising quantum system currently being used towards the implementation of practical quantum computers. Trapped ion quantum gates have traditionally been implemented using laser radiation and this has led to world leading entangling gate fidelities. Nevertheless, when laser beams are involved, they have to be carefully stabilised in frequency, amplitude and phase and need to be aligned to the µm level. This poses a significant challenge when scaling to practical quantum computers which may feature millions or billions of qubits. An alternative is the use of more robust microwave technology which allow implementing quantum gates with easy to stabilise microwave or rf fields which are broadcast across all qubits from global emitters. This opens up a promising path to building quantum computers and has motivated the idea of quantum computing with global fields [1]. Instead of implementing quantum gates with lasers, global rf-fields allow the execution of quantum gates via the application of voltages to a microchip as shown in Fig. 1, a mechanism resembling the operation of classical transistors. Here, in stark contrast to other proposals for quantum information processing with trapped ions, the number of radiation fields (such as lasers or microwave fields) required for quantum gate implementation does not scale with the number of qubits. The number of fields scales with the number of different types of gates to be performed, making it inherently very scalable. This new approach to quantum information processing relies on the seminal work by Mintert and Wunderlich in 2001 who showed that combining a static magnetic field gradient with externally applied long-wavelength radiation creates a sizable effective Lamb-Dicke parameter [2]. This permits to drive transitions that also change the motional state of the ion therefore enabling the execution of entangling gates with trapped ions. However, their scheme requires the use of states with different magnetic moments, ruling out the use of a so-called clock qubit. Therefore, naturally occurring magnetic field fluctuations would constitute a limit to the achievable quantum gate fidelities such an approach may permit. A promising avenue to circumvent this drawback makes use of microwave "dressed states" [3,4] where one can quantum engineer an effective clock qubit that is highly protected from magnetic field fluctuations while giving rise to a state-dependent force in the presence of a static magnetic field gradient. Microwave dressed states can be created in the $^2S_{1/2}$ ground-state hyperfine manifold of an $^{171}Yb^+$ ion [3,4].

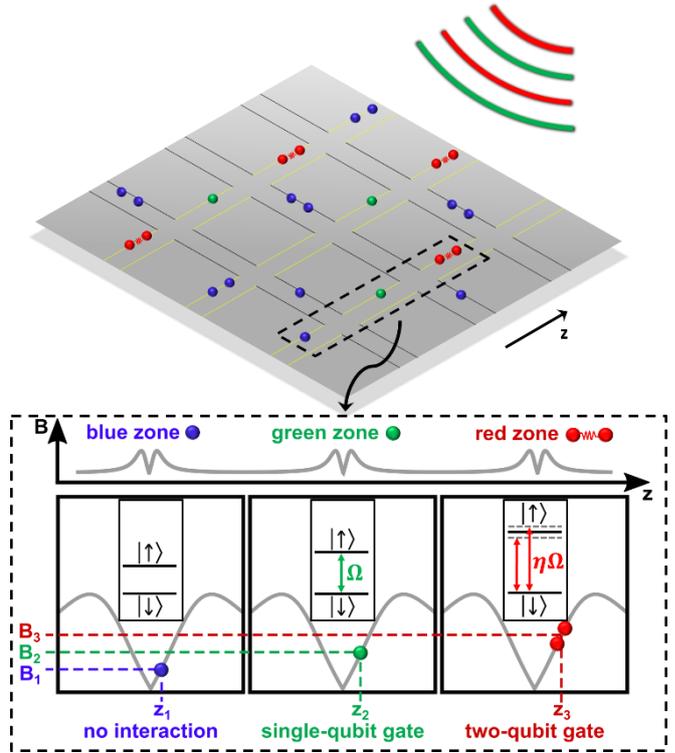

Figure 1: Ions are confined in a two-dimensional X-junction surface trap architecture which contain a single gate region in every X-junction. The inset shows three such gate regions. Voltages applied to local trap electrodes in each gate region place the ion in the correct part of the local B-field gradient making that ion resonant with a particular set of global radiation fields, effectively executing a particular quantum gate (Figure reproduced with permission from ref. [1]).

This $^2S_{1/2}$ manifold consists of four states, of which $|0\rangle \equiv |F = 0\rangle$ and $|0'\rangle \equiv |F = 1, m_F = 0\rangle$ are first order insensitive to magnetic fields, while $|+1\rangle \equiv |F = 1, m_F = +1\rangle$ and $|-1\rangle \equiv |F = 0, m_F = -1\rangle$ are magnetic field sensitive. Coupling the magnetic field sensitive states $|+1\rangle$ and $|-1\rangle$ to the magnetic field insensitive state $|0\rangle$ by two resonant microwave fields, as shown in Fig. 2(a) leads to a three-level dressed state system. Setting the two dressing field Rabi frequencies to be equal , moving to the interaction picture and performing the rotating wave approximation gives rise to a new Hamiltonian with three eigenstates, so called dressed states. Using one of these eigenstates, $|D\rangle = \frac{1}{\sqrt{2}}(|+1\rangle - |-1\rangle)$ along with the atomic state $|0'\rangle$ allows creating a quantum engineered clock qubit. This qubit is resilient to magnetic field noise demonstrated by a nearly three order of magnitude increase in decoherence time compared to a bare state qubit. Furthermore qubit splitting can be tuned by the application of a static

magnetic field making it ideal for the method of quantum information processing with global fields. We have demonstrated that it can be easily manipulated with a single rf field tuned to either the |0'> to |-1> or the |0'> to |+1> transition [4]. STIRAP was involved in the first realisation of such dressed states [3] as well as in a subsequent experiment to demonstrate a much simpler qubit manipulation method [4]. In both experiments, the ion is initalized in state |0> and the population is transferred via a π pulse to |+1>. Subsequently, the first half of a STIRAP process is carried out in order to prepare the dressed state |D> as shown in Fig. 2(c). Then, both microwave fields are kept at constant amplitude in order to allow for the execution of quantum gates during time $t_h$ within the dressed state qubit basis. Subsequently, population is transferred to |-1> via the second half of the STIRAP process. A final π pulse on the |-1> - |0> transition subsequently maps the dressed state qubit into the atomic dark state – bright state

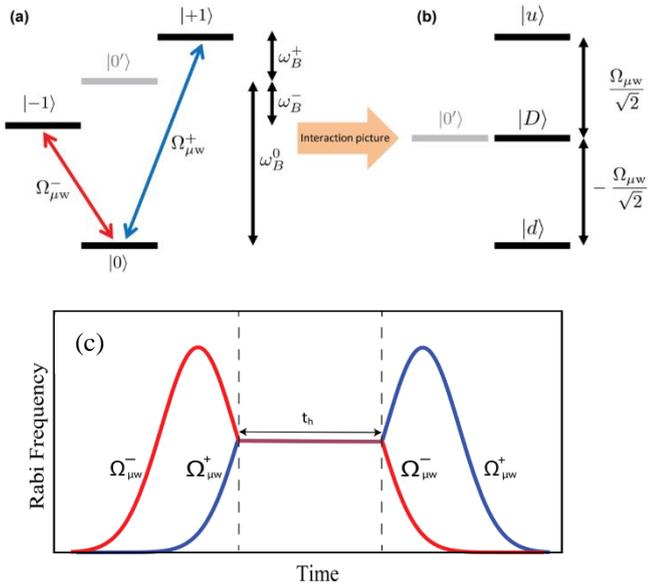

manifold for detection. It has been verfied that the population transfer is robust in terms of fluctuations in the pulse area, an advantage of using STIRAP [3].

Figure 2: (a) Energy level diagram of the $^2S_{1/2}$ ground-state hyperfine manifold of $^{171}Yb^+$ ion and (b) the resultant energy diagram in the dressed state qubit basis. (c) Illustration of the STIRAP process where the microwave fields are ramped adiabatically in a particular order that transfers population from |+1> to the dressed state |D>. The fields are then held at equal Rabi frequencies for a hold time $t_h$, during which quantum gates in the dressed basis can be performed. Finally, the second half of the STIRAP process is carried out, transferring any population in |D to |−1> (Figure reproduced with permission from ref. [5]).

## Current and Future Challenges

While the use of STIRAP indeed provided a powerful tool for the preparation of dressed states, the contrast of the Rabi oscillations between the dressed state qubit states (Fig. 3 in ref. [2], Fig. 3 in ref. [3]) was found to be limited by the transfer fidelity of only ≈ 0.93 for the STIRAP process. Furthermore, the use of a magnetic field sensitive transition in the preparation of the STIRAP sequence may give rise to detrimental decoherence. With an alternative approach, namely the preparation of population in the |0'> state via a clock transition π-pulse followed by turning on the dressing fields, we were able to demonstrate dressed state qubit Rabi oscillations with a contrast of 0.99(1) [5]. This method would indeed be capable of achieving high enough fidelities towards fault-tolerant quantum information processing. However, this type of mapping between the dressed state qubit and the atomic qubit does not preserve coherences because the phase relationship between the two qubit states is scrambled when turning off the dressing fields before the final mapping step that is part of this particular mapping method. While this is acceptable at the end of the execution of a quantum algorithm, it precludes coherent mapping back and forth between the two qubit types during the algorithm. This observation demonstrates the importance of coherent control methods such as STIRAP for quantum information processing compared to the simple pulse method demonstrated in ref. [4]. Considering the practical implementation of quantum information processing with global fields as discussed in ref. [5], it is clear that retaining the qubit as a dressed qubit (|0'>,|D>) throughout the full computation may be challenging – because shuttling operations will be required in between quantum gate operations. We have therefore developed a new qubit mapping technique that incorporates high-fidelity mapping of both populations and coherences [7]. While there are plenty of quantum control methods available for two level systems, few coherent control methods are available for multilevel systems. We exploited the equivalence between multilevel systems with SU(2) symmetry and spin-1/2 systems to develop a technique for generating new multilevel control methods derived from commonly used two-level methods [7]. We demonstrated this technique by proposing and implementing two multi-level coherent control methods which have been derived from the well-known two level techniques, rapid adiabatic passage [8] and composite pulses [9]. Both methods do not just map populations but also coherences. We measured the average infidelity of the process in both cases to be around $10^{-4}$. To the best of our knowledge, these infidelities constitute one of the best fidelities ever realized with any coherent control method. We also demonstrated significant robustness to pulse area error and realised a significant mapping speedup over traditional adiabatic mapping techniques by using a resonant composite pulse sequence [7].

## Advances in Science and Technology to Meet Challenges

Increases in the speed of qubit mapping can be realized by increasing relevant Rabi frequencies as well as experimenting with different coherent control methods. It is obvious that many coherent control methods are intimately related and their particular choice depends on the exact atomic system to be considered along with other practical considerations. Our work in ref. [7] may even provide an inspiration to expand STIRAP to higher level systems. While the focus in this chapter is on the implementation of qubit mapping, it is important to point to the potential of coherent control methods to increase fidelities for quantum gates allowing to minimize a wide range of error terms. Coherent control methods can also be used to make quantum gate methods more resilient to parameter fluctuations. Indeed, we recently demonstrated a dramatic improvement in making two qubit quantum gates more resilient to fluctuations in operational parameters and motional heating [10].

## Concluding Remarks

Because of the exquisite control possible in the generation of multi-tone microwave and rf signals afforded by advances in mobile phone and radar technologies, coherent control methods such as STIRAP, rapid adiabatic passage and composite pulses form a powerful tool to enhance fidelities in the quest to build practical quantum computers. These techniques have the potential to help reducing the size of a practical quantum computer significantly and to enable the practical construction of such a device.

**Acknowledgments –** The work discussed here was carried out by members of our group and collaborators, in particular, the authors of Ref. 1, 4, 5, 6, 7, and 10. We acknowledge support from the European Commission's FET-Flagship on Quantum Technologies project 820314 (MicroQC), the U.K. Quantum Technology hub for Networked Quantum Information Technologies (No. EP/M013243/1), and the U.K. Quantum Technology hub for Sensors and Metrology (No. EP/M013294/1), the U.S. Army Research Office under Contract No. W911NF-14-2-0106 and the University of Sussex.

### A5.1 STIRAP in rare-earth ion doped crystals

*Thomas Halfmann*
Technical University of Darmstadt, Germany

**Status**

Realistic future applications of quantum information technologies require media, which offer large storage capacity, scalability, robust handling, and the potential to integrate them into larger data processing architectures. Hence, while quantum information science started from the background of atomic physics, quickly also solid media attracted the attention of researchers. However, typical bulk crystals with broad band structures suffer from ultra-fast decoherence processes, which is detrimental for coherent storage or processing of quantum information. Specific classes of "atom-like" solid media combine the above advantages of solids (i.e., large density, scalability, robustness, capability of integration) and gases (i.e., spectrally narrow transitions and long coherence times). Examples for such media are colour centers, semiconductor quantum dots, or rare-earth ion doped crystals (REICs). In the latter case, rare-earth dopand ions are embedded during the crystal growth process in a host lattice. The specific level structure of rare-earth ions shields the valence electron from the host lattice, which enables spectrally very sharp optical absorption peaks in the solid. REICs are already well-known for many decades as efficient laser crystals and classical optical memories.

Roughly 20 years ago research started on adiabatic interactions in REICs, with pioneering work on electromagnetically-induced transparency (EIT) in a $Pr^{3+}:Y_2SiO_5$ crystal (termed Pr:YSO from now on) [1]. Similar to STIRAP, also EIT relies on adiabatic passage, driven by two radiation fields in a three-level $\Lambda$-type scheme. Thus, the first developments on EIT in a REIC were followed soon by successful demonstrations of rapid adiabatic passage (RAP), several variants of STIRAP [2-6] (see Figure 1), applications of cyclic STIRAP processes for classical logic operations [7], and the implementation of an optical memory by stopping light pulses with EIT [8,9]. The latter set new benchmarks for long storage times and large storage efficiencies in an EIT-driven light storage protocol, applicable as a quantum memory.

An EIT memory typically aims at storing a weak pump pulse (usually termed "probe" in the EIT protocol), which is coincident with an intense Stokes pulse (usually termed "control" in the EIT protocol). The control pulse modifies the speed of light in the medium, such that the probe pulse is decelerated and adiabatically stored in an atomic coherence, i.e., a superposition of the two bare ground states in the medium. The final goal is to store quantum information encoded in quantum states of light (i.e., single photons). Hence, the pulse configuration in EIT differs from conventional STIRAP, where both pulses are strong and delayed. Nevertheless, the EIT memory protocol resembles fractional STIRAP (see the contribution by K Bergmann for a definition of fractional STIRAP), enabling the generation of an arbitrary atomic coherence by switching off the pump and Stokes pulse during the interaction, or changing the intensity or delay of the two driving pulses. Such, the information encoded on the probe pulse is stored as a spatio-temporal variation of atomic coherences in the quantum memory, i.e., a "spin wave".

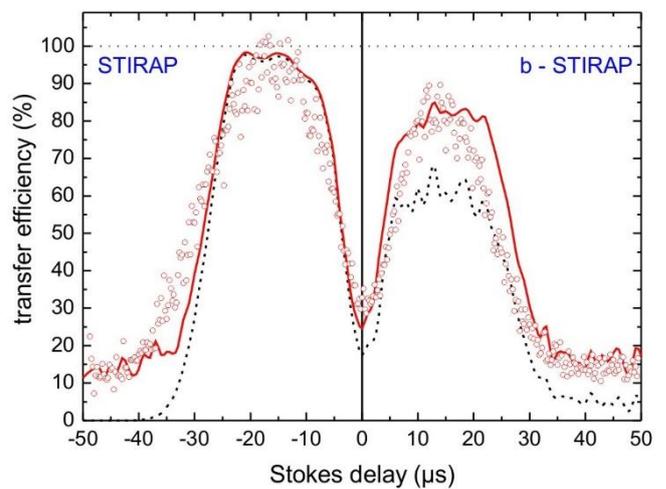

Figure 1 — Transfer efficiency between hyperfine states in Pr:YSO vs. delay between Stokes (S) and pump (P) pulse by STIRAP (conventionally via the dark state) and b-STIRAP (via the bright state) (from [2]). Red, hollow circles show experimental data. Solid lines shows simulations (red: full simulation with laser frequency jitter, black: without jitter). Note, that b-STIRAP works only well for long lifetime of the intermediate state, while this is irrelevant for STIRAP.

**Current and Future Challenges**

An EIT-based quantum memory protocol in REICs typically involves the following steps: (i) Preparation of the inhomogeneously broadened medium by optical pumping schemes. (ii) Storage of a probe pulse by stopped light with EIT and generation of a spin wave during. (iii) Rephasing or dynamical decoupling of the spin wave to compensate for inevitable dephasing in the inhomogeneous manifold. (iv) Readout by back-conversion of the spin wave to an optical signal. Adiabatic passage is applied via EIT in the storage and retrieval stage, but also provides advantages in the rephasing stage. This potential was already demonstrated by the superior performance of RAP for rephasing in a REIC memory, compared to $\pi$-pulses.

In the simplest configuration, dynamical decoupling typically involves a large number of identical radiofrequency pulses on the spin transition between the ground states. The aim is a fast and cyclic inversion of the population distribution in the atomic coherences, such that the net dephasing remains as small as possible. The spin wave is temporally refocused then for an efficient readout process. However, radiofrequency sequences are slow, which limits the repetition rate of the write/read processes in the memory. It would be desirable to accompany the optical write/read stages by optical rephasing sequences on the two-photon Raman transition (rather than on the radiofrequency spin transition). This would provide an all-optical memory protocol, and with shorter pulses at increased repetition rate. STIRAP could play an important role here, as it enables adiabatic population switching between the spin ground states, which is required for rephasing.

Moreover we note, that the frequency bandwidth of an adiabatically-driven memory is rather small. EIT (or fractional STIRAP) use a spectrally sharp two-photon transition between the ground states in a Λ-type scheme. Therefore, the memory frequency bandwidth at reasonable optical depth is typically a few 100 kHz only in REICs. This is not yet sufficient to provide a truly broadband memory, which must enable storage of short data pulses at high repetition rate. Hence, it would be important to broaden the frequency bandwidth of the adiabatic passage process for memory applications.

### Advances in Science and Technology to Meet Challenges

Applications of STIRAP for optical rephasing, or also in the write/read steps of a quantum memory protocol require improvements of the adiabatic passage process. As an obstacle for rephasing of atomic coherences, STIRAP is very sensitive to variations in the initial population distribution, i.e., its deviation from a pure initial state. This is a problem for applications of STIRAP to rephase a quantum memory, as the latter aims at storing the state of a single photon in atomic coherences with a very small amplitude in the excited state (corresponding to excitation by a single quantum).

As already theoretically proposed, it is possible to improve the robustness of adiabatic passage processes by application of composite pulse sequences. The latter approach originally stems from nuclear magnetic resonance, but meanwhile also found its way into varies other fields, such as trapped ions, quantum information processing, nonlinear optics, or optical retarders. It replaces a single excitation pulse (or pulse pair, as in STIRAP and EIT) by a sequence of pulses. The relative phases of the pulses serve as control parameters to steer the quantum system on a robust path through Hilbert space towards a desired target state.

Recent experiments in REICs demonstrated composite versions of STIRAP [10], compensating for variations in the driving laser intensities and pulse delays. The composite sequences enabled stable and highly-efficient population transfer in a much enlarged range of experimental parameters. While the latter demonstration aimed at variations in specific parameters, appropriate composite sequences will permit robustness with regard to the fluctuation of any arbitrary parameter. Thus, it seems a straightforward extension of composite pulses to develop specific composite sequences of STIRAP, which could be robust with regard to variations of the initial population distribution, e.g., to invert any arbitrary initial atomic coherence.

As composite sequences can be applied to optimize the performance of coherent excitation with regard to any arbitrary parameter, composite adiabatic passage by STIRAP or EIT could also broaden the frequency bandwidth of the process. It was already shown by implementations of composite approaches in applied optics, that improved narrow-, pass-, or broadband operation is possible by such approaches. However, we must take the basic difference between STIRAP for population transfer and EIT (or fractional STIRAP) for storage of an optical pulse into account: Composite STIRAP essentially cycles a quantum system back-and-forth between two ground states by sequences of pump-Stokes pulse pairs with discrete phase jumps inbetween. On the other hand, the storage of a single data pulse by EIT must be implemented during a single cycle. Hence, the composite approach with several separated pulse pairs is not applicable in the write/read steps of an EIT memory. Nevertheless, it may be feasible to vary the phase of the Stokes (control) pulse during the storage process. This is related to the concepts of optimal control theory, shortcuts to adiabaticity, or single-shot shaped pulses (SSSPs), which use a continuous variation of the pulse phase and/or intensity as control parameters to drive robust excitation pathways.

### Concluding Remarks

STIRAP and related adiabatic processes are already thoroughly investigated methods to coherently manipulate atomic populations and coherences also in REICs. The developments of new versions of adiabatic passage processes in recent years, in particular in combination with composite pulses, exhibits an encouraging route to boost the use of STIRAP also for applications in quantum memories. Most probably, these developments will optimize the rephasing/dynamical decoupling stage of the quantum memory, possibly also the preparation sequences by optical pumping, and may also exhibit potential to improve the write/read processes, which are at the heart of the protocol. However, all of this will require variations of the initial concept of STIRAP, to match it to the demands of a quantum memory protocol.

### A5.2 Adiabatic population transfer in diamond nitrogen vacancy centers

*Hailin Wang*
University of Oregon

## Status

Negatively-charged nitrogen vacancy (NV) centers in diamond have recently emerged as a promising solid-state spin qubit system for quantum information processing[1,2]. NV centers feature long spin decoherence time, along with highly efficient state preparation and single-shot optical readout. While high-fidelity quantum control of individual NV spin qubits via microwave transitions has been well established, adiabatic population transfer such as STIRAP provides an avenue for optical spin control. In addition, the adiabatic process can in principle be robust against inevitable variations in relevant experimental parameters.

The need for STIRAP in NV centers is further motivated by recent interests in the development of mechanical or phononic quantum networks of robust spin qubits as an experimental platform for spin-based quantum computers[3]. Spin-mechanical interactions of NV centers can take place via either ground-state or excited-state strain coupling, for which long wavelength acoustic vibrations lead to energy shifts or state-mixings of the relevant energy levels. The robust room-temperature ground-state spin coherence of NV centers dictates that the ground-state strain coupling is extremely weak. In comparison, much stronger strain coupling can take place via the orbital degrees of freedom of the NV excited states[4,5]. The STIRAP process can enable the use of the excited-state strain coupling for spin-mechanical interactions in a phononic quantum network, while still employing the ground spin states as the qubits. In this case, a key limiting factor is the nonadiabatic excitation of the NV excited states. Because of the rapid optical spontaneous emission from these excited states, even a relatively small population in the excited states can lead to an optically induced decoherence rate that is large compared with the relevant spin decoherence rate.

The basic STIRAP process has been experimentally realized in a Λ-type three-level system in a single NV center (see Fig. 1a)[6]. This study also demonstrates that the STIRAP process can be robust against spectral diffusion of NV optical transitions, which would otherwise be difficult to overcome, as long as the adiabatic condition is satisfied. Accelerated STIRAP processes in a NV center have also been demonstrated with specially-designed temporal shapes of optical pulses[7]. The temporal pulse shaping leads to superadiabatic transitionless driving (SATD)[8]. Note that replacing one of the dipole optical transitions in the

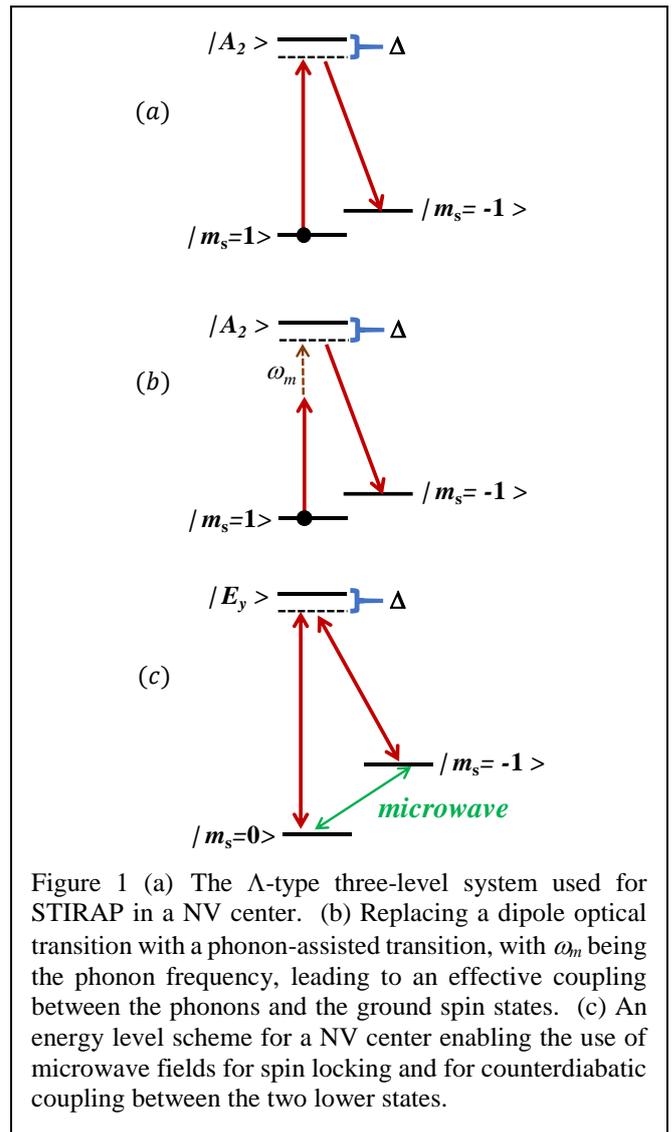

Figure 1 (a) The Λ-type three-level system used for STIRAP in a NV center. (b) Replacing a dipole optical transition with a phonon-assisted transition, with $\omega_m$ being the phonon frequency, leading to an effective coupling between the phonons and the ground spin states. (c) An energy level scheme for a NV center enabling the use of microwave fields for spin locking and for counterdiabatic coupling between the two lower states.

Λ-type three-level system in Fig. 1a with a phonon-assisted transition, as shown in Fig. 1b, leads to an effective interaction between the phonons and the ground spin states[5]. This spin-mechanical interaction is mediated by the excited-state strain coupling.

## Current and Future Challenges

Both adiabatic passage experiments discussed above are limited by nonadiabatic excitations of the NV excited states. A STIRAP process follows adiabatically a dark instantaneous eigenstate, which is controlled by two external driving pulses with a counterintuitive pulse sequence. This dark state is a coherent superposition of the two lower states in a Λ-type three-level system. The finite lifetime of the spin coherence between the two lower states leads to the decay of the dark state, thereby inducing nonadiabatic excitations. Diamond with a natural abundance of $^{13}C$ nuclei, which act as a nuclear spin bath, features a relatively short spin dephasing time of one to several μs, even though much longer spin dephasing times can be obtained in isotopically-purified diamond.

Shortcuts to adiabatic passage (STA) can be used to speed up relatively slow STIRAP processes. One approach employs specially-designed temporal pulse shapes to achieve SATD, as discussed earlier[8]. Another approach drives a direct coupling between the two lower states with an auxiliary pulse. This so-called counter-diabatic coupling can in principle keep the system in the dark instantaneous eigenstate by cancelling precisely nonadiabatic transitions to other instantaneous eigenstates[9]. For NV centers, neither of these approaches, however, can effectively suppress the nonadiabatic excitation of the NV excited states to the desired level. The current SATD approach allows small, but still considerable excitation of the excited states. The counterdiabatic coupling relies on the coherence of the two lower states, which is still limited by the spin dephasing induced by the nuclear spin bath.

## Advances in Science and Technology to Meet Challenges

In addition to the use of isotopically-purified diamond, there are a number of experimental approaches that can be exploited to circumvent effects of spin dephasing. Time-domain approaches based on dynamical decoupling can suppress effects of spin dephasing with a sequence of π-pulses. This fast pulse sequence, however, is incompatible with the adiabatic passage. Alternatively, the spin dephasing process can be suppressed with a spectral domain or so-called spin-locking approach by using a microwave field to dress the electron spin states. In this case, the energy gap between the dressed spin states can protect the spins from magnetic fluctuations induced by the nuclear spin bath[10]. The experimental challenge is thus to realize STIRAP with the dressed spin states. With strain or magnetic-field induced excited-state mixing, we can engineer a Λ-type three-level system in a NV center such that the two lower states are coupled by a microwave transition (see Fig. 1c), which can be used for spin-locking as well as counterdiabatic coupling. The combination of the spin-locking and the counterdiabatic coupling can in principle suppress the optically-induced decoherence rate induced by the nonadiabatic excitation of the excited state to far below the intrinsic spin decoherence rate.

The SATD demonstrated in the earlier experimental study accelerates the adiabatic passage process, but is still subject to considerable nonadiabatic excitations of the NV excited-states. Earlier theoretical work has attempted to design temporal pulse shapes that can minimize the nonadiabatic excitation. Further theoretical efforts are still needed to explore if the use of both temporal and spectral shaping of the optical pulses can suppress the nonadiabatic excitation to the desired or an arbitrarily small level. Alternatively, optical pulse shaping techniques can be combined with the counterdiabatic coupling to cancel the residual excited-state population, though this will increase the complexity of the experimental implementation. It will be interesting to see what experimental techniques or combinations of experimental techniques can achieve the desired high fidelity for adiabatic population transfer in NV centers. For applications of NV centers in a phononic quantum network, these techniques also need to be compatible with the operations of the quantum network.

## Concluding Remarks

With the experimental realization of STIRAP in NV centers, the next milestone is to achieve high fidelity adiabatic passage such that the optically-induced decoherence rate due to the residual excited-state population can be suppressed to levels below the relevant spin decoherence rate. The resulting adiabatic passage can be used for optical spin control and perhaps more importantly for spin-mechanical interactions mediated by the excited-state strain coupling. The latter application can play an essential role in the development of phononic quantum networks for spin-based quantum computers.

**Acknowledgments –** This work is supported by AFOSR and by NSF under grants No. 1606227, No. 1641084, and No. 1719396.

## A5.3 Applications of STIRAP in superconducting quantum circuits

G. S. Paraoanu
Aalto University

### Status

Advances in nanofabrication and cryogenic microwave measurements during recent years have allowed the realization of electrical devices such as quantum dots and superconducting circuits. These artificial structures behave in accordance with quantum physics, displaying similar discrete energy-level structures as the atoms, therefore one expects that the same state-preparation protocols apply. In particular, in circuit electrodynamics, precursors of STIRAP experiments include the observation of the Autler-Townes effect [1] and phenomena based on the formation of dark states, such as coherent population trapping and electromagnetically induced transparency, see e.g. [2] and references therein.

STIRAP in a superconducting circuit (a transmon) was first realized in an experiment reported in 2016 [3]. For this, the first three levels $|g\rangle$, $|e\rangle$, $|f\rangle$ of this artificial atom were employed; the transitions between consecutive levels were addressed by the counterintuitive sequence $|e\rangle \xrightarrow{\text{Stokes}} |f\rangle$ and $|g\rangle \xrightarrow{\text{pump}} |e\rangle$, and a transfer between the states levels $|g\rangle$ and $|f\rangle$ with fidelity exceeding 80% was achieved. All the basic features of STIRAP appeared clearly, e.g. the insensitivity to single-photon detunings and shape of the pulses, and the possibility of reversing the process. Of course, the method works for any kind of superconducting qubit, not only the transmon; STIRAP with a phase qubit (transfer fidelity 67%) has been subsequently demonstrated [4].

Moreover, in circuit quantum electrodynamics STIRAP offers an interesting tool for creating highly nonclassical states. It offers a route for state preparation when direct transitions are forbidden by Hamiltonian symmetries. Combinations of Rabi pulses - used to initialize the system in a superposition of the ground state and first excited state – and STIRAP also lead to the creation of nontrivial states. This was already demonstrated experimentally [3] and further explored theoretically in [5]. In [6], a transmon embedded in a 3D aluminium cavity was used to create Fock states and superpositions of Fock states of the cavity field. Also in this case, STIRAP is useful to couple states between which there is no direct transition. To understand the procedure, recall that the Jaynes-Cummings model in the dispersive regime consists of two ladders, with rungs indexed as $|g0\rangle$, $|g1\rangle$, $|g2\rangle$, … and $|e0\rangle$, $|e1\rangle$, $|e2\rangle$, … correspondingly. A $\Lambda$-structure can be identified as formed by the states $|e0\rangle$, $|e1\rangle$, and $|g1\rangle$ and STIRAP can be run with the pump driving the $|e0\rangle \rightarrow |e1\rangle$ transition and with the Stokes on the $|e1\rangle \rightarrow |g1\rangle$ transition. Thus, a single-photon Fock state in the cavity can be created by the protocol $|g0\rangle \xrightarrow{\pi} |e0\rangle \xrightarrow{\text{STIRAP}} |g1\rangle$, a two-photon Fock state by

$$|g0\rangle \xrightarrow{\pi} |e0\rangle \xrightarrow{\text{STIRAP}} |g1\rangle \xrightarrow{\pi} |e1\rangle \xrightarrow{\text{STIRAP}} |g2\rangle$$

and so on, by further concatenating one can get any Fock state with fixed number of photons (and the qubit in the ground state), see Fig. 1

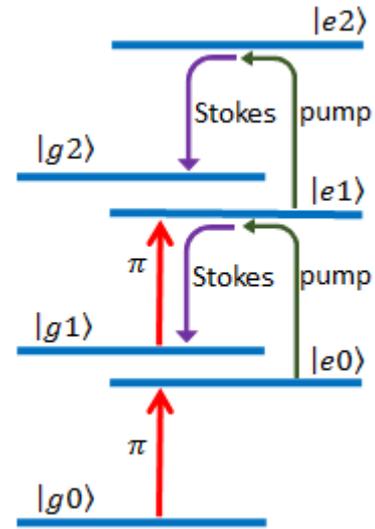

**Figure 1** Schematic of the pulse sequence for creating Fock states using STIRAP.

In the experiment [6], the fidelities were 89%, 68%, and 43% for the one-photon, two-photon, and three-photon Fock states. It is also straightforward to see that superpositions of Fock states can be created by employing a $\pi/2$ pulse instead of $\pi$ in the beginning,

$$|g0\rangle \xrightarrow{\frac{\pi}{2}} (|g0\rangle + |e0\rangle)/\sqrt{2} \xrightarrow{\text{STIRAP}} (|g0\rangle + |g1\rangle)/\sqrt{2} .$$

The experiment demonstrates again that combinations of Rabi pulses and STIRAP can lead to the creation of interesting states, as was done previously for the first three states of the transmon.

### Current and Future Challenges

For quantum processing of information, adiabatic protocols are in general very interesting because they offer robustness against variation in parameters such as the shape of the pulse and single-photon detunings. In the context of quantum processing of information using superconducting processors this would address certain technical limitations in present-day circuits that preclude the realization of high-fidelity gates. Indeed, the precise shaping of the microwave pulses is limited by the bandwidth of the arbitrary waveform generator

(typically of the order of 400 MHz); as the pulse propagates, it gets further deformed due to impedance mismatching between components (for example the connection to the chip, realized by bonding wires of finite inductance, is one such place). It is also known that the frequency of superconducting devices can have variations on the timescale of the experiment.

On the other hand, adiabatic protocols have a major disadvantage: in order to satisfy the adiabatic theorem, the time-scale of the operation has to be long in order to keep the wavefunction as close as possible to the instantaneous value. For STIRAP, this typically means a duration of many Rabi cycles. But systems such as solid-state based superconducting qubits are severely limited by decoherence, and although we have witnessed significant progress in recent times, utilizing an operation that consumes the precious resource of time of coherence is not a good long-term strategy.

Presently, various alternatives to go around this limitation are being explored. There are situations when the population of the intermediate state at intermediate times is acceptable, with the only restriction that at the end of the protocol the full population transfer is achieved. State preparation in superconducting circuits based on such protocols has been achieved. As in the adiabatic case discussed previously, the technique uses Rabi pulses to initialize certain superpositions, then applies a fast non-Abelian operation instead of STIRAP [7]. During this operation, the system does not follow the instantaneous eigenstate of the Hamiltonian, yet the parallel transport condition is still fulfilled.

Surprisingly, there is also a way to force the system to follow the instantaneous Hamiltonian eigenvector by employing so-called transitionless or superadiabatic STIRAP. This protocol (dubbed saSTIRAP) was implemented successfully in a superconducting circuit [8] by adding to the standard STIRAP a two-photon pulse specially crafted such that the spurious nonadiabatic excitations are cancelled at all times. The key element in this scheme is the generation of a complex Peierls coupling between the initial and the target states. Under a suitable choice of phases of the two STIRAP pulses and the two-photon pulse, this complex coupling becomes purely imaginary, thus realizing precisely the so-called counterdiabatic Hamiltonian. Using this scheme, the transfer of population with fidelity around 90% has been achieved even for very short pulses, in a regime where STIRAP alone would not function. The scheme has also been shown to be robust under errors in phases and amplitudes. To increase the fidelity, the main stumbling block is the existence of ac Stark shifts produced especially by the intense two-photon pulse. At the peak of this pulse, the energy levels seen by the STIRAP pulses can be detuned considerably even between the initial and target state (a detuning against which STIRAP is not robust). However, it is possible to adjust in time-domain the frequency of the STIRAP pulses in order to follow the effective energy levels and eliminate this detuning [9].

In fundamental research, STIRAP might be the ideal technique for revealing the structure of certain exotic states. Superconducting circuits have been used recently to investigate the ultrastrong coupling limit. Consider the Rabi Hamiltonian

$$H = \hbar\omega_0 a^\dagger a + \frac{\hbar\nu}{2}\sigma_z + g\,\sigma_x(a^\dagger + a),$$

which models a generic interacting qubit-oscillator system. The standard Jaynes-Cummings Hamiltonian and the associated ladder of eigenvalues are obtained from the Rabi Hamiltonian under the rotating wave approximation. However, if $g$ is of the order of $\omega_0$ and $\nu$, the rotating wave approximation fails and effects due to the Bloch-Siegert terms become apparent. The eigenvalues of the Rabi Hamiltonian also have an interesting structure, which can be unraveled by noticing that the Hamiltonian commutes with the parity operator $P = \sigma_z(-1)^{a^\dagger a}$, $[P, H] = 0$. For example the ground state, which in the case of the Jaynes-Cumming model is simply $|g, 0\rangle$, acquires an admixture of high-number of photon states, respecting the parity condition. For example, the next order term will contain the state $|g, 2\rangle$, with two "virtual" photons. STIRAP offers a way to directly put in evidence these photons, by extracting them as "real" photons into the resonator [10]. To do so, an additional energy level $|b\rangle$, highly detuned from the cavity, would be used. The initial state is therefore $|b0\rangle$, the intermediate state is the ground state of the Rabi Hamiltonian, and thus the final state is $|b2\rangle$. Thus the two virtual photons present in the ground state will have a tangible consequence – the appearance of two photons in the resonator. A key element in this scheme is the fact that the ground state has matrix elements both with $|b0\rangle$ and $|b2\rangle$, since it contains the $|g, 0\rangle$, $|g, 2\rangle$ components with zero and two photons. In the Jaynes-Cummings limit this process would be manifestly forbidden, since the ground state under the rotating wave approximation contains only $|g, 0\rangle$. We have seen before that in the case of the dispersively-coupled qubit with the 3D cavity, STIRAP was able to "move" the excitation from the qubit into the cavity [6]; in the case of ultrastrongly coupled qubit-cavity, STIRAP is making use of the binding energy which is already there in the ground state.

## Advances in Science and Technology to Meet Challenges

The superadiabatic STIRAP and the non-Abelian holonomic protocols described below are part of a larger class of methods referred to as shortcuts to adiabaticity. An interesting protocol in this class is also the method based on Lewis-Riesenfeld invariants, which has not been demonstrated yet. It is also not known which of these protocols are more resilient under the effects of decoherence. At present, to study this, machine-learning approaches are well suited. Moreover, it is possible to design quantum gates based on these principles, as already proposed in [9]. At the technical level, to implement shortcuts to adiabaticity in a superconducting qubit setup one needs to generate pulses with specific shapes and have exquisite control over their phase. Presently, waveform and continuous-signal generators which preserve the phase coherence between different channels are being developed commercially.

In the case of STIRAP used as a method to probe the ultrastrong regime, the challenge is designing such an experiment. Because the auxiliary mode needs to be decoupled from the cavity, a system with high anharmonicity such as the flux qubit could be beneficial.

## Concluding Remarks

In practical setups, at this point it is not clear which shortcut-to-adiabaticity techniques are best suited when taking into account real experimental limitations. The advantage of the invariant-based methods is the use of only two pulses, with a special time-dependence. On the other hand, the state transfer fidelity might be limited since the amplitude pulse is limited by the boundary condition. In the case of the ultrastrongly coupled systems, applying STIRAP to extract photons in the resonator could lead to a novel source of correlated photons, which can find applications in microwave electronics and nanophysics.

**Acknowledgments** –This work was done under the "Finnish Center of Excellence in Quantum Technology" (Academy of Finland project 312296).

# B1 Quantum information

### B1.1 STIRAP in quantum information

*Nikolay V. Vitanov*
St Kliment Ohridski University of Sofia

## Status

Due to its resilience to spontaneous emission and robustness to experimental errors, STIRAP has become a popular tool in quantum information. Several examples in this respect are presented below and more details can be found in a recent review [1].

*Quantum gates.* It is natural to form a qubit from the two end states 1 and 3 of STIRAP. The most general unitary transformation of this qubit reads

$$|1\rangle \rightarrow a\,|1\rangle + b\,|3\rangle, \qquad |3\rangle \rightarrow -b^*\,|1\rangle + a^*\,|3\rangle,$$

where a and b are two complex parameters ($|a|^2 + |b|^2 = 1$). One may consider fractional STIRAP [1] (see the contribution by K Bergmann for a definition of fractional STIRAP) as a candidate to realize this gate operation. Indeed, when the system is initially in state 1 fractional STIRAP can produce a coherent superposition of states 1 and 3 by interrupting the evolution at an appropriate intermediate time when both the pump and Stokes pulses are present (see Ref. [2] in the contribution by K Bergmann). However, if the system is initially in state 3 and all fields are on resonance, fractional STIRAP would produce a superposition of all three states. Hence we do not have a qubit gate. Nonetheless, STIRAP can still be used to construct robust single-qubit gates. One possibility is to use a large single-photon detuning. Then the middle state 2 can be eliminated adiabatically and the $\Lambda$ system is reduced to an effective two-state system of states 1 and 3. In this case fractional STIRAP will act as an SU(2) gate. Alternatively, robust rotation gates can be produced by a sequence of two (inverted and regular) fractional STIRAP processes [1], see Fig. 1(c). If the ratio $\Omega_P(t)/\Omega_S(t)$ tends to $\cot(\alpha)$ initially and $\tan(\alpha)$ in the end, then this sequence produces a robust rotation gate of angle $2\alpha$.

Of particular interest to quantum information is the tripod version of STIRAP because of its two dark states [2], see Fig. 1(a) and (b). In particular, it was recognized that the phase factors associated with the two dark states during the evolution are of non-Abelian nature, and the ensuing mixing angle between the two dark states is of geometric origin; this angle can be controlled by the pulse delays [1]. Kis and Renzoni [2] proposed a robust rotation gate by application of two STIRAP processes in the tripod system of Fig. 1(a), with the pulse sequence of Fig. 1(d). The qubit is formed of states 1 and 3, while state 4 is an ancilla state. In the adiabatic limit, this sequence of pulses produces a qubit rotation. The Kis-Renzoni gate was demonstrated in an experiment with trapped $^{40}Ca^+$ ions by Toyoda *et al* [3]. The gate operation is shown in Fig. 2 where the population is seen to oscillate vs the phase $\varphi$ between the qubit states with only negligible population in the other two states. Rousseaux *et al* [4] extended these ideas to an *N*-pod – a linkage of *N* lower states coupled to a single excited state. They showed that a double-STIRAP sequence, as the one in Fig. 1 (c), can produce a Householder reflection in the subset on *N* lower states. Householder reflections are a powerful tool for construction of arbitrary quantum gates of qudits (*d*-state systems).

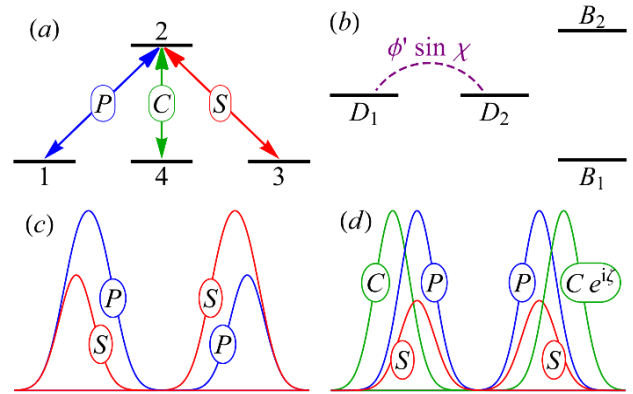

**Figure 1.** (a) Tripod linkage. In addition to the pump (P) and Stokes (S) fields an additional state 4 is coupled to the middle state 2 by a control pulse (C). (b) The tripod system transformed in the adiabatic basis has two bright states $B_1$ and $B_2$ decoupled from the two degenerate dark states $D_1$ and $D_2$, one of which is initially populated. In the adiabatic limit all couplings can be neglected except the one between the two dark states. (c) Pulse sequences for rotation gates with two fractional-STIRAP processes, and (d) two tripod-STIRAP processes. Adapted from Ref. [1].

STIRAP and fractional STIRAP have been the engines also in proposals for two-qubit quantum gates. One of the first proposals for a two-qubit conditional phase gate considered ions trapped in a cavity and used a combination of STIRAP and environment-induced quantum Zeno effect [5]. The latter keeps the qubits in a decoherence-free subspace. The method avoids both spontaneous emission, because of STIRAP, and cavity loss because no photon is present in the cavity at any time. Various other proposals are listed in Ref. [1].

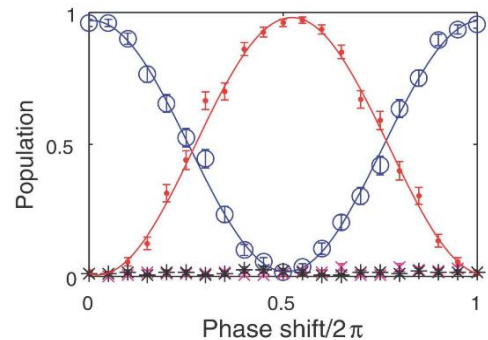

**Figure 2.** Geometric phase gate demonstration. Populations of states 1 (blue hollow circles), 3 (red filled circles), 4 (magenta crosses) and 2 (black asterisks). Adapted from Ref. [3].

*Entangled states.* A number of authors have proposed using STIRAP to construct many-qubit entangled states. In two of the most ubiquitous quantum-information platforms — trapped ions and trapped atoms — STIRAP allows one to perform qubit manipulations without populating the noisy common bus mode, ie the vibration mode shared by the trapped ions or the cavity mode shared by the trapped atoms. Linington and Vitanov [6] proposed a method for generation of arbitrary-sized Dicke states in a chain of trapped ions, which are equally-weighted coherent superpositions of collective states of qubits that share the same number of excitations. The ion qubits are cooled to their vibrational ground state, and then a vibrational Fock state with $m$ phonons is prepared. Next, the system is driven from this state to the desired Dicke state by multistate STIRAP via a multiqubit dark state by two delayed pulses applied simultaneously on all ions, the first on the carrier transition, and the second on the red-sideband transition. Noguchi *et al* [7] experimentally demonstrated a modified version of this proposal with global red- and blue-sideband pulses in chains of two and four trapped $^{40}Ca^{+}$ ions. Simon *et al* [8] experimentally demonstrated phase-coherent transfer by multistate STIRAP of a spin wave (quantized collective spin excitation) from one ensemble of $^{133}Cs$ atoms to another via an optical resonator serving as a quantum bus, which, benefiting from the features of STIRAP, was only virtually populated. An entangled state with one excitation jointly stored in the two ensembles was deterministically created by fractional STIRAP. Among the many other proposals for creating entangled states by STIRAP and STIRAP-inspired methods [1] we mention cavity-QED schemes that map atomic Zeeman coherences onto photon states and generate entangled photon multiplets and atom-photon entanglement in a two-mode optical cavity; generation of many-particle entangled states of dipole-dipole interacting Rydberg atoms by using the dipole blockade effect; and a method to adiabatically transfer field states between two partly overlapping cavities via an atom passing through them [1].

*Quantum algorithms.* Daems and Guerin [9] proposed to use multistate fractional STIRAP for adiabatic implementation of Grover's quantum search algorithm. The database is an ensemble of $N$ identical three-level atoms trapped in a single-mode cavity and driven by two lasers, and the marked atom has an energy gap between its two ground states. Starting from an entangled state fractional STIRAP populates, with high efficiency, the marked state in time that scales as $N^{1/2}$, thereby achieving the same speed-up as the discrete Grover algorithm.

## Current and Future Challenges

The main challenge for the application of STIRAP in quantum information is the high fidelity of operation required, with the admissible error of operation being less than 0.01%. The straightforward implementation of STIRAP with Gaussian pulse shapes makes it possible to reach 95-99% efficiency with reasonably modest pulse areas and reasonably smooth pulse shapes. One can boost the pulse areas to very large values, thereby further improving adiabaticity. However, in many systems this a) may not be technically possible, b) may lead to unwanted coupling to additional states and/or Stark shifts, c) makes the entire process slower and hence may collide with the various decoherence sources. Decreasing the error by additional two orders of magnitude requires radiation sources with excellent coherence properties and sophisticated coherent control approaches, some of which are listed below.

## Advances in Science and Technology to Meet Challenges

Three possible approaches to meet the challenge of high fidelity of operation are listed below.

(i) One possibility is to optimize the pulse shapes in order to suppress the non-adiabatic couplings [1]. The efficiency of pulse shaping for adiabatic techniques has been demonstrated recently in a half level-crossing experiment with a single trapped ion where an error close to 0.01% has been reached (see the contribution by W K Hensinger and S Weidt and Ref. [7] there).

(ii) Another approach is to use ideas from composite pulses and apply composite STIRAP — a sequence of several STIRAP processes wherein the pump and Stokes pairs have well defined phases which are used as control parameters to cancel the non-adiabatic errors. This approach was demonstrated in a proof-of-principle experiment with a doped solid (see the contribution by T. Halfmann and Ref. [10] there), in which the fidelity has been limited by large inhomogeneous broadening. However, the application of composite STIRAP to other physical systems, eg trapped ions, trapped atoms and superconducting qubits, still promises to achieve the ultrahigh fidelity needed for quantum computation.

(iii) The "shortcuts to adiabaticity" method promises very high fidelity with modest pulse areas. However, it introduces an additional field on the transition $1 \leftrightarrow 3$, which must have a precise phase relation to the other two fields, a precise pulse shape and a precise pulse area (equal to $\pi$) [1]. Modifications of this method may still be interesting to pursue.

## Concluding Remarks

STIRAP is a textbook technique in quantum control. Due to its resilience to some types of decoherence it is very attractive for quantum information processing and it has been successfully applied in a number of related experiments already. If nonadiabatic losses can be suppressed beyond the fault-tolerant threshold STIRAP will certainly become a basic tool for quantum gate operations and entanglement generation.

**Acknowledgements** – The author acknowledges support from the European Commission's FET-Flagship on Quantum Technologies project 820314 (MicroQC).

# B2 Matter waves

## B2.1 Spatial Adiabatic Passage

*J. Mompart[1] and Th. Busch[2]*

1Universitat Autònoma de Barcelona

2OIST Graduate University, Okinawa

### Status

Techniques to coherently control the spatial degrees of freedom of trapped matter waves are a fundamental building block of the area of quantum technologies. Apart from directly moving external trapping potentials, quantum transport of trapped matter waves can also be performed via direct tunneling through the potential barriers that separate the different traps. This latter process, however, is strongly dependent on the external parameter values and, in general, gives rise to very sensitive Rabi-type oscillations of the localised populations. However, an efficient transfer of population between distant traps can be achieved using a generalization of STIRAP to real space, the so-called spatial adiabatic passage (SAP) process [1-3].

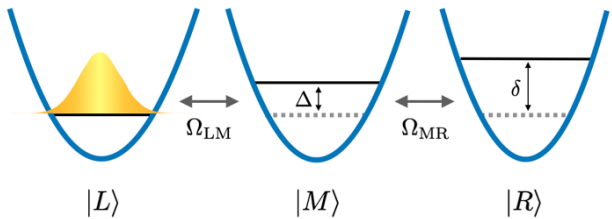

Figure 1: Schematic of a triple-well potential where the localized states in each well ($|L\rangle$, $|M\rangle$, $|R\rangle$) are coupled through the tunnelling amplitudes $\Omega_{LM}$ and $\Omega_{MR}$. The energy difference between the localized states in the left and middle well is given by $\Delta$ and between the left and right well by $\delta$. The two-photon resonance condition in STIRAP corresponds to $\delta = 0$ in SAP.

In SAP the atomic states from STIRAP are replaced by three localized states of a triple well potential, $\psi_1 \leftrightarrow |L\rangle$, $\psi_2 \leftrightarrow |M\rangle$ and $\psi_3 \leftrightarrow |R\rangle$ (see Fig. 1), with the aim to efficiently and robustly transfer a single particle from e.g., the left to the right well. The couplings between the localised states are realized by tunneling interactions, which are assumed to be individually controllable in time by, for example, changing the distances between the traps. The SAP dynamics are therefore driven by the same Hamiltonian as in STIRAP, see Eq. (2), with $\Omega_{LM} \leftrightarrow \Omega_P$ and $\Omega_{MR} \leftrightarrow \Omega_S$ now representing the tunneling amplitudes between the localized states of the left and the middle and of the middle and the right wells. The detunings correspond to differences in the eigenstate energies between the left well and the others and are given by $\Delta$ and $\delta$ for states in the middle and the right well, respectively. The diagonalization of the SAP Hamiltonian then gives rise to the same dark state as in Eq. (4). In fact, it is easy to see that this state has to exist, assuming that the localised eigenstates are the ground states of the trap: the dark state is the natural solution for the delocalized, first excited, eigenstate of the system. It therefore has to be an odd function with exactly one zero-crossing, which for symmetry reasons has to lie in the center well. By manipulating the height of the barriers or the separation between the wells one can then apply first a tunneling *pulse* between middle and right wells and later on, with an appropriate time delay T, a tunneling *pulse* between the left and middle wells, following the counter-intuitive sequence of STIRAP. The mixing angle $\vartheta$ in Eq. (4) then changes from 0 to $\pi/2$, and if the adiabaticity conditions are fulfilled, see Eqs. (6) and (7), it is possible to efficiently and robustly transfer the single particle from the left to the right well.

While the above is a straightforward generalization of STIRAP, the fact that SAP deals with the external (localized) degrees of freedom of trapped particles and STIRAP deals with the internal ones implies a number of differences between both techniques. In particular, the signature of both STIRAP and SAP is the fact that the intermediate state is not being populated during the whole process for $\delta=0$. This is of crucial importance in STIRAP since, usually, this intermediate state is a fast decaying one, at variance with SAP where the middle state is as stable as the localized states of the outermost wells. In addition, the fact that in SAP the middle state is not populated seems to indicate that the local continuity equation associated to the Schrödinger equation fails. This paradoxical issue was addressed in [4], however it is still a source of debate. For a detailed comparative between SAP and STIRAP techniques see [3].

Additionally, it was soon realized that SAP could be extended beyond being a direct analogue to STIRAP, as it allows access to many more degrees of freedom, such as multi-dimensional configurations, many particle systems or different particle statistics. In particular, for single particles (neutral or charged), few interacting particles or Bose-Einstein condensates (BECs), proposals have been developed for tasks such as vibrational state and velocity filtering, quantum tomography, interferometry, atomtronics, the generation of orbital angular momentum states and matter-wave Fock state emitters. Combining different degrees of freedom and extending the ideas behind SAP to new ones is one of the future directions of the area.

While the two initial works in this area considered neutral atoms in optical dipole potentials [1] and electrons in quantum dots [2] for the localised states, and despite the significant theoretical interest that SAP has attracted, no experimental demonstration of the effect for matter waves does exist yet. It is worth noting, however, that due to the wave-like nature of the SAP processes, they can be extended to classical wave systems, e.g, light and sound, and have been experimentally demonstrated with light beams in systems of three coupled waveguides [5-7], see Fig. 2 and section A3.2.

### Current and Future Challenges

Even though one of the main challenges is still the experimental implementation and observation of SAP with matter waves, its potential for being extended to include various other degrees of freedom and settings holds its biggest promise. For ultracold atoms one of the main challenges is to identify or design systems where the three tunnel-coupled states are not necessarily the ground states of the potential, in order to create high fidelity quantum state preparation settings. This is particularly interesting for many-particle systems, where non-classical correlations and entanglement can be created this way [8]. However, as the densities of typical spectra for many-particle systems quickly become too large for adiabatic processes to be realistic, careful stability analyses and the development of shortcut algorithms becomes important. While a fundamental shortcut to adiabaticity for STIRAP/SAP processes is known [9], it requires complex tunneling frequencies, which are not straightforward to implement. Techniques from optimal control, and for many-particles also from the emerging area of machine learning, are therefore also important to explore. Finding and optimizing shortcuts that are experimentally realistic is an important task to make SAP useful in the future. The combination of SAP with supersymmetry techniques is another promising possibility that has only recently been considered in the context of coupled optical guidelines [10].

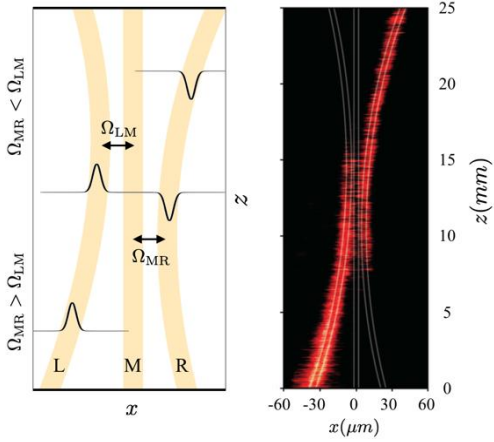

Figure 2 Left: Sketch of three single-mode coupled waveguides arranged in the SAP geometry with the spatial dark supermode (in black). Right: Top view experimental image of a triple waveguide system performing SAP from the left to the right waveguide. Reprinted with permission from [6]. Copyright 2012 IEEE.

Another avenue of progress is the extension to non-equilibrium states or complex eigenstates, such as angular momentum states, which are a natural source of complex tunneling amplitudes. SAP combined with complex tunneling could be used to implement tunneling-induced geometric phases, simulate artificial magnetic fields, or implement entangling two-qubit gates.

A third big future challenge is given by the possibility to move SAP into momentum space. The basic idea is to create systems where an energy spectrum possesses several roton minima, and the system can tunnel between different dynamical states. Such dispersion relations can be created, for example, in spin-orbit coupled BECs, where the position of the roton minimum can be adjusted through the external parameters inducing the spin-orbit coupling.

While SAP processes are currently mostly discussed as self-standing, including them into protocols and connecting them to other quantum dynamics is another important area to investigate in the future. This could, for example, involve using them for certain tasks inside complex quantum information processing devices, or as highly robust parts in quantum metrology applications.

Finally, a larger number of techniques known in STIRAP have not been examined for their use in SAP. These include, for example, the use of composite pulses, see Section B1, or the extension to systems with a larger number of states. While STIRAP schemes using the bright eigenstates are known [11], these are challenging to realize in quantum-optical systems due to the presence of spontaneous emission. In spatial systems, where this problem does not exist, they are promising candidates for future extensions of SAP. Even more, it is possible to envisage hybrid schemes combining STIRAP with SAP in such a way that, for example, one arm of the three-level system couples internal degrees of freedom by means of a laser pulse and the other couples external degrees of freedom by means of a tunneling *pulse*.

**Advances in Science and Technology to Meet Challenges**

The challenges for SAP development are of experimental and of theoretical nature.

On the theoretical side the move to many-particles systems requires more advanced numerical techniques to be available, and a deeper understanding of the underlying processes to design suitable shortcut or control protocols. For implementations, more realistic simulations in full 3D and with realistic noise and loss will need to be performed.

Experimentally the challenges are different for each implementation of SAP. For atomic systems the development of flexible traps with higher stability for single atoms is already on its way, however getting into the ranges required by SAP is still a challenge. Furthermore, increasing the couplings between dipole traps usually leads to changes in the spectra of the individual states, which would have to be appropriately compensated for.

For electrons in quantum dots currently the most challenging issue is the requirement of resonance, which demands to make and couple highly-identical quantum dots. This is still a hard engineering task but can be expected to be solved in the future.

The main obstacle to see more proof-of-concept experiments and applications in optical waveguides is the presence of losses and the requirement to make the dynamics adiabatic, leading to larger devices.

**Concluding Remarks**

Over the past decade SAP has proven to be an extension of STIRAP that has allowed to extend the initial idea to many new settings with degrees of freedom not available in optical systems. As external potentials for individual quantum particles are becoming more and more controllable, SAP can be expected to play a significant role for high fidelity quantum state engineering tasks beyond single particle movement. Go SAP!

**Acknowledgments** – TB acknowledges support from OIST Graduate University. JM acknowledges support from the Spanish Ministry of Economy and Competitiveness

# B3 Manipulation of molecules

### B3.1 Populating Rydberg States of Molecules by STIRAP

*Timothy J. Barnum[1], David D. Grimes[2,3,4], Robert W. Field[1]*

[1]Department of Chemistry, Massachusetts Institute of Technology
[2]Department of Chemistry and Chemical Biology, Harvard University
[3] Department of Physics, Harvard University
[4] Harvard-MIT Center for Ultracold Atoms

## Status

Rydberg states of atoms and molecules have been studied for decades by the atomic, molecular, and optical physics community due to their special properties, which include extreme sensitivity to electric fields, strong dipole-dipole interactions, and long radiative lifetimes. Researchers continue to find new and more sophisticated ways to employ Rydberg states, ranging from manipulation of their external degrees of freedom through deceleration and trapping [1] to state-selective production of molecular ions for precision measurement [2]. Spectroscopic investigation of molecular Rydberg states, in particular, offers insights into the rich intramolecular dynamics of molecular Rydberg states as well as a sensitive probe of the energy level structure and electric properties of the molecular ion-core. Rydberg states exhibit huge transition dipole moments ($\mu \sim 1$ kiloDebye at $n \sim 40$, where n is the principal quantum number) in the microwave and millimeter-wave regions, allowing investigators to utilize high-resolution microwave spectroscopy [3].

STIRAP excitation of atomic Rydberg states has been demonstrated for several species by double resonance laser excitation in a ladder level scheme [4]. Efficient population transfer to these highly excited electronic states has enabled a number of novel applications. Sparkes et al. used STIRAP population of Rydberg states followed by pulsed field ionization to generate intense, sub-ns electron bunches for gas-phase ultrafast electron diffraction experiments [5]. Taking advantage of the strong dipole-dipole interactions between Rydberg states, a number of proposals exist for combining Rydberg-Rydberg interactions with coherent excitation via STIRAP to generate multiparticle entangled states and to perform quantum computation and simulation [6]; experimental work is just beginning to demonstrate the potential of these ideas [7]. In addition, experimental implementations of STIRAP on atomic Rydberg systems have advanced our understanding of STIRAP due to both the efficiency of ionization-based detection of Rydberg states for precise quantification of the population transfer and the exceptional sensitivity of Rydberg states to external fields. In particular, studies of the influence of blackbody radiation, multi-level effects, and the time dependence of the intermediate state population in STIRAP have all been stimulated by Rydberg experiments [8].

In contrast to Rydberg states of atoms, the study of Rydberg states of *molecules* is relatively less mature. Crucial to the ability to perform high-resolution microwave spectroscopy, deceleration and trapping, and other modern Rydberg experiments, is the long lifetime of the Rydberg state of interest. The radiative lifetimes of Rydberg states are intrinsically long, and scale as $n^3$, such that a state with $n \sim 40$ typically has a radiative lifetime $> 10$ μs. In molecules, this long lifetime is curtailed by the presence of ubiquitous non-radiative decay pathways, especially predissociation. Specifically, low orbital angular momentum ($\ell$) Rydberg states exhibit core-penetration, which means the wavefunction of the Rydberg electron experiences short-range interactions with valence electronic states that are localized near the ion-core. In contrast, high orbital angular momentum states are prevented from interacting with the ion-core at short range due to the centrifugal barrier, thus exhibiting much slower predissociation rates than the low-$\ell$ states. Therefore, the primary challenge for molecular Rydberg experiments is the simple issue: how does one access long-lived, high-$\ell$ Rydberg states while avoiding the rapidly predissociating low-$\ell$ states? STIRAP offers a path forward.

Initial efforts to access high-$\ell$ Rydberg states via STIRAP have focused on proof-of-principle experiments for an optical-millimeter-wave implementation of STIRAP in atomic systems [9]. In those experiments, UV radiation creates transition amplitude from a low-lying valence state to a Rydberg state, while millimeter-wave (mmW) radiation provides transition amplitude between the intermediate Rydberg state and a final Rydberg state with one additional unit of orbital angular momentum, thus avoiding direct population of the lossy low-$\ell$ Rydberg state. Uniquely, the populations of both the intermediate and final states were simultaneously monitored as a function of the delay between pump and Stokes pulses by chirped-pulse Fourier transform mmW spectroscopy, enabling a more complete characterization of the population transfer. The initial experiments reported in [9] failed to achieve STIRAP in a buffer gas cooled beam of Ba atoms as a result of the use of a low-coherence pulsed dye laser. Nevertheless, enhanced population transfer during overlap of the optical and mmW pulses suggested significant two-photon coupling despite the nearly three orders of magnitude difference in the frequencies of the two photons.

More recent work on optical-mmW STIRAP in the Field group has focused on a similar system with more favorable characteristics [10], including larger transition dipole moments and the use of a high-power pulse amplified CW laser system to generate the optical radiation. Simulation results for this experimental scheme are shown in the left panel of Fig. 1. These simulations demonstrate the characteristic enhanced population transfer to the final state in the counter-intuitive pulse timing. The population transfer is limited to less than 60% predominantly by the square pulse shape of the mmW field, which hinders adiabatic following. Simulation results with a smooth turn-off of the mmW pulse, but otherwise identical parameters, show that population transfer of nearly 75% is possible. The remaining inefficiency in this scheme can be attributed to the incomplete cancellation of the Doppler shifts and non-uniform spatial profiles for the radiation fields, two challenges discussed further in the following sections. The right panel of Fig. 1 shows the simulated population transfer to the final state for the parameters previously described, but including a 1 ns intermediate state lifetime, representative of molecular predissociation. Significant population transfer to the final state only at the counter-intuitive pulse timing demonstrates the characteristic insensitivity of STIRAP to intermediate state decay. These simulations show promise in the application of optical-mmW STIRAP to both an atomic system and a molecular system with fast predissociation of the intermediate level.

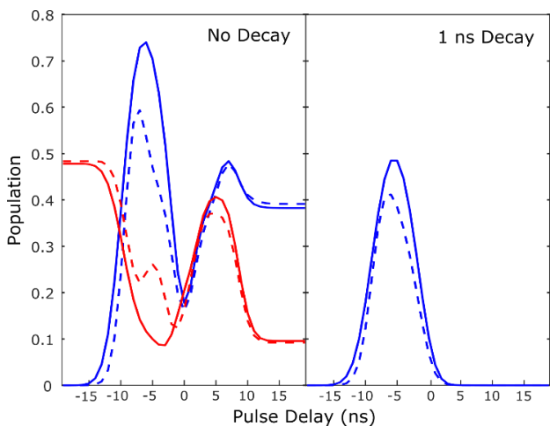

Fig. 1: Left panel: Simulated population transfer from the initial 4s5p to the final 4s28f state (blue lines) via the intermediate 4s30d state (red lines) of Ca by optical-mmW STIRAP. The dashed lines show results for an experimentally achievable square wave mmW pulse, while the continuous lines are for a hypothetical mmW pulse with a temporally smooth amplitude profile. Right panel: Simulated population transfer to the final state when the intermediate state lifetime is reduced to 1 ns, representative of predissociation. All other parameters are identical to the simulation in the left panel.

## Current and Future Challenges

Due to the high internal energy of Rydberg states of atoms and molecules, STIRAP efficiency is often limited by the availability of high-coherence, high-power laser sources in the UV region of the spectrum. It has also been noted that excitation efficiency can depend sensitively on the spatial profile of a pulsed beam; beam-shaping strategies as discussed in [4, 5] will be useful in future experiments to achieve uniformly high excitation efficiencies.

Optical-mmW STIRAP presents several unique challenges to achieving efficient population transfer. One of the most significant differences from optical-optical STIRAP is the enormous difference in the frequencies of the two photons. In a Doppler-broadened medium, the Doppler shift of the pump radiation can be compensated by the Doppler shift of the Stokes radiation by proper choice of geometry, provided that the two photons have similar frequencies. In optical-mmW STIRAP, this frequency ratio is $\geq 10^3$. This means that the Doppler shift of the optical photon effectively creates a range of two-photon detunings across the sample. This challenge can be mitigated through the use of cold samples and/or well-collimated beams; in [9], the use of a cryogenic buffer gas cooled beam dramatically reduces the transverse velocity spread of samples relative to that of supersonic jet sources.

## Advances in Science and Technology to Meet Challenges

Both all-optical and optical-mmW STIRAP will benefit from advances in pulsed laser technology to produce sources with high-power and longer pulse durations. STIRAP efficiency scales favorably with pulse duration due to improved adiabatic following. Moreover, the lower peak power of long pulses reduces undesired multiphoton ionization, an additional pathway for population loss in pulsed laser excitation of Rydberg states.

One of the most stringent limitations of optical-mmW STIRAP is the inability to produce high-power, smooth-shaped mmW pulses. At lower microwave frequencies, pulse shaping can be accomplished electronically by arbitrary waveform generation and upconversion with low frequency shaped pulses followed by amplification. Advances in arbitrary waveform generators and millimeter-wave diode technology in general are occurring rapidly as the mmW region of the spectrum is being developed for communication, radio-astronomical, and imaging applications. Technological improvements in the mmW regime will immediately benefit optical-mmW STIRAP experiments.

## Concluding Remarks

The study of Rydberg states of atoms and molecules is at once an old field and a frontier research area in 21[st] century physics. Fundamental studies on the application of STIRAP to *molecular* Rydberg states will lift molecules to the same level of understanding and control as Rydberg atoms. The application of STIRAP to Rydberg states of atoms and molecules will continue to stimulate developments in areas that range from quantum computation to chemical physics.

## Acknowledgments

We gratefully acknowledge the many members of the Field group who have contributed to the "Rydberg project," and financial support from the National Science Foundation (Grant No. CHE-1361865) and the Air Force Office of Scientific Research (Grant No. FA9550-16-1-0117). TJB was supported by a National Science Foundation Graduate Research Fellowship under Grant No. 1122374.

### B3.2 Controlling molecular beams with magnetic fields and STIRAP– *Mark G. Raizen[1] and Edvardas Narevicius[2]*

[1] Univ. of Texas at Austin
[2] Weizmann Institute of Science

### Status

The development of the molecular beam has enabled the production of intense sources of atoms and molecules, stimulated by the quest for controlling chemical reactions and the study of quantum chemistry. These beams are created by a high-pressure inert gas that is emitted through a shaped nozzle, which cools upon expansion to produce a very directional and mono-energetic source. Desired atoms and molecules can be entrained in the carrier gas so that they attain the same beam characteristics. Such intense beams have been used in many experiments such as crossed beams, but until recently, the limitations of molecular beams have been their large mean velocity in the laboratory frame, and further cooling and compression of phase space was not possible. Controlling molecular beams could open many new scientific directions, such as the study of quantum chemistry and the development of cooling techniques that could be far more general and powerful than laser cooling and evaporative cooling. The first step in this effort is to develop a decelerator in order to bring atoms and molecules to rest. One approach, applicable to polar molecules and Rydberg-state atoms, is a series of pulsed electric field plates. The other approach, described in this roadmap article is magnetic slowing and stopping with a series of pulsed magnetic fields. The latter is widely applicable to any atom or molecule that is paramagnetic. This method was used to slow and stop atoms and molecules, and to merge two crossed molecular beams and observe quantum pathways for chemical reactions [1-5]. The pulsed electromagnetic coils were initially implemented as a serial decelerator, and later were used to produce a moving magnetic trap that traps the atoms and molecules in the moving frame, and brings them to rest adiabatically [6].

### Current and Future Challenges

Successful operation of the magnetic decelerator requires that the atoms and molecules be in the correct internal state, a so-called "low-field seeker," in order for them to be trapped in the magnetic field minimum during the deceleration process. In order to prepare atoms in this state, resonant lasers typically are used for optical pumping which can be accomplished with high efficiency. The situation with molecules is much more complicated, due to their many rotational and vibrational states, making optical pumping a major challenge. Until now, the only option has been to select a small subset of molecules that happen to be in the correct internal state, at the expense of flux, but this is a serious problem that hampers future progress. The solution appears to be STIRAP, as proposed in a recent publication, and may be the only viable way to produce optically-pumped molecules. This use of STIRAP can enable efficient optical pumping of high-spin systems as well as molecules, since it minimizes the number of spontaneous emission events which populate the excited rotational, vibrational and electronic states [7]. The other challenge is how to further cool the translational motion of atoms and molecules after they are brought to rest. Since laser cooling is not a practical option in most cases, especially for molecules, a new approach is required. A solution was recently proposed, magneto-optical (MOP) cooling [8], which is a variation on earlier work that realized a one-way wall for atoms, a practical realization of Maxwell's demon [9-10]. This new method relies on a cycle of optical pumping followed by a one-dimensional magnetic kick to compress phase space. The realization of MOP cooling will also rely critically on a novel implementation of STIRAP to drive stimulated transitions between magnetic sublevels for a particular velocity distribution of atoms. This form of STIRAP has not been implemented to the best of our knowledge, and is explained in detail in [8]. The realization of MOP cooling will enable cooling of many molecules, a result of far-reaching significance.

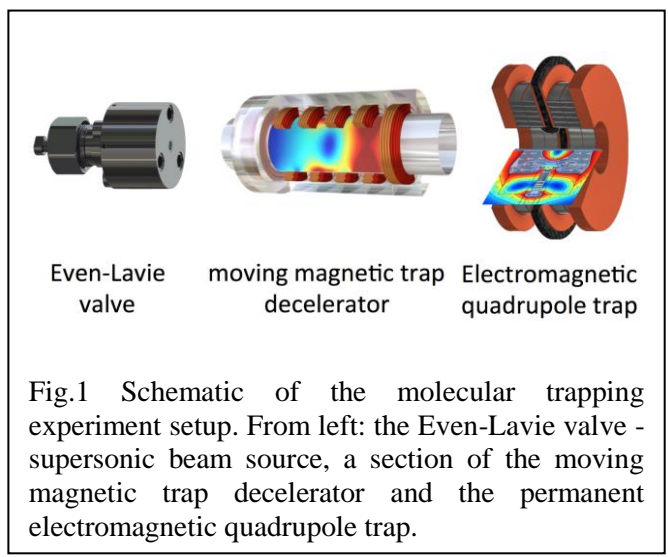

Fig.1 Schematic of the molecular trapping experiment setup. From left: the Even-Lavie valve - supersonic beam source, a section of the moving magnetic trap decelerator and the permanent electromagnetic quadrupole trap.

### Advances in Science and Technology to Meet Challenges

Future progress in magnetic control of molecular beams will rely on emerging technologies. One direction will be to improve the performance of the molecular beam and the efficiency of entrainment, where there is still much room for improvement despite the fact that it is a mature technology. For example, we recently found that by using argon as a carrier gas and cooling the skimmer to freeze the carrier atoms upon impact, we can avoid skimmer clogging and multiple scattering. This innovation leads to a much improved beam quality and could enable a higher repetition rate of the nozzle and magnetic slower. There is a need for more realistic numerical simulations of the molecular beam, a complex hydrodynamic system, as well as the process of entrainment of desired atoms and molecules, so that both can be optimized. Likewise, the development of a new experimental probe of the local density and velocity field in the molecular beam could validate the numerical simulations and enable further progress. Another important direction for future work is to create stable electro-magnetic traps for storing atoms and molecules. We will also need to develop strategies for further cooling of atoms and molecules after they are trapped, and for merging several species at high phase-space density. The method of MOP cooling, described above, relies on optical pumping which may not always be possible. For example, in the case of atomic hydrogen (and its two isotopes, deuterium, and tritium) the optical pumping transition is in the far-UV near 121 nm, not accessible with current laser technology. Furthermore, the large single-photon recoil velocity for hydrogen atoms would cause undesirable heating. An alternative approach is sympathetic cooling of a desired species by collisions with another atom, such as lithium, which can be cooled independently. Another possibility is to attempt direct evaporative cooling on a dense ensemble of magnetically trapped molecules such as oxygen. The success of any collisional-based cooling approach will depend on the ratio between the elastic cross section and inelastic processes, which would need to be tested experimentally. Sympathetic cooling may also be used to cool anti-hydrogen atoms, and theoretical calculations are in progress to evaluate feasibility.

## Concluding Remarks

Controlling and cooling the translational motion of atoms and molecules, both in molecular beams and after they are stopped and trapped, hinges on the ability to control their internal state via optical pumping. It is expected that STIRAP holds the key to internal state manipulation and control. The opportunities for basic science and applications are truly exciting, and the future appears very bright.

Acknowledgments – We acknowledge support from the W. M. Keck Foundation and the Sid W. Richardson Foundation (MGR).

### B3.3 Application of STIRAP for polarizing angular-momentum states

*Marcis Auzinsh[1] and Dmitry Budker[2]*

[1] University of Latvia
[2] Johannes Gutenberg University, Mainz and University of California, Berkeley

### Status

Optical polarization of atoms and molecules preparing them in a well-defined angular-momentum quantum state is [1] known since the pioneering work of A. Kastler in the late 1940s [2]. Till this day, optical polarization of angular-momentum states continues to be important for applications in science and quantum technologies.

In general, polarization of angular momentum states is characterized by the multipole expansion of the density matrix [3,4]. The first-rank moment is called orientation. A density matrix corresponding to pure orientation along $z$ corresponds to populations of the $m$ sublevels proportional to $m$. Since negative populations are non-physical, rank-zero should always be present also. The second-rank moment is called alignment. Alignment along $z$ corresponds to populations of the Zeeman sublevels proportional to $m^2$.

Especially attractive for a variety of applications are the so-called stretched (maximum-projection) states. These are angular momentum states where only one magnetic sublevel with $m = J$ or $-J$ is populated. These stretched states are often used in experiments in quantum optics, magnetometry and spin squeezing. In some applications, these states have an advantage of being immune to the relaxation in spin-exchange collisions. They can also be useful for cooling atoms and molecules via interaction with spatially and temporally varying magnetic fields, see [5] and references therein. Such maximally polarized angular-momentum states can also be used to control chemical reactions, see [6] and references therein.

In some particular cases, a straightforward and efficient method to create polarized angular-momentum states is by their coherent interaction with two or more light fields each tuned to cause a transition between levels in atoms or molecules. For example, N. Mukherjee and R. Zare [6] demonstrated how, for a case of rotational-vibrational states in the HCl molecule, starting from the $v = 0$, $J = 0$ state, a STIRAP scheme can be used to transfer the population to the $v = 2$, $J = 2$, and $m = 0$ state with 100% efficiency for mutually parallel excitation-light polarizations and $v = 2$, $J = 2$, and $m = \pm 1$ for mutually perpendicular excitation-light polarizations. Orientation of the $v = 2$, $J = 2$ state, with $m$ either +2 or -2 can also be achieved using two pulses of circularly polarized light of the same helicity. In the same paper it was mentioned that polarization of molecules in higher vibrational levels can be achieved using additional light fields in a ladder scheme [7] with sequential STIRAP.

### Current and Future Challenges

Although the approach described above to polarize angular-momentum states by coherent processes is efficient, it has serious limitations. All coherent processes represent unitary (reversible) evolution. The optical polarization in the examples described above works because the system is initially in a $J = 0$ state with only one magnetic sublevel $m = 0$. In this case, a STIRAP process efficiently transfers the population from this single magnetic sublevel to another single magnetic sublevel or a coherent superposition of several sublevels corresponding to a pure state. In this way a highly polarized excited state is created. This method does not work if in the initial state has $J > 0$ with more than one incoherently populated sublevel.

If the initial state has more than one magnetic sublevel ($J>0$), there are general limitations on how coherent processes including STIRAP can manipulate atomic or molecular states. To illustrate this, let us consider an angular momentum state $J$ that can be described by magnetic-sublevel population, without coherences. Let us see what happens if these populations undergo unitary evolution caused by some coherent process and end up in a state that again does not have coherences.

A quantum state with no coherences between magnetic sublevels can be represented with a diagonal density matrix $\rho$ [4]. A unitary transformation cannot change the eigenvalues of the matrix. In the case of a diagonal matrix, these eigenvalues are diagonal elements of the matrix – the populations of magnetic sublevels. When the system coherently interacts with radiation, we can expect that populations of magnetic sublevels to be rearranged. But it is impossible to combine the population of multiple magnetic sublevels into one [5]. This means that after the coherent interaction with radiation, the population of any magnetic sublevel will not be larger than the population of the magnetic sublevel with largest population before the interaction.

Thus, if in the initial state we have equally populated magnetic sublevels with no coherences, coherent processes cannot create population of any of the excited state magnetic sublevels larger than the population of a ground-state magnetic sublevel. For example, this restriction does not allow to create a stretched state by coherent interaction alone if we start from a ground state other than one with $J = 0$.

Reformulating this conclusion in a more general form: coherent processes usually describe the evolution of a closed quantum system and cannot change the entropy of the system. To demonstrate this, we use Neumann's entropy $S$ [4] for a quantum system expressed in the units of Boltzmann constant. For a diagonal density matrix with eigenvalues $\lambda_i$, we have

$$S_{atom} = -\text{tr}(\rho \ln \rho) = -\sum_i \lambda_i \ln \lambda_i, \qquad (1)$$

where $\lambda_i$ is the eigenvalue of density matrix $\rho$ which corresponds to the population of the respective magnetic sublevel. The entropy in the angular momentum states where only one magnetic sublevel is populated (for example, a stretched state) is zero. In all other cases it is positive. So, whenever we increase the polarization of a state, we decrease its entropy – we increase the order of the system.

In a general case, coherent processes cannot efficiently (i.e., without losing substantial part of the population) create highly polarized states. These processes must be used in combination with spontaneous emission which is an incoherent process. To achieve complete polarization for a state with an arbitrary angular momentum $J$, in an ideal case, it is sufficient to have one spontaneous emission per atom combined with coherent evolution of atomic states [5]. The challenge is to find a way to do this most efficiently.

## Advances in Science and Technology to Meet Challenges

Based on the entropy consideration, it was argued in [5] that the minimum number of spontaneous emissions needed to create a fully polarized state corresponds to just one spontaneous photon per atom (or molecule). This is in contrast to conventional optical pumping [2], where polarization of an initially unpolarized system with $2J + 1$ sublevels, each with initial population $p = 1/(2J + 1)$, typically requires $\sim J$ spontaneous emissions per atom.

The challenge is to find an efficient and at the same time practical method to create a highly polarized state where we combine coherent and incoherent interactions of the atoms or molecules with radiation.

The way to meet this challenge is to use coherent processes to arrange the atoms among the sublevels in a way maximizing entropy removal from the system by each spontaneous decay.

There are well-known methods for transferring populations between sublevels coherently, i.e., via absorption and stimulated emission only, without depending on spontaneous emission. For example, lower- and upper-state populations may be swapped through the technique of adiabatic fast passage (AFP), in which an optical pulse is applied with its frequency swept through resonance. Another approach is STIRAP, which uses a two-pulse sequence to transfer atoms from a populated state to an initially unpopulated state with the aid of an intermediate state, without ever developing an appreciable population in the intermediate state [8]. Thus, spontaneous decay from the intermediate state can be eliminated, even if the lifetime of the intermediate state is short.

In [5] several schemes to achieve efficient polarization of atomic state were proposed. In the first suggested approach, the sublevel populations are "folded" in half prior to each spontaneous decay. This means that the number of populated magnetic sublevels before each spontaneous decay event is reduced by factor of about two. The folding can be achieved by transferring atoms between the sublevels of a ground $J$ manifold and an upper $J'$ manifold using AFP.

Alternatively, when the upper-state lifetime is short, it may be advantageous to implement the method using an additional shelving state. This folding scheme reduces the average number of spontaneous decays per atom from $\sim J$, obtained in conventional optical pumping, to $\sim \log_2(2J)$. For a folding scheme with shelving state to work, the coherent interaction with light must be able to swap populations between ground and excited state when both levels are populated [5]. Unfortunately, this rules out the traditional STIRAP because this process is "unidirectional," typically with high losses of the population initially present in the final state of STIRAP. We need to use either a two-photon AFP method [9] or, for example, the recently discussed bi-directional STIRAP. A survey and systematic comparison of bi-directional STIRAP methods will be given elsewhere.

The third scheme, which uses two shelving states, can further reduce the average number of spontaneous decays per atom down to close to one, which, as discussed above, is the minimum possible number needed for producing a completely polarized state. However, there is a trade-off when using this scheme: Even though each atom is required to decay just once, only a fraction $1/(2J + 1)$ of the population decays in each step, so that $\sim 2J$ sequential decays are needed to complete the sequence [5]. Thus, this scheme may not be feasible if the upper-state lifetime is long and the total interaction time is limited. The advantage of the scheme with two shelving states is that it allows to use conventional unidirectional STIRAP to exchange population between states coherently.

## Concluding Remarks

The methods that combine coherent manipulation of atomic states and spontaneous decay can also be generalized beyond just populating the stretched state. Indeed, once the population has been combined in one state, it can be moved, with an appropriate sequence of coherent population transfers, to any other state, or a coherent superposition thereof.

A rather complete theoretical understanding of the issues discussed in this note appears to be around the corner, while the challenge for the forthcoming work is experimental implementation and application of these ideas. The technological requirements to the lasers and other parts of the experimental systems seem challenging [5].

Polarization of angular-momentum states can be thought of as cooling of internal degrees of freedom. It would be interesting to connect this with cooling of the motional degrees of freedom, where, in recent years, cooling maximally utilizing stimulated processes has also been considered, see [10] and references cited therein.

It would be very interesting to find a unified description of these methods, as well as perhaps find a way to couple the internal and external degrees of freedom (as it is done, for example, in Zeeman slowers and magneto-optical traps), so that efficient polarization can also cool other degrees of freedom and vice versa.

### Acknowledgments
We are grateful to K. Bergmann and G. Genov for critical reading of the manuscript. DB acknowledges the support by the German Federal Ministry of Education and Research (BMBF) within the Quantum-Technologies program (FKZ 13N14439). MA acknowledges the support by the Ministry of Education and Science of Latvia within the project 1.1.1.2/VIAA/1/16/068.

# B4   Isomers in nuclear physics

### B4.1 Nuclear coherent population transfer

*Adriana Pálffy and Christoph H. Keitel*

Max Planck Institute for Nuclear Physics, Heidelberg

**Status**

Beyond atomic physics, atomic nuclei also present level schemes coupled by electromagnetic transitions suitable for coherent population transfer via the STIRAP technique. The goal of such an endeavor would be to achieve control of nuclear population. This is of special interest due to the existence of long-lived nuclear states or so-called nuclear isomers, which can store large quantities of energy over long periods of time [1]. Controlled release of the energy stored in the metastable state could provide an efficient energy storage solution, a goal with a large potential impact on society.

The direct decay of a nuclear isomeric state is typically strongly hindered. Possible depletion schemes therefore involve the excitation of the isomer to a triggering state lying above the isomer. Ideally, this triggering level can also decay directly or via a cascade to the ground state, thus releasing the stored energy [2]. This sketches an effective Λ-type three-level system as illustrated in Fig. 1(a), with the triggering state 3 as upper level and two ground states: the nuclear isomer as state 1 and state 2 which can be either directly the ground state or rapidly decay to it. By shining just one pump radiation pulse to promote the nucleus from the isomeric state 1 to the triggering level 3, one can expect that part of the nuclei will decay to state 2 (and further to the ground state if applicable) releasing their stored energy via spontaneous emission. However, such nuclear state control would depend on branching ratios of incoherent processes and its efficiency would therefore be low. The STIRAP technique presents the advantage that it can transfer population from state 1 to state 2 ideally with up to 100% efficiency.

Compared to transitions in atomic or molecular systems, nuclear transitions have much higher energies, typically in the x-ray or gamma-ray regime. The availability of suitable coherent light sources is limited. The operational X-ray Free Electron Laser (XFEL) sources at LCLS in the United States of America or at SACLA in Japan are limited to 10, respectively 20 keV photon energy and only partial temporal coherence. To bridge the gap between x-ray laser frequency and nuclear transition energies, a key proposal is to combine moderately accelerated target nuclei and novel x-ray lasers [3]. This allows for a match of the x-ray photon and nuclear transition frequency in the nuclear rest frame. Using this scenario, XFEL pulses could be used to implement quantum optics schemes for nuclear two- or three-level systems.

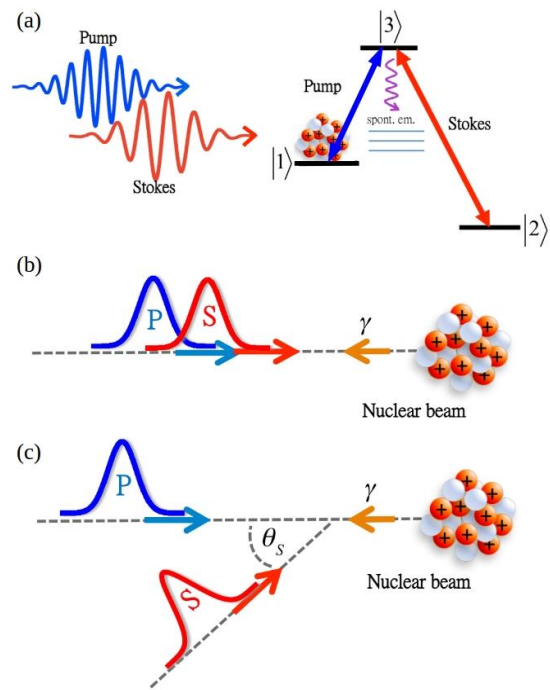

Figure 1- (a) Nuclear three-level system with initial population in the isomeric state 1. Depending on the availability of a two-color XFEL, one can envisage (b) a two-color setup with copropagating beams or (c) a one-color scheme with crossed beams in the laboratory frame. Reprinted with permission from Ref. [5]. Copyright 2013 by the American Physical Society.

A first theoretical study on the interaction of coherent x-ray laser pulses with a two-level nuclear system comprising the ground and first excited nuclear states in $^{223}$Ra was presented in Ref. [3]. The control of nuclear state population by STIRAP was addressed for the first time in Refs. [4,5]. Nuclear coherent population transfer via STIRAP requires two overlapping x-ray laser pulses to interact with a nuclear Λ-level scheme. Together with the accelerated ion beam, STIRAP therefore requires a three-beam setup [5] either in collinear geometry or envisaging two crossed photon beams meeting at an angle the ion beam [see Figs. 1(b) and (c)]. The copropagating beams scheme requires the existence of a two-color XFEL providing the P and S fields with different frequencies, but is less challenging as far as the spatial and temporal beam overlap is concerned. The crossed -beam setup can use the same XFEL frequency for both P and S pulses, however at the cost of a more complicated beam geometry. The three-beam setup STIRAP studies [4,5] were followed by more detailed works on nuclear STIRAP by a train of coincident pulses [6] and nuclear-state engineering in tripod systems using x-ray laser pulses [7].

What efficiency can be reached for nuclear STIRAP and what are the prospects for isomer depletion and control of nuclear state population? Refs. [4,5] could theoretically show that for nuclear transition energies of several hundred keV up to 1 MeV, fully temporally coherent XFEL sources with intensities of $10^{18} - 10^{20}$ W/cm² could transfer the nuclear population with efficiency up to 100% from one nuclear ground state to the other in the chosen three-level systems in $^{154}$Gd or $^{168}$Er isotopes. However, in order to reach maximum population transfer, the three-beam system needs careful optimization of the P and S pulse parameters. Fig. 2 gives an example of the role of the time delay between the two pulses for the required laser intensity in the crossed-beam setup. For higher intensities the typical wide plateau as the signature of STIRAP is clearly visible.

Allegedly, it appears to be more challenging to apply the STIRAP scheme to a realistic case of isomer depletion. The depletion of $^{97}$Tc and $^{113}$Cd isomers discussed in Refs. [4,5] shows that the required laser intensities for 100% nuclear coherent population transfer are up to four orders of magnitude larger than the ones calculated for $^{154}$Gd or $^{168}$Er due to weak transition probabilities for the considered transitions. One should however keep in mind that typically triggering levels high above isomeric states, that would present the advantage of larger linewidths and smaller required laser intensities, are less well known. A detailed analysis of nuclear data in the search for the best candidate is therefore required for successful isomer depletion.

### Current and Future Challenges

As discussed already in Section A1, a key requirement for STIRAP is the temporal coherence of the two pulses. The complete coherence of the two laser beams ensures also their mutual coherence, which is paramount for the creation of the dark state and the success of STIRAP.

This is by no means trivial for x-ray beams, which display rather poor temporal coherence – for instance, for a pulse duration of 100 fs at the LCLS, the estimated coherence time is only 0.2 fs.

A second challenging aspect is that a three-beam setup with coherent x-ray beams and strong ion acceleration is experimentally nothing but trivial. First, one needs a facility which can provide simultaneously the x-ray and ion beams with the required parameters. Second, the spatial and temporal coincidence of the three beams needs to be achieved.

Finally, better knowledge of nuclear level schemes involving isomeric states would be important in order to apply the STIRAP technique to nuclear three-level systems which can provide an isomer depletion scheme.

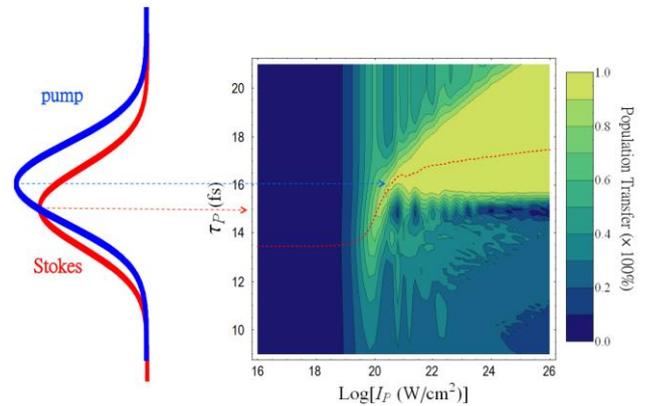

Figure 2 — The laser peak-intensity and pulse-delay dependent coherent population transfer for $^{154}$Gd for the crossed-beam scenario with seeded XFEL pulses. The Stokes peak position is fixed at 15 fs and the $I_S/I_P$ intensity ratio is 0.81, corresponding to a Rabi frequency ratio of 0.9. Reprinted figure with permission from Ref. [5]. Copyright 2013 by the American Physical Society.

### Advances in Science and Technology to Meet Challenges

There are two current approaches to improve the temporal coherence of XFEL pulses. One of them relies on an XFEL oscillator using diamond mirrors to create an x-ray cavity for the high-frequency pulse [8]. This would lead to temporal coherence times equal to the pulse duration length and much narrower bandwidth. A second approach envisages a so-called seeded XFEL. The x-ray pulse is in this case generated by using a monochromatized XFEL pulse to seed a second undulator and generate longitudinally coherent x-ray pulses. While the XFEL oscillator is still in its case study and planning phase, seeded XFEL radiation has already been demonstrated [9].

The simultaneous requirement of three beams remains up to date challenging. A potentially feasible scenario involves the use of table-top plasma-based acceleration to generate the nuclear beam, while the x-ray pulses would be produced by an XFEL source. Plasma ion acceleration was successfully demonstrated with compact shaped-foil-target systems, foil-and-gas targets and via radiation pressure acceleration and microlens beam focusing [10]. The relevant parameters for the accelerated ion beam are the divergence and ion velocity spread, with poor beam qualities reflecting upon the STIRAP efficiency. Dedicated feasibility studies should be performed to establish the minimal beam requirements for STIRAP depending on the chosen setup.

## Concluding Remarks

Nuclear coherent population transfer would bring the techniques of coherent control to the nuclear physics realm and potentially provide means to efficiently release the energy stored in nuclear isomers. Nuclear STIRAP is however experimentally challenging due to the stringent requirements on x-ray coherent beams together with ion beam acceleration. With the development of table-top coherent beam solutions and the commissioning of several new XFEL sources worldwide, it is however likely that first attempts to demonstrate the STIRAP technique on nuclear systems will soon become reality.

**Acknowledgments –** The authors would like to acknowledge the essential contribution of Wen-Te Liao to the development of the nuclear STIRAP project.